\documentclass{article}
\usepackage{setspace}
\usepackage{bm}
\usepackage{url}
\usepackage{cite}
\usepackage{amsmath,amssymb}
\usepackage{algorithm,algorithmic}
\usepackage{booktabs,enumitem}
\usepackage{multirow,array,stfloats}
\usepackage{epsfig,subfigure,epstopdf}
\DeclareMathOperator*{\argmax}{arg\,max}

\providecommand{\norm}[1]{\lVert#1\rVert}
\renewcommand{\vec}[1]{\mathbf{#1}}

\usepackage[preprint]{spconfa4}

\copyrightnotice{978-1-7281-1331-9/20/\$31.00 ©2020 IEEE}

\begin{document}\sloppy

% Title.
% ------
\title{
%\spaceskip=0.2em\relax
3D Dynamic Point Cloud Inpainting Via Temporal Consistency on Graphs
%\vspace{-0.05in}
}
%
% Address.
% ---------------
\name{Zeqing Fu, Wei Hu{\small $~^{\ast}$}, Zongming Guo\vspace{-0.1in}
\thanks{Corresponding author: Wei Hu (forhuwei@pku.edu.cn). This work was supported by National Natural Science Foundation of China [61972009] and Beijing Natural Science Foundation [4194080].}}
\address{Wangxuan Institute of Computer Technology, Peking University}

\maketitle

\begin{abstract}
\vspace{-0.03in}
With the development of 3D laser scanning techniques and depth sensors, 3D dynamic point clouds have attracted increasing attention as a representation of 3D objects in motion, enabling various applications such as 3D immersive tele-presence, gaming and navigation. However, dynamic point clouds usually exhibit holes of missing data, mainly due to the fast motion, the limitation of acquisition and complicated structure. Leveraging on graph signal processing tools, we represent irregular point clouds on graphs and propose a novel inpainting method exploiting both intra-frame self-similarity and inter-frame consistency in 3D dynamic point clouds. Specifically, for each missing region in every frame of the point cloud sequence, we search for its self-similar regions in the current frame and corresponding ones in adjacent frames as references. Then we formulate dynamic point cloud inpainting as an optimization problem based on the two types of references, which is regularized by a graph-signal smoothness prior. Experimental results show the proposed approach outperforms three competing methods significantly, both in objective and subjective quality.
\end{abstract}
\vspace{-0.03in}

\vspace{-0.1in}
\section{Introduction}
\label{sec:intro}
\vspace{-0.12in}

3D dynamic point cloud is a natural representation of arbitrarily-shaped 3D objects in motion. It consists of a sequence of point clouds, each of which is a set of points. Each point corresponds to a measurement point, which contains 3D geometric coordinates and possibly attributes such as color and normal. 
The acquisition of dynamic point clouds becomes convenient with the development of depth sensing and 3D laser scanning techniques\footnote{Examples include Microsoft Kinect, Intel RealSense, LiDAR, arrays of color plus depth (RGBD) video cameras~\cite{Loop13}, {\it etc.}}, thus enabling a variety of applications such as navigation in autonomous driving, animation, gaming and virtual reality~\cite{Tulvan16}.

\begin{figure}[h]
	\vspace{-0.1in}
	\centering
	\includegraphics[width=0.43\textwidth]{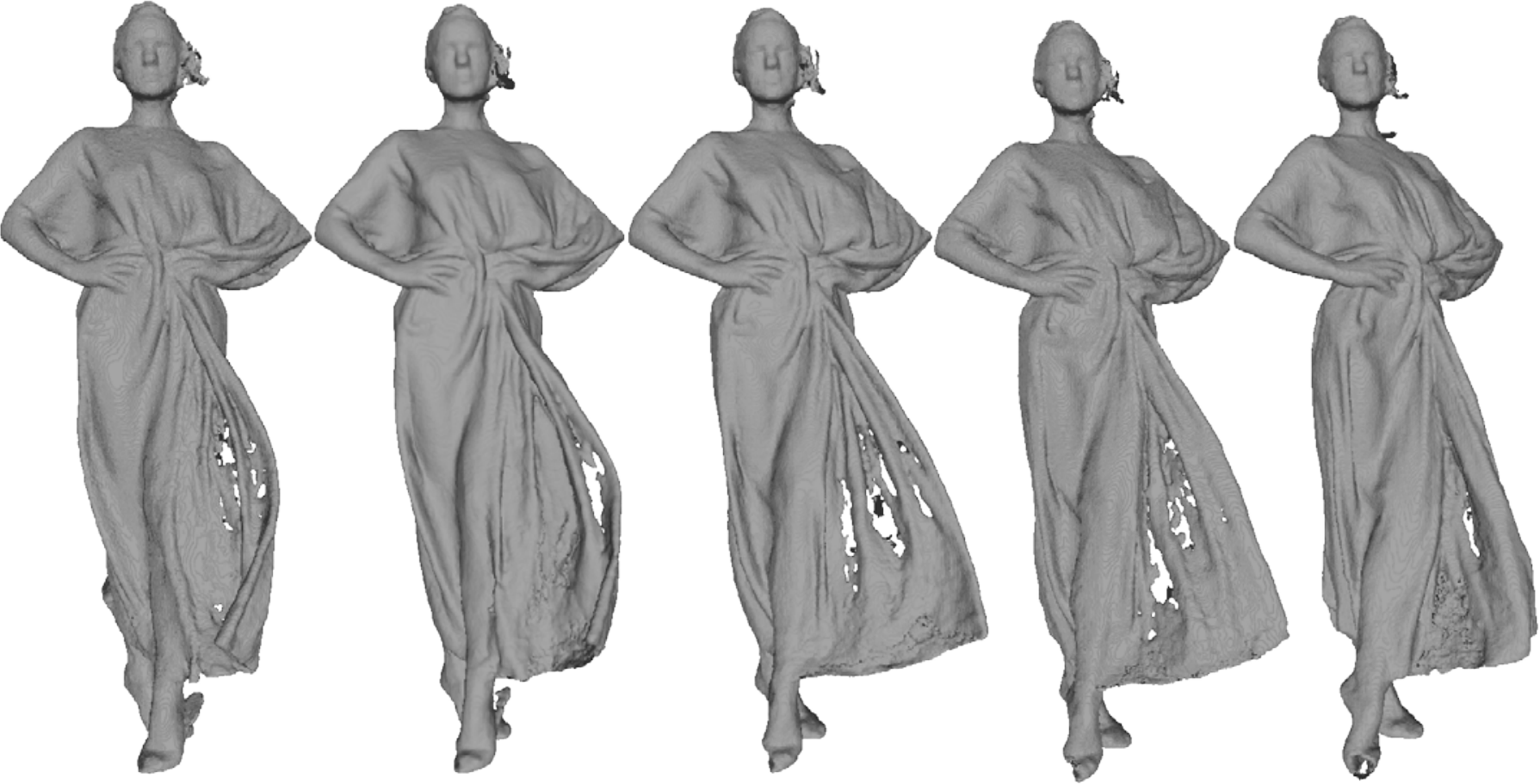}
	\vspace{-0.15in}
	\caption{Several frames of the 3D dynamic point cloud \textit{Longdress} with holes, captured at different moments.}
	\label{fig:intro}
	\vspace{-0.2in}
\end{figure}

Nevertheless, 3D dynamic point clouds often exhibit holes of missing data inevitably, as shown in Fig.~\ref{fig:intro}. This is mainly due to complicated object structure, fast object motion, inherent limitations of the acquisition equipments and incomplete scanning views. Besides, the object itself may have some missing regions ({\it e.g.}, dilapidated heritage). Therefore, it is necessary to inpaint incomplete point clouds prior to the subsequent applications. 
Nevertheless, 3D dynamic point cloud inpainting is challenging to address, because each frame is irregularly sampled with possibly different numbers of points. Further, there is no explicit temporal correspondence between points over time. 
% \footnote{For example, 3D laser scanning is less sensitive to the objects in dark colors. This is because the darker it is, the more wavelengths of light the surface absorbs and the less light it reflects. Thus the laser scanning devices are unable to receive enough reflected light for dark objects to recognize.}

However, the direct inpainting of 3D dynamic point cloud has been largely overlooked so far in the literature, while several approaches have been proposed for static point clouds. 
These methods mainly include two categories according to the cause of holes: 1) restore holes in the object itself such as heritage and sculptures~\cite{Chalmoviansk03,Sahay12,Sahay15,Setty15,Setty18,Dinesh17,Dinesh18}, and 2) inpaint holes caused by the limitation of scanning devices~\cite{Wang07,Quinsat15,Lozes14,Lozes15,Lin17,Muraki17,Fu18,Fu18TIP}. For the first class, the main hole-filling data source is online database, as the holes are often large. Sahay {\it et al.}~\cite{Sahay15} project the point cloud to a depth map, search a similar one from an online depth database via dictionary learning for inpainting, while the projection process inevitably introduces geometric loss. Instead, Dinesh {\it et al.}~\cite{Dinesh17,Dinesh18} search best matching regions in the object itself based on the smallest rotation difference to fill the missing area. The results still suffer from geometric distortion due to the simple data source.

\begin{figure*}[hb]
	\vspace{-0.15in}
	\centering
	\includegraphics[width=0.85\textwidth]{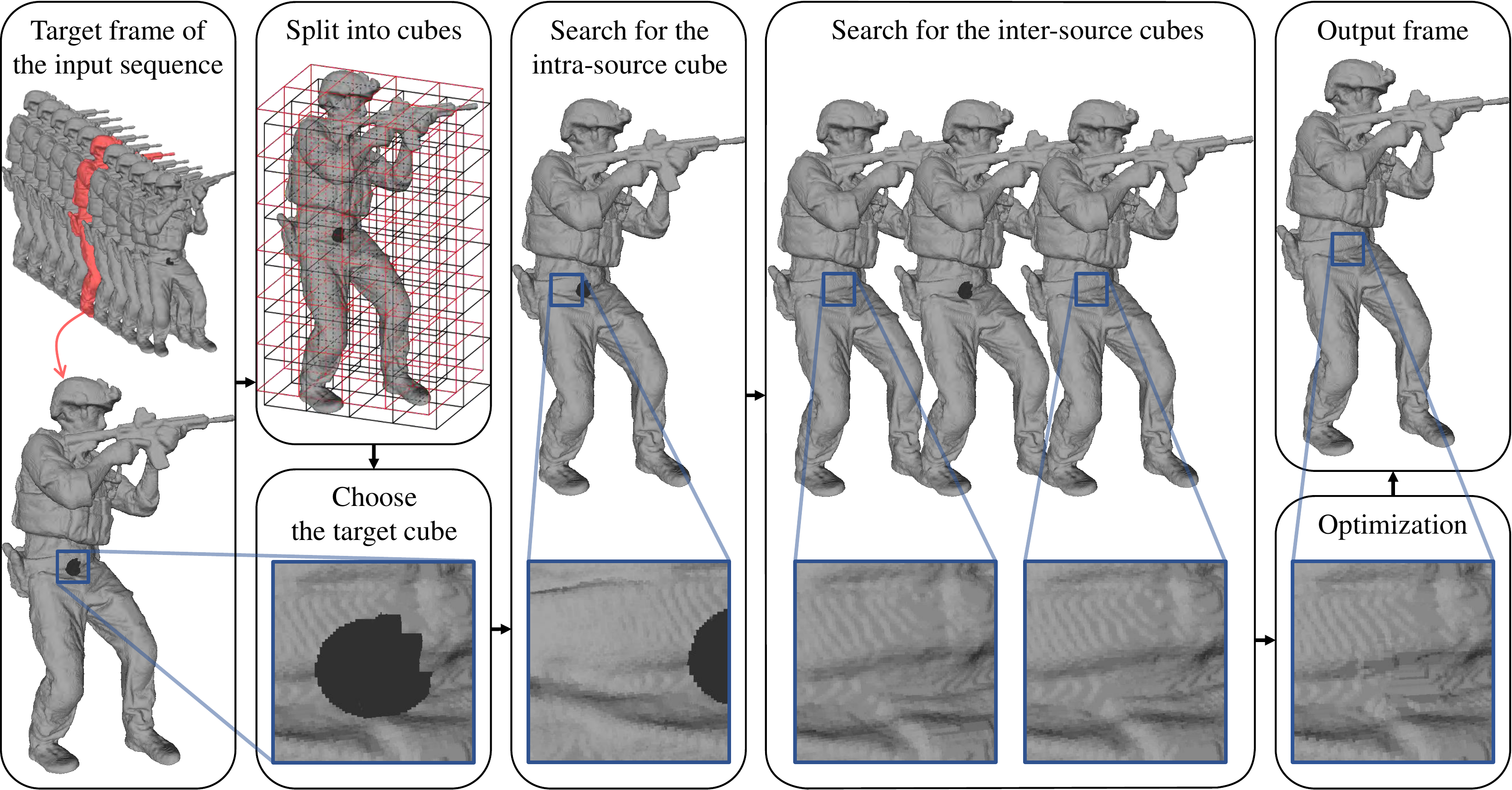}
	\vspace{-0.1in}
	\caption{The framework of the proposed 3D dynamic point cloud inpainting method.}
	\label{fig:pipe}
	\vspace{-0.17in}
\end{figure*}

The other class of methods focus on holes generated due to the limitations of scanning devices, which is smaller and more fragmentary than the aforementioned ones in general. Wang {\it et al.}~\cite{Wang07} and Quinsat {\it et al.}~\cite{Quinsat15} create a triangle mesh from the input point cloud and fill holes in the mesh for final interpolation. These methods rely on the quality of mesh construction though. Lozes {\it et al.}~\cite{Lozes14,Lozes15} deploy partial difference operaters to solve an optimization problem on the point cloud. Muraki {\it et al.}~\cite{Muraki17} generate a surface to fit the vicinity of the hole and interpolate the surface for inpainting. Due to the reference from local neighborhood only, their results tend to be more planar than the ground truth, and artifacts are likely to occur around the boundary in complicated structure. Fu {\it et al.}~\cite{Fu18,Fu18TIP} exploit the non-local similarity in the point cloud, which inpaints holes from their most similar regions based on normals of points.

%Besides, some works for static point clouds deal with particular point cloud data such as geometrically regular point clouds of buildings in~\cite{Friedman12}, flattened bar-shaped holes in the human body data in~\cite{Wu15}, and dynamic holes with static objects in~\cite{Chen16}, which are unsuitable for general cases though. 

Nevertheless, all above methods are not tailored for {\it dynamic} point clouds. If we apply them to a point cloud sequence frame by frame, the inpainting process of each frame is independent to each other, which neglects the inter-frame correlation and is thus sub-optimal. 
To this end, as an extension to the inpainting method in~\cite{Fu18TIP}, we exploit both \textit{intra-frame self-similarity} and \textit{inter-frame consistency} in the geometry for dynamic point cloud inpainting. 
For each target region which contains a hole in the target frame, our key idea is to search its most similar region in the target frame as well as its corresponding regions in the adjacent frames as source regions, and then inpaint the hole from these source regions.

Due to the irregularity of point clouds, it is difficult to search similar regions and fill holes according to the source regions. Hence, we resort to Graph Signal Processing~\cite{Shuman13}, and represent point clouds on graphs naturally. In particular, for each target frame in the input point cloud sequence with holes, we first segment it into cubes of the same size and choose the target cube with missing area. Secondly, we search the most similar cube to the target cube in the target frame as the \textit{intra-source cube} as in~\cite{Fu18TIP}. Thirdly, we search the corresponding cubes of the target cube in the previous and subsequent frames as the \textit{inter-source cubes}. The correspondence is set based on searching for a cube that contains the most nearest neighbors of the points in the target cube in the relative location. Next, we formulate the hole-filling step as an optimization problem, which is based on both intra- and inter-source cubes and regularized by a graph-signal smoothness prior~\cite{Liu15,Dong16,Hu18} for the target cube. Finally, we acquire the closed-form solution of the optimization problem, leading to the inpainted result.
Experimental results show that we outperform separate inpainting from state-of-the-art methods. 

%The outline of the paper is as follows. We first review graph signal processing tools in Section~\ref{sec:graph}. Next, in Section~\ref{sec:method}, we introduce the proposed method, including preprocessing, intra- and inter-source cube matching and problem formulation. Experimental results and conclusions are presented in Section~\ref{sec:results} and \ref{sec:conclude}, respectively.

\vspace{-0.1in}
\section{Spectral Graph Theory}
\label{sec:graph}
\vspace{-0.1in}

%We provide a review on basic concepts in spectral graph theory~\cite{Chung96}, including graph, graph Laplacian, graph-signal smoothness prior, which will be utilized in the proposed dynamic point cloud inpainting.

%\vspace{-0.12in}
%\subsection{Graph and Graph Laplacian}
%\label{subsec:knn}
%\vspace{-0.06in}
We consider an undirected graph $ \mathcal{G}=\{\mathcal{V},\mathcal{E},\mathbf{W}\} $ composed of a vertex set $ \mathcal{V} $ of cardinality $|\mathcal{V}|=N$, an edge set $ \mathcal{E} $ connecting vertices, and a weighted \textit{adjacency matrix} $ \mathbf{W} $. $ \mathbf{W} $ is a real symmetric $ N \times N $ matrix, where $ w_{i,j} $ is the weight assigned to the edge $ (i,j) $ connecting vertices $ i $ and $ j $. For example, $ K $-Nearest Neighbor ($ K $-NN) graph is a commonly used undirected graph, which is constructed by connecting each point with its nearest $ K $ neighbors.

The Laplacian matrix is defined from the adjacency matrix~\cite{Chung96}. Among different variants of Laplacian matrices, the \textit{combinatorial graph Laplacian} used in~\cite{Shen10,Hu12,Hu15} is defined as $ \mathcal{L}:=\mathbf{D}-\mathbf{W} $, where $ \mathbf{D} $ is the \textit{degree matrix}---a diagonal matrix where $ d_{i,i}=\sum_{j=1}^N w_{i,j} $.

%\vspace{-0.12in}
%\subsection{Graph-Signal Smoothness Prior}
%\label{subsec:prior}
%\vspace{-0.06in}
Graph signal refers to data residing on the vertices of a graph. For example, if we construct a $ K $-NN graph on the point cloud, then the normal or the coordinate of each point can be treated as graph signal defined on the $ K $-NN graph. This will be discussed further in the proposed spatio-temporal graph construction in Section~\ref{subsec:constru}. A graph signal $ \vec{z} $ defined on a graph $ \mathcal{G} $ is smooth with respect to the topology of $ \mathcal{G} $ if
\vspace{-0.1in}
\begin{equation}
	\sum\limits_{i \sim j}w_{i,j}(z_i - z_j)^2 < \epsilon, ~~\forall i,j,
	\label{eq:prior}
    \vspace{-0.13in}
\end{equation}
where $ \epsilon $ is a small positive scalar, and $ i \sim j $ denotes two vertices $ i $ and $ j $ are one-hop neighbors in the graph. In order to satisfy (\ref{eq:prior}), $ z_i $ and $ z_j $ have to be similar for a large edge weight $ w_{i,j} $, and could be quite different for a small $ w_{i,j} $. Hence, (\ref{eq:prior}) enforces $ \vec{z} $ to adapt to the topology of $ \mathcal{G} $, which is thus coined \textit{graph-signal smoothness prior}.

As $ \vec{z}^T \mathcal{L} \vec{z} = \sum\limits_{i \sim j}w_{i,j}(z_i - z_j)^2 $ \cite{Spielman04}, (\ref{eq:prior}) is concisely written as $ \vec{z}^T \mathcal{L} \vec{z} < \epsilon $ in the sequel. This prior will be deployed in our problem formulation of point cloud inpainting as a regularization term, as discussed in Section~\ref{subsec:formula}.

\vspace{-0.1in}
\section{The Proposed Inpainting Method}
\label{sec:method}
\vspace{-0.1in}
Leveraging on spectral graph theory, we introduce the proposed point cloud inpainting method from both intra-frame self-similarity and inter-frame consistency. 
The input is a point cloud sequence of $q$ frames denoted by $ \mathbf{S}=\{\mathbf{P}_1,\mathbf{P}_2,...,\mathbf{P}_q\} $, where $ \mathbf{P}_f, f=1,...,q $ denotes each frame of point cloud in the sequence. As shown in Fig.~\ref{fig:pipe}, we inpaint each target frame $ \mathbf{P}_f $ with holes in order. 
Firstly, we split $ \mathbf{P}_f $ into fixed-sized cubes as processing units in the subsequent steps and choose the target cube with missing area. 
Secondly, we search for the most similar cube to the target cube in $ \mathbf{P}_f $ as the \textit{intra-source cube}. 
Thirdly, we search for the corresponding cubes to the target cube in $ \mathbf{P}_{f-1} $ and $ \mathbf{P}_{f+1} $, which are referred to as the \textit{inter-source cubes}. 
Fourthly, we formulate dynamic point cloud inpainting as an optimization problem, which poses the graph-signal smoothness prior via the intra-source cube and enforces consistency with the inter-source cubes. 
We derive the closed-form solution of the optimization problem, leading to the resulting cube. 
Finally, we replace the target cube with the resulting cube as the output.

\vspace{-0.12in}
\subsection{Preprocessing}
\label{subsec:prepro}
\vspace{-0.06in}
For each target frame of point cloud $ \mathbf{P}_f=\{\vec{p}_1,\vec{p}_2, ...\} $ with $ \vec{p}_i\in \mathbb{R}^3 $ meaning the coordinates of the $ i $-th point in the point cloud, we first split $ \mathbf{P}_f $ into overlapping cubes $ \{\vec{c}_1,\vec{c}_2, ... $ $ \} $ with $ \vec{c}_i\in \mathbb{R}^{M^3 \times 3} $ ($ M $ is the size of the cube), as the processing unit of the proposed inpainting algorithm.
%$ M $ is empirically set according to the coordinate range of $ \mathbf{P}_f $ ($ M=20 $ in our experiments), while the overlapping step is empirically set as $ \frac{M}{4} $. This is a trade-off between the computational complexity and ensuring enough geometry information available to search for source cubes.
Then we choose the cube with missing data as the target cube $ \vec{c}_t $.
%Further, in the presence of a hole larger than the cube size, we will divide the hole into several small holes, and then inpaint the small holes in the inward order.
Next, we obtained the intra-source cube $ \vec{c}_s $ as in~\cite{Fu18TIP} based on the direct component (DC) and the anistropic graph total variation (AGTV) of the normals
of points in the cube. Further, we also perform structure matching ({\it i.e.}, coarse registration) for $ \vec{c}_s $ and $ \vec{c}_t $ so as to match the relative locations as in~\cite{Fu18TIP}, which includes both translation and rotation as a simplified Iterative Closest Points (ICP) operation~\cite{Besl02,Chetverikov02}. This leads to the final intra-source cube, denoted as $ \hat{\vec{c}}_s $, which will be adopted in the final inpainting step.

\vspace{-0.12in}
\subsection{Inter-frame Cube Matching}
\label{subsec:inter}
\vspace{-0.06in}

\begin{figure}[h]
    \vspace{-0.1in}
	\centering
	\includegraphics[width=0.45\textwidth]{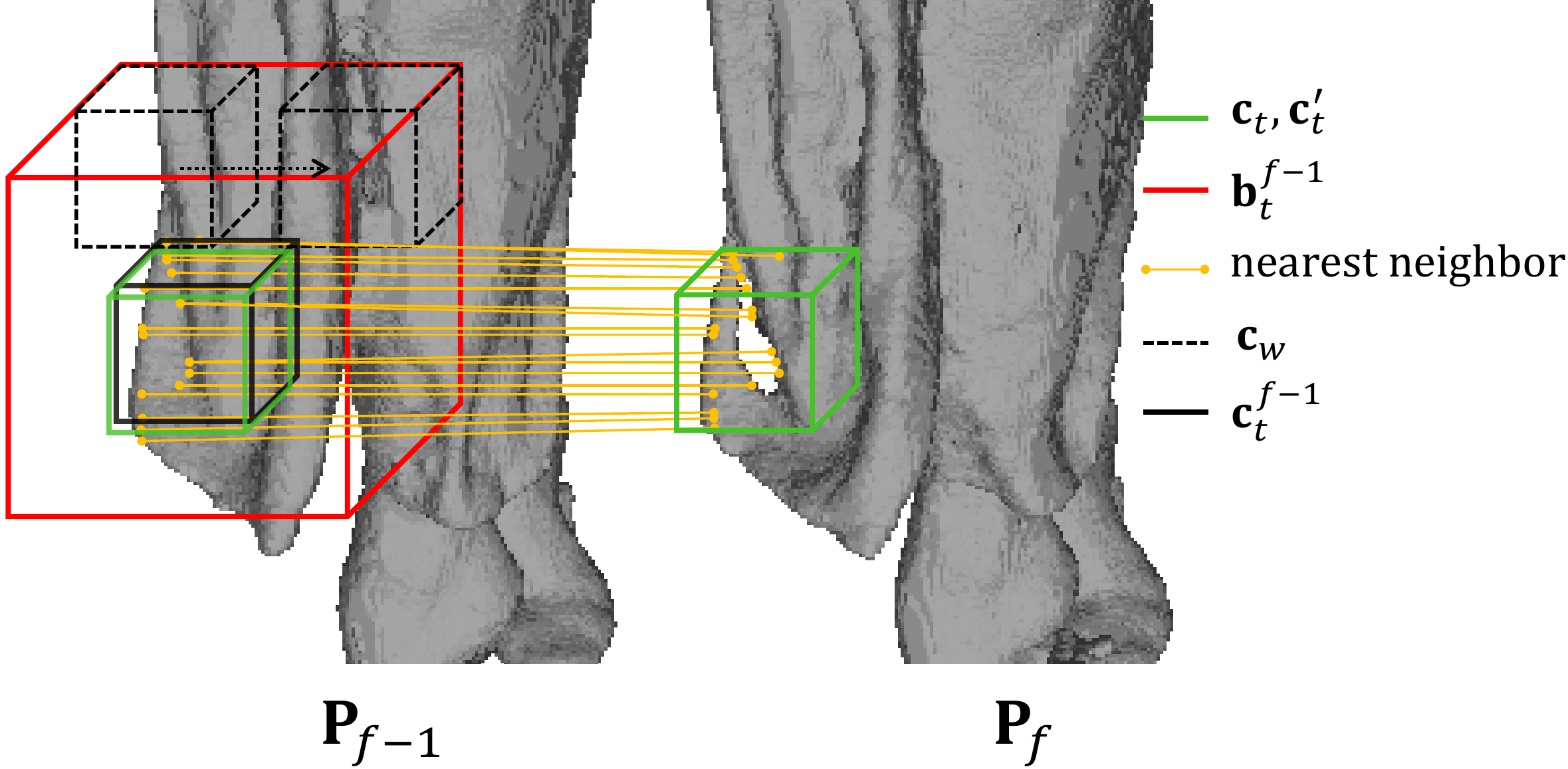}
    \vspace{-0.17in}
	\caption{The inter-source cube searching.}
	\label{fig:box}
    \vspace{-0.12in}
\end{figure}

Considering that the inpainted results of dynamic point clouds should be coherent among consecutive frames, it is necessary to explore the temporal correspondence between neighboring frames in a point cloud sequence. Unlike videos, dynamic point clouds are irregular, thus the temporal correspondence is challenging to define.

In order to efficiently explore the temporal correspondence in dynamic point clouds for coherent inpainting, we propose to find corresponding cubes for $ \vec{c}_t $ both in $ \mathbf{P}_{f-1} $ and $ \mathbf{P}_{f+1} $, which are denoted by $ \vec{c}_t^{f-1} $ and $ \vec{c}_t^{f+1} $ respectively as the inter-source cubes. Note that, the inter-frame consistency can be generalized to several previous and subsequent frames, instead of one forward and one backward as in our method.

Specifically, inspired by the observation that a set of points representing the same region have little variation in consecutive frames, we find the temporal correspondence via searching the nearest neighbor of each point in the target cube. As shown in Fig.~\ref{fig:box}, we first find a co-located cube $ \vec{c}'_t\in \mathbb{R}^{M^3 \times 3} $ in $ \mathbf{P}_{f-1} $ as $ \vec{s}(\vec{c}'_t)=\vec{s}(\vec{c}_t) $, where $ \vec{s}(\cdot) $ denotes the coordinate of the centering point of a cube. 
% This means $ \vec{c}'_t $ is at the same position as $ \vec{c}_t $ in the relative location.

Then we construct a bounding box $ \vec{b}_t^{f-1}\in \mathbb{R}^{H^3 \times 3} $ ($ H $ is the size of the box) around $ \vec{c}'_t $ as $ \vec{s}(\vec{b}_t^{f-1})=\vec{s}(\vec{c}'_t) $, which serves as the search range. Next, we search the nearest neighbor in $ \vec{b}_t^{f-1} $ of each point in $ \vec{c}_t $ in terms of the relative location.
Then we employ a sliding cubic window $ \vec{c}_w\in \mathbb{R}^{M^3 \times 3} $ with stride $ 1 $ in the bounding box $ \vec{b}_t^{f-1} $ to search for the inter-source cube $ \vec{c}_t^{f-1}\in \mathbb{R}^{M^3 \times 3} $ in $ \mathbf{P}_{f-1} $:
\vspace{-0.1in}
\begin{equation}
    \vec{c}_t^{f-1}=\argmax_{\vec{c}_w} V(\vec{c}_w),
    \vspace{-0.13in}
\end{equation}
where $ V(\vec{c}_w) $ is the number of the nearest neighbors of $ \vec{c}_t $ in $ \vec{c}_w $ in terms of the relative location. That is, $\vec{c}_t^{f-1}$ contains the most nearest neighbors of points in $ \vec{c}_t $.

We further perform the same structure matching on $ \vec{c}_t^{f-1} $ as the way we deal with $ \vec{c}_s $ in Section~\ref{subsec:prepro}, which leads to the final inter-source cube in $ \mathbf{P}_{f-1} $, denoted as $ \hat{\vec{c}}_t^{f-1} $. The final inter-source cube in $ \mathbf{P}_{f+1} $, denoted by $ \hat{\vec{c}}_t^{f+1} $, is searched in the same way as in $ \mathbf{P}_{f-1} $. Thus we obtain two source cubes as the temporal reference, which will be adopted in the final inpainting step.

\vspace{-0.12in}
\subsection{Spatio-Temporal Graph Construction}
\label{subsec:constru}
\vspace{-0.06in}
To take advantage of both intra-frame self-similarity and inter-frame consistency, we construct a spatial-temporal graph on the target cube $ \vec{c}_t $ as the following.

We first build {\it spatial} connectivities within $ \vec{c}_t $. 
As there exists a hole in $ \vec{c}_t $, we approximate the connectivities via the similarities in its intra-source cube $ \hat{\vec{c}}_s $.   
We choose to build a $ K $-NN graph mentioned in Section~\ref{sec:graph}, based on the affinity of geometric distance among points in $ \hat{\vec{c}}_s $. 
Specifically, the edge weight $ w_{k,l} $ between nodes $k$ and $l$ in $ \hat{\vec{c}}_s $ is assigned as
\vspace{-0.12in}
\begin{equation}
    w_{k,l}=
	\begin{cases}
		~\text{exp}\{-\frac{\norm{\vec{p}_k-\vec{p}_l}_2^2}{2\sigma^2}\}, & k \sim l\\
		~0, & \text{otherwise}
	\end{cases}
	\label{eq:knnweight}
	\vspace{-0.1in}
\end{equation}
where $k \sim l$ means nodes $k$ and $l$ are $K$ nearest neighbors and thus connected. 
$\sigma$ is a weighting parameter (empirically $\sigma=1$ in our experiments). 
This is based on the assumption that geometrically closer points are more similar in general. 
% which can better describe the sparsity of the overall graph gradient in normals of the points on the graph.

We then construct {\it temporal} connectivities between $ \vec{c}_t $ and its previous frame $ \hat{\vec{c}}_t^{f-1} $, as well as temporal connectivities between $ \vec{c}_t $ and its subsequent frame $ \hat{\vec{c}}_t^{f+1} $. 
Due to the hole in $ \vec{c}_t $, the connectivities for known points and unknown points in $ \vec{c}_t $ are linked in different manners. 
Specifically, we link each known point in $ \vec{c}_t $ with its corresponding point in $ \hat{\vec{c}}_t^{f-1} $. 
To circumvent the unavailability of unknown points in $ \vec{c}_t $, we approximate the temporal connectivities by the similarities between their corresponding points in the intra-source cube $ \hat{\vec{c}}_s $ and their corresponding points in the inter-source cube $ \hat{\vec{c}}_t^{f-1} $. 
The temporal correspondence is based on the the nearest neighbor in the relative location in the cube.

\begin{figure*}[hb]
	\vspace{-0.1in}
	\centering
	\subfigure[f=4]{
		\begin{minipage}[b]{0.098\textwidth}
			\includegraphics[height=0.09\textheight]{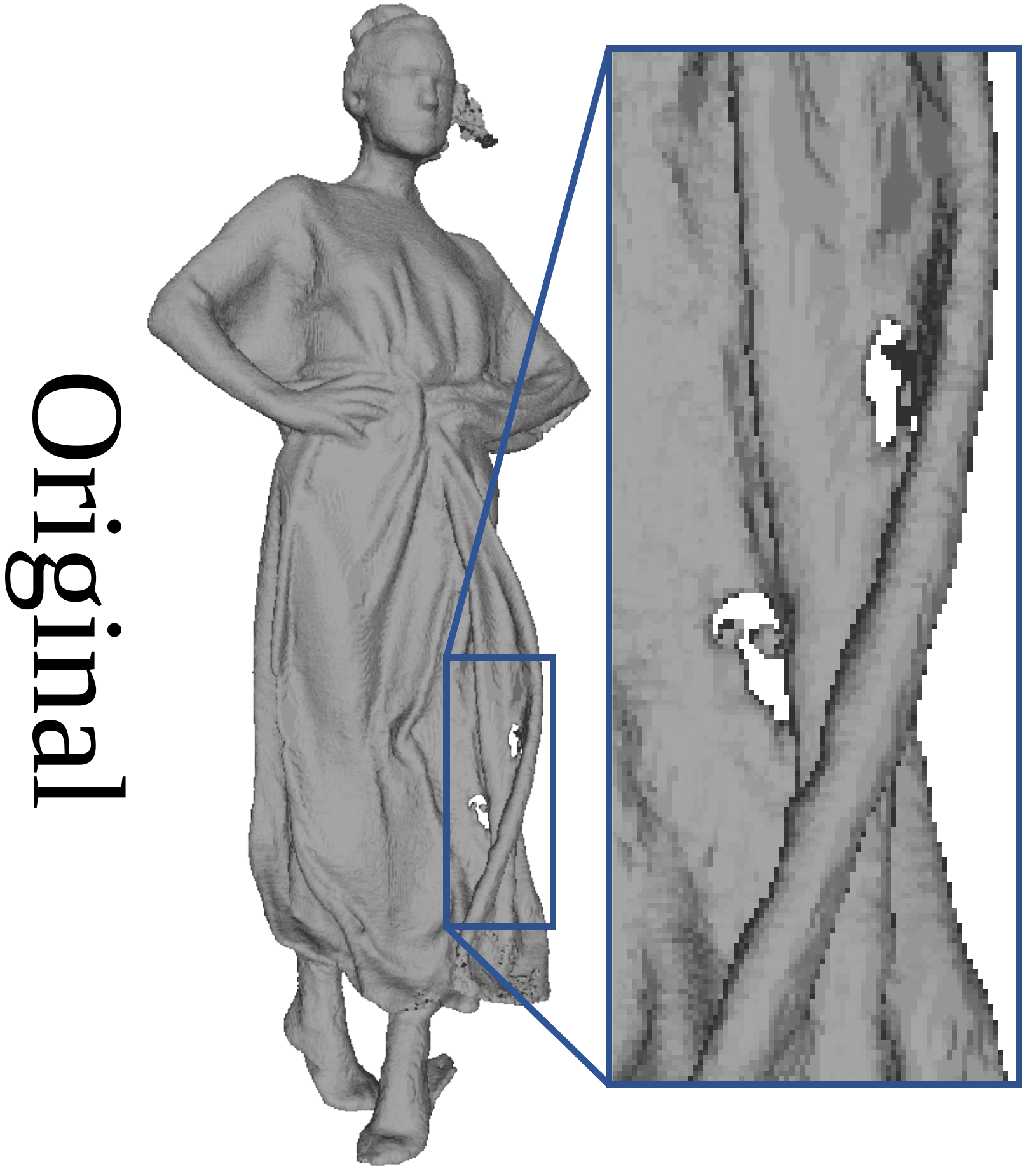} \\
			\includegraphics[height=0.09\textheight]{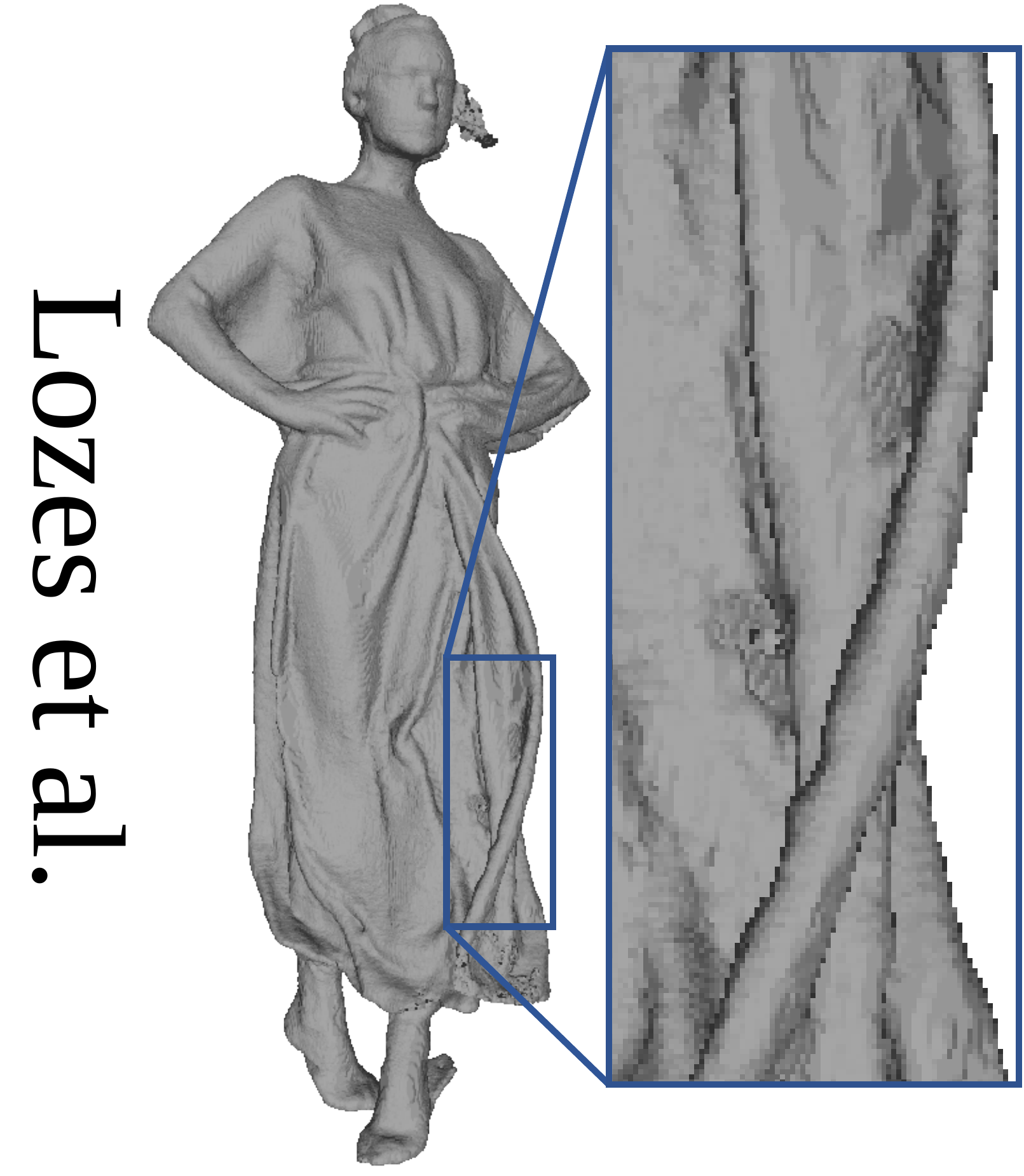} \\
			\includegraphics[height=0.09\textheight]{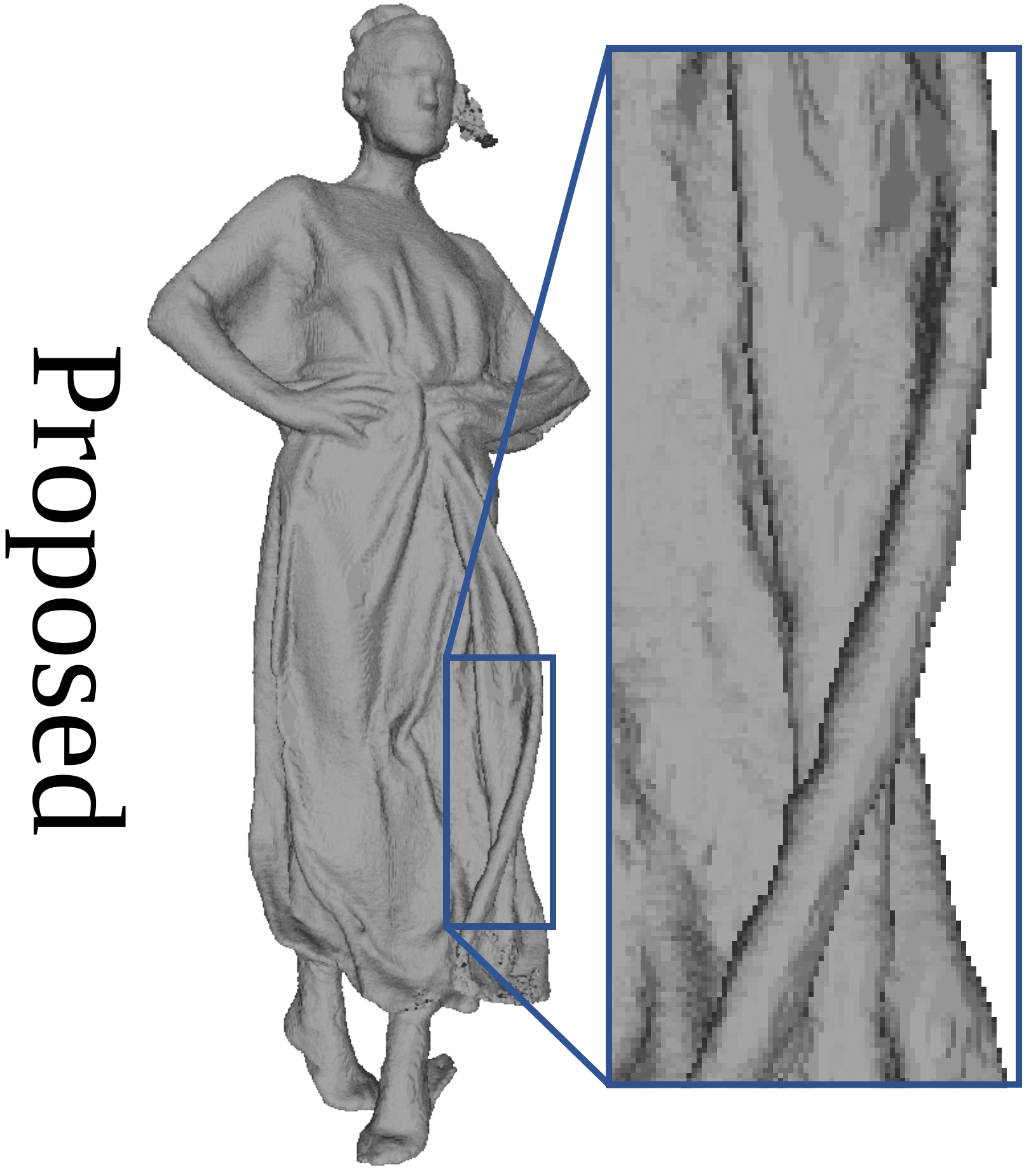} \\
		\vspace{-0.15in}
		\end{minipage}
	}
	\hspace{0.05in}
	\subfigure[f=5]{
		\begin{minipage}[b]{0.094\textwidth}
			\includegraphics[height=0.09\textheight]{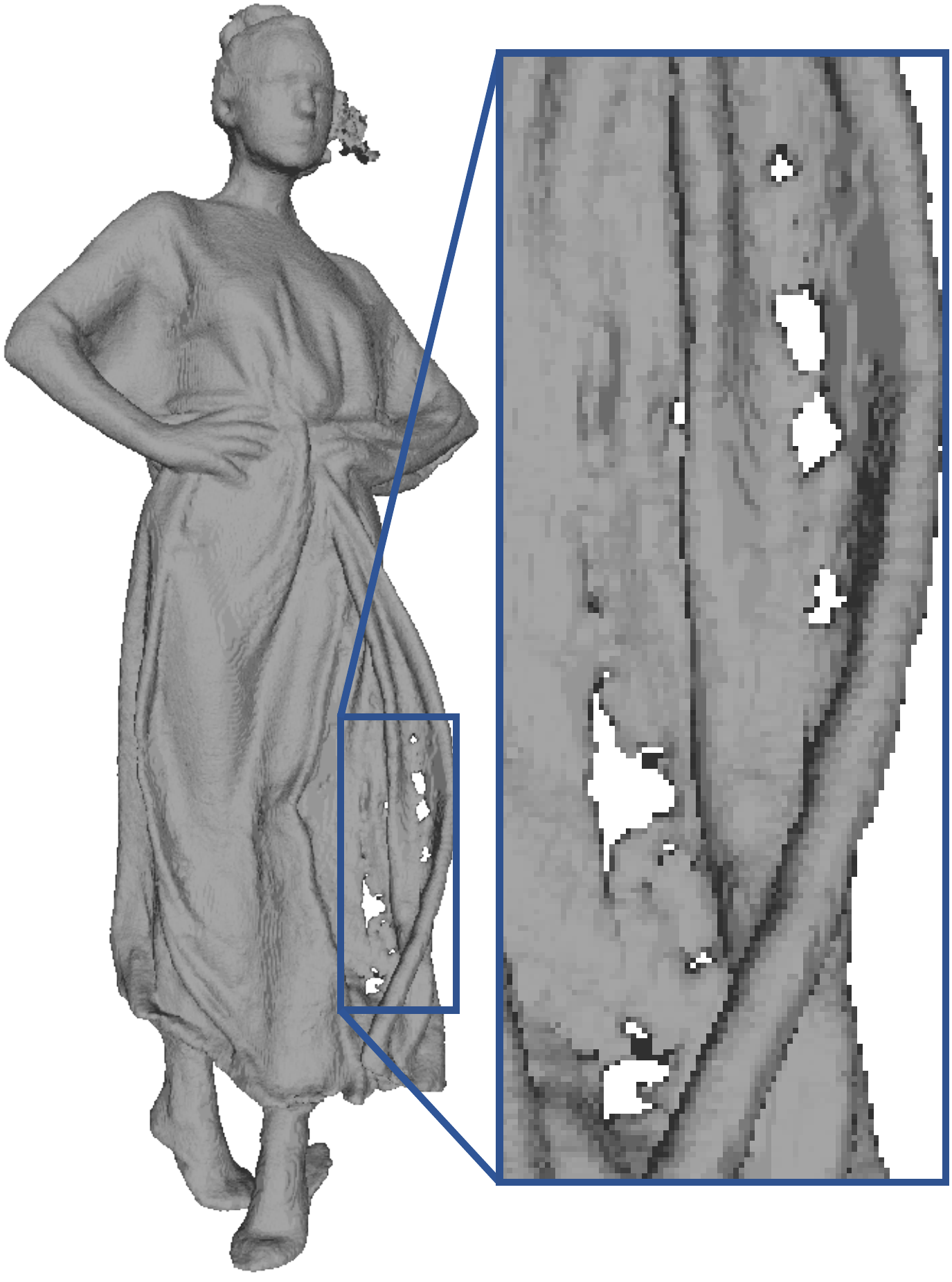} \\
			\includegraphics[height=0.09\textheight]{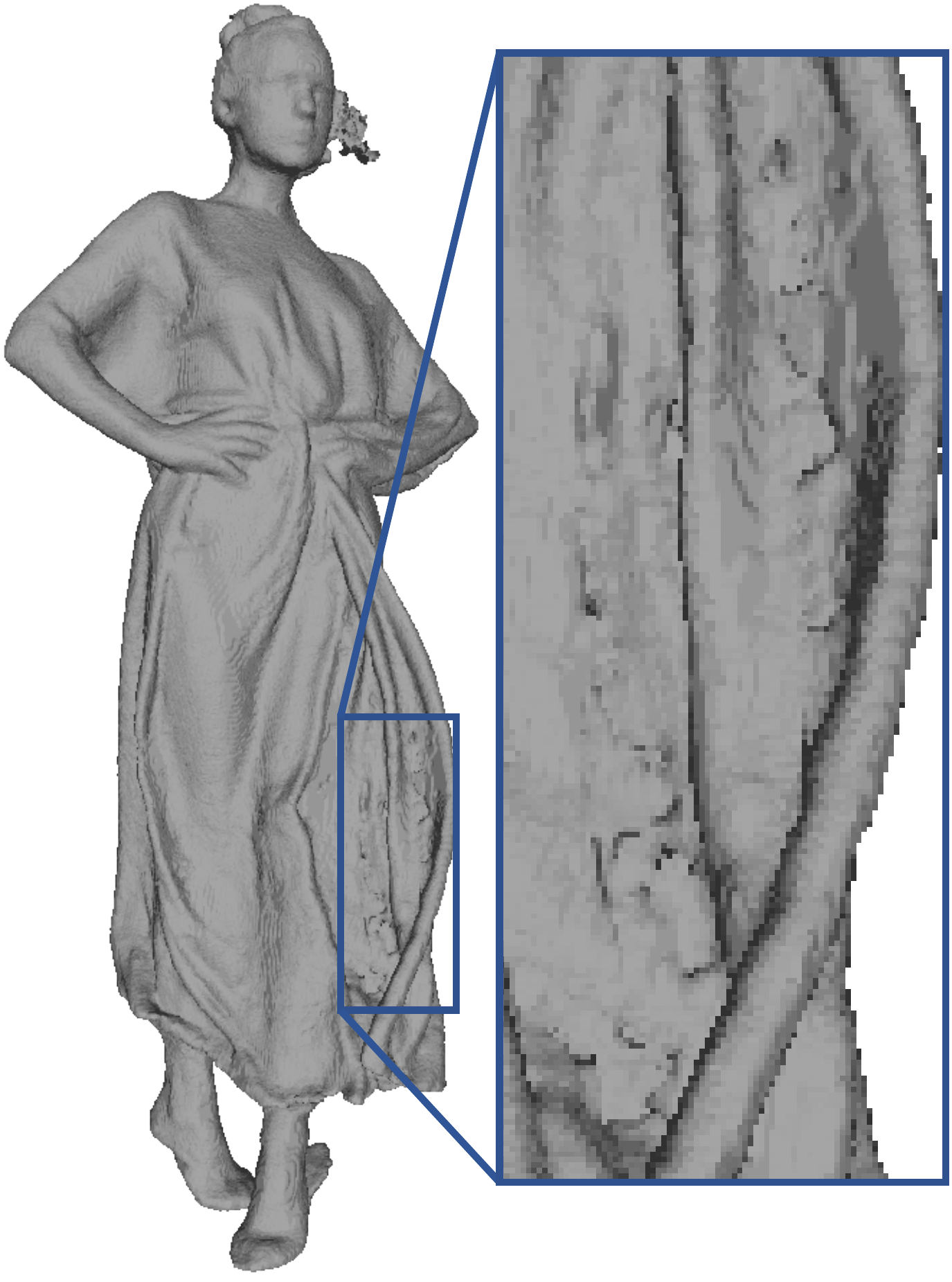} \\
			\includegraphics[height=0.09\textheight]{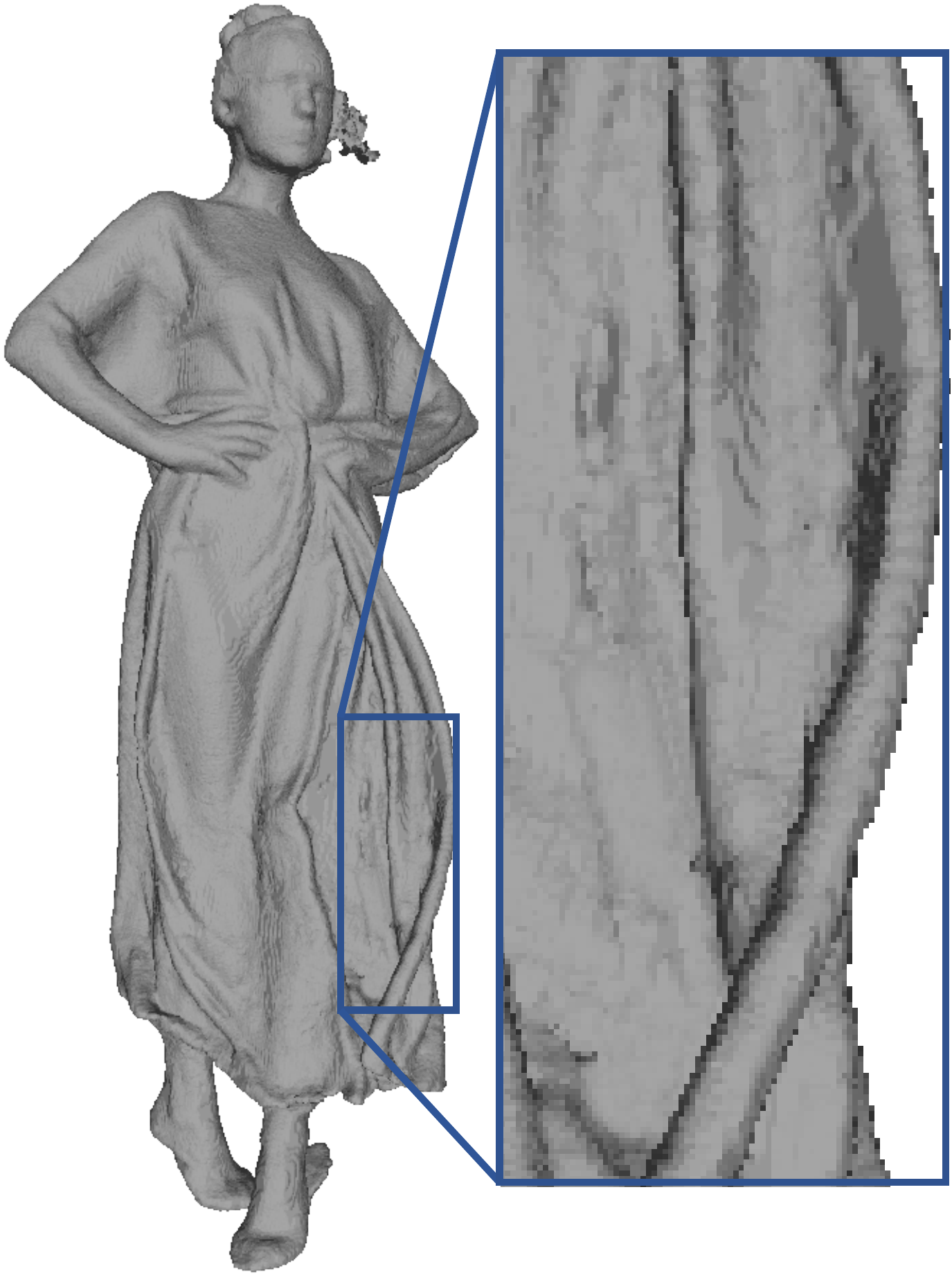} \\
		\vspace{-0.15in}
		\end{minipage}
	}
	\hspace{-0.05in}
	\subfigure[f=6]{
		\begin{minipage}[b]{0.094\textwidth}
			\includegraphics[height=0.09\textheight]{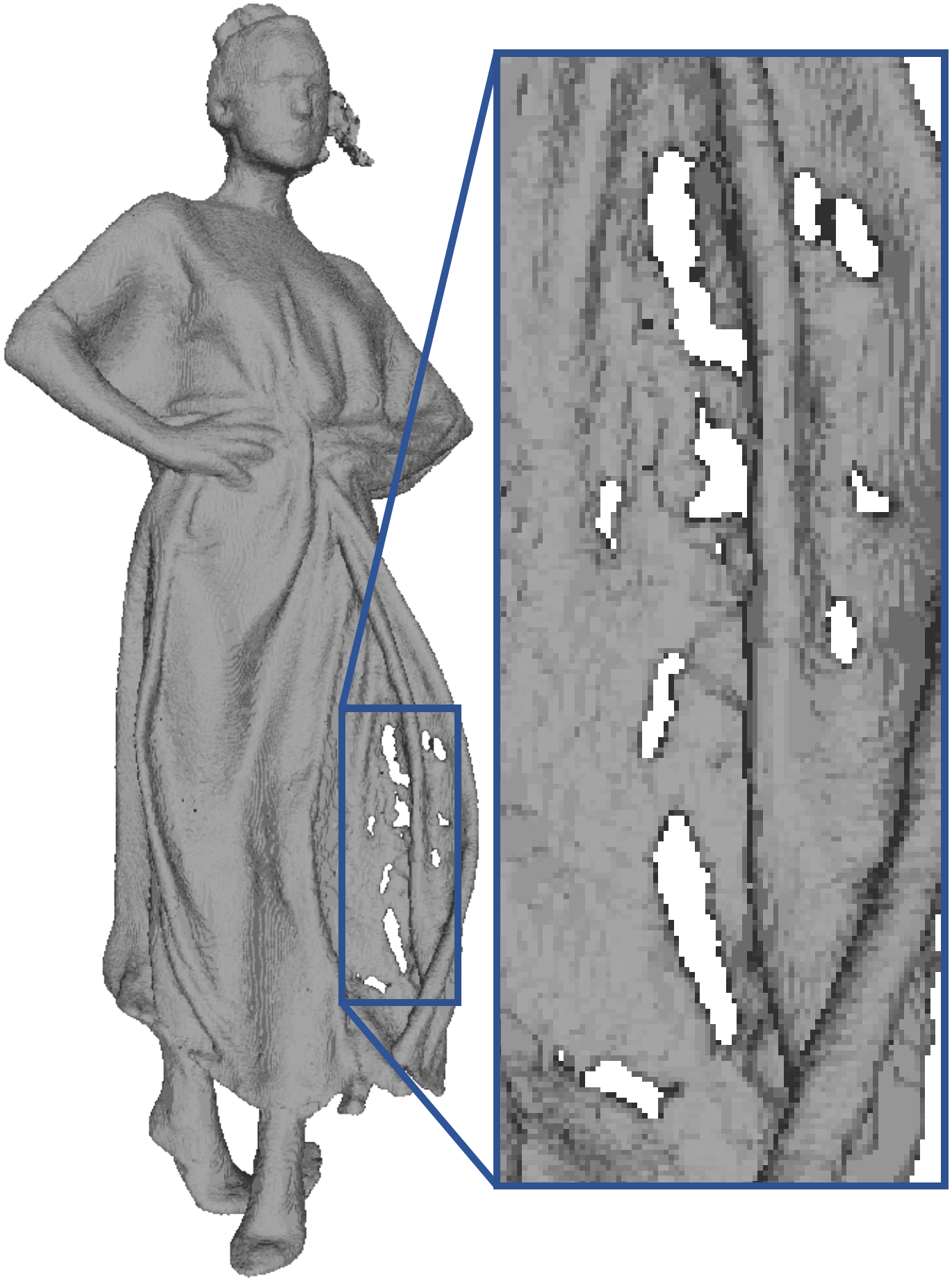} \\
			\includegraphics[height=0.09\textheight]{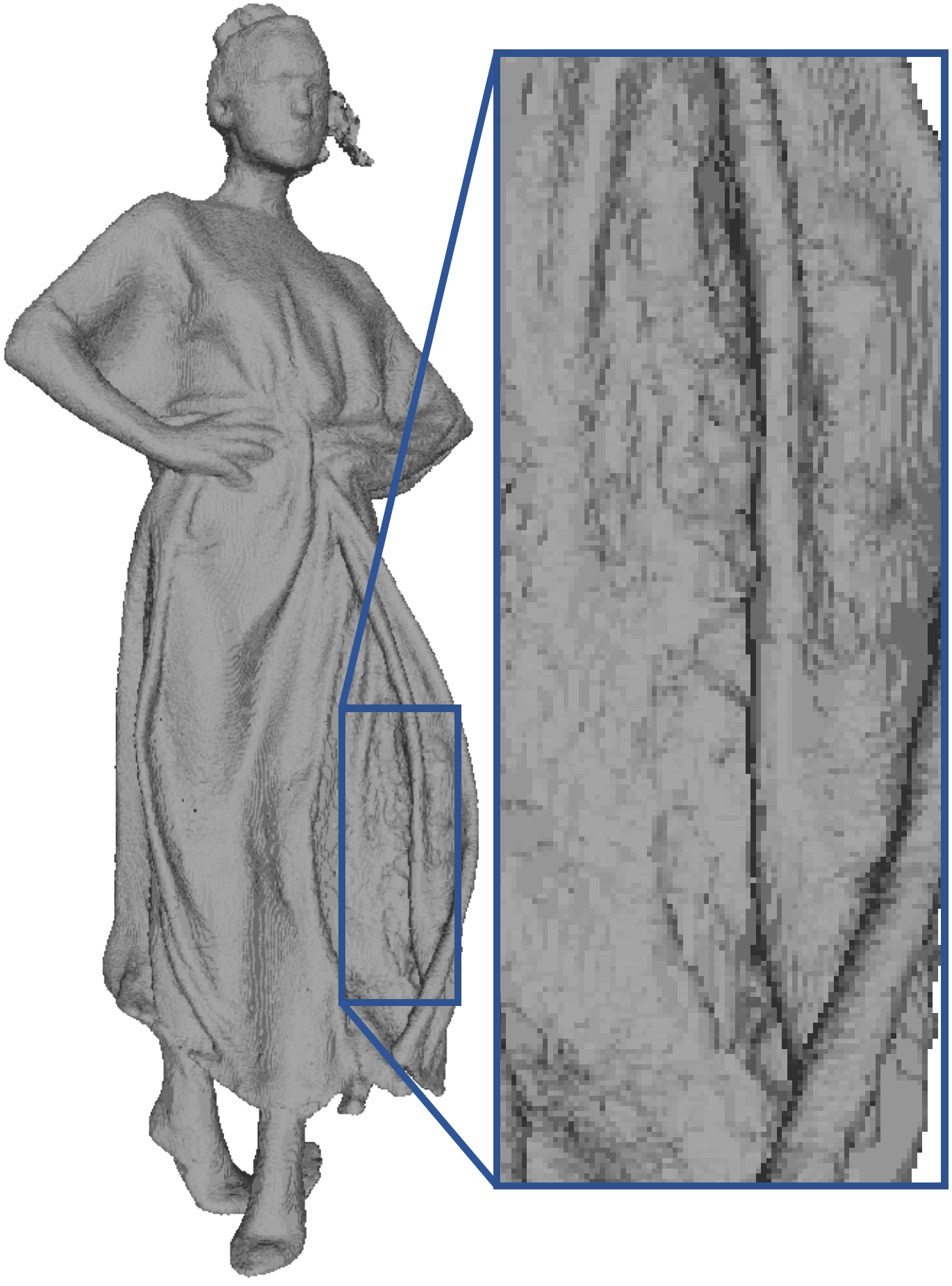} \\
			\includegraphics[height=0.09\textheight]{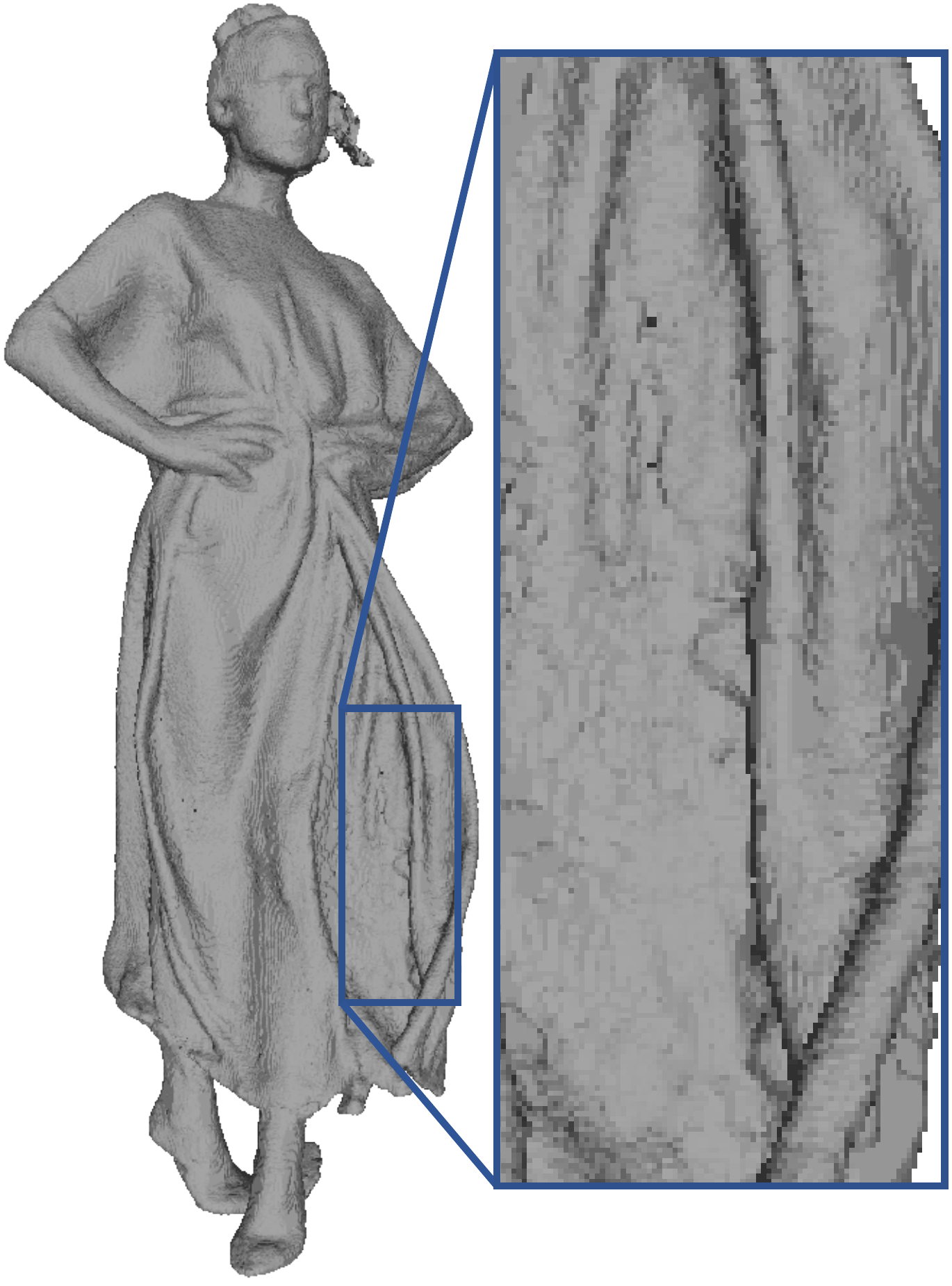} \\
		\vspace{-0.15in}
		\end{minipage}
	}
	\hspace{-0.05in}
	\subfigure[f=7]{
		\begin{minipage}[b]{0.095\textwidth}
			\includegraphics[height=0.09\textheight]{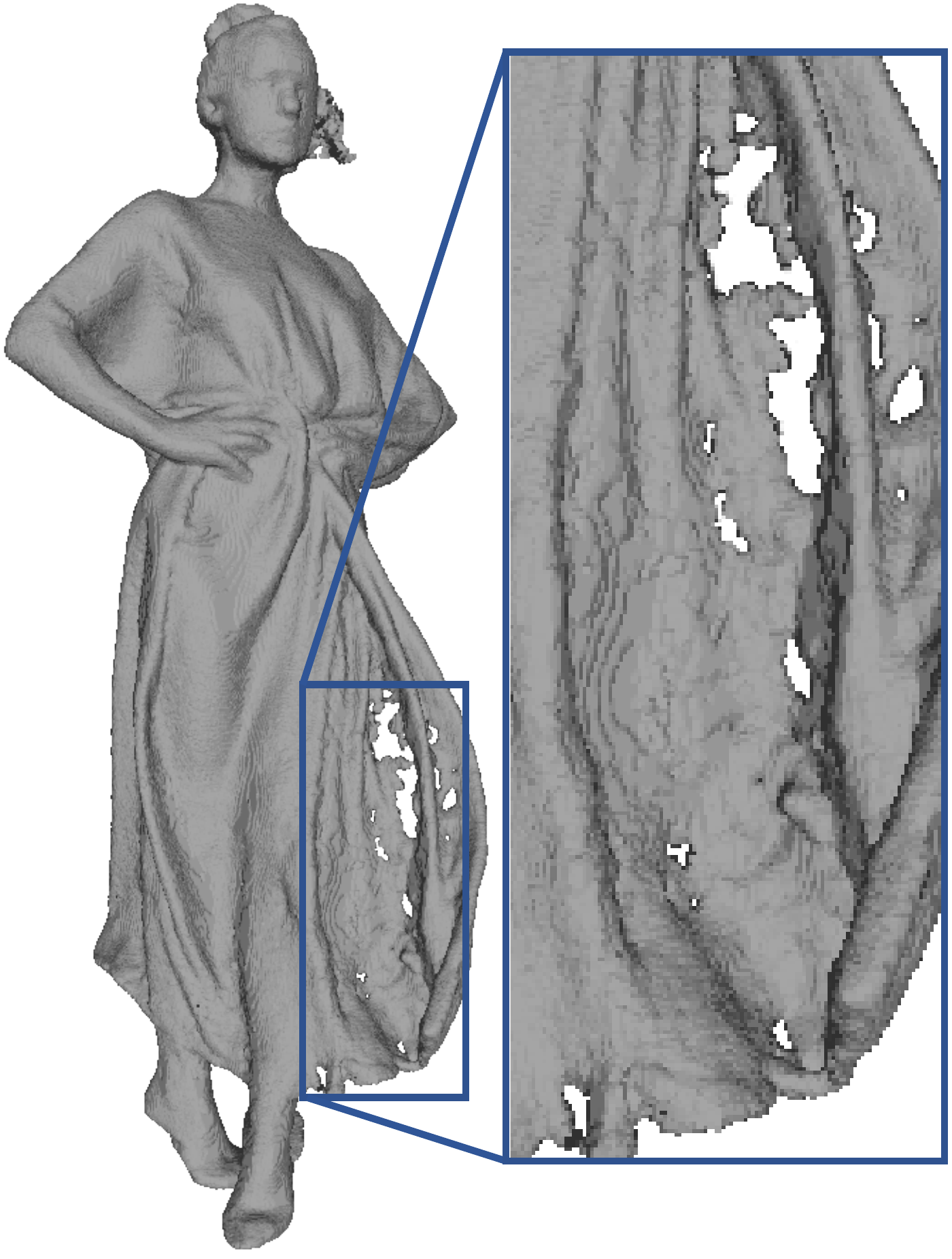} \\
			\includegraphics[height=0.09\textheight]{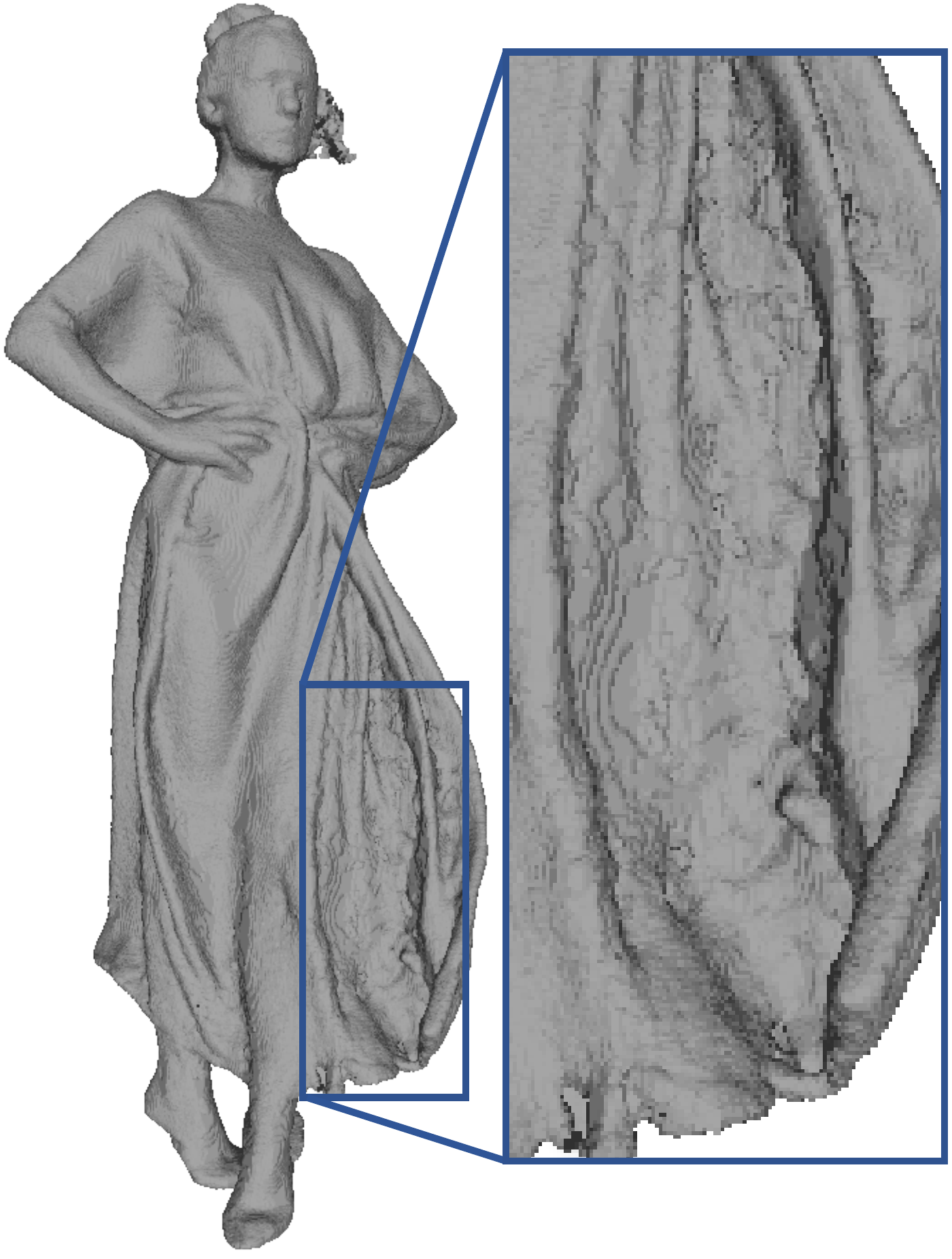} \\
			\includegraphics[height=0.09\textheight]{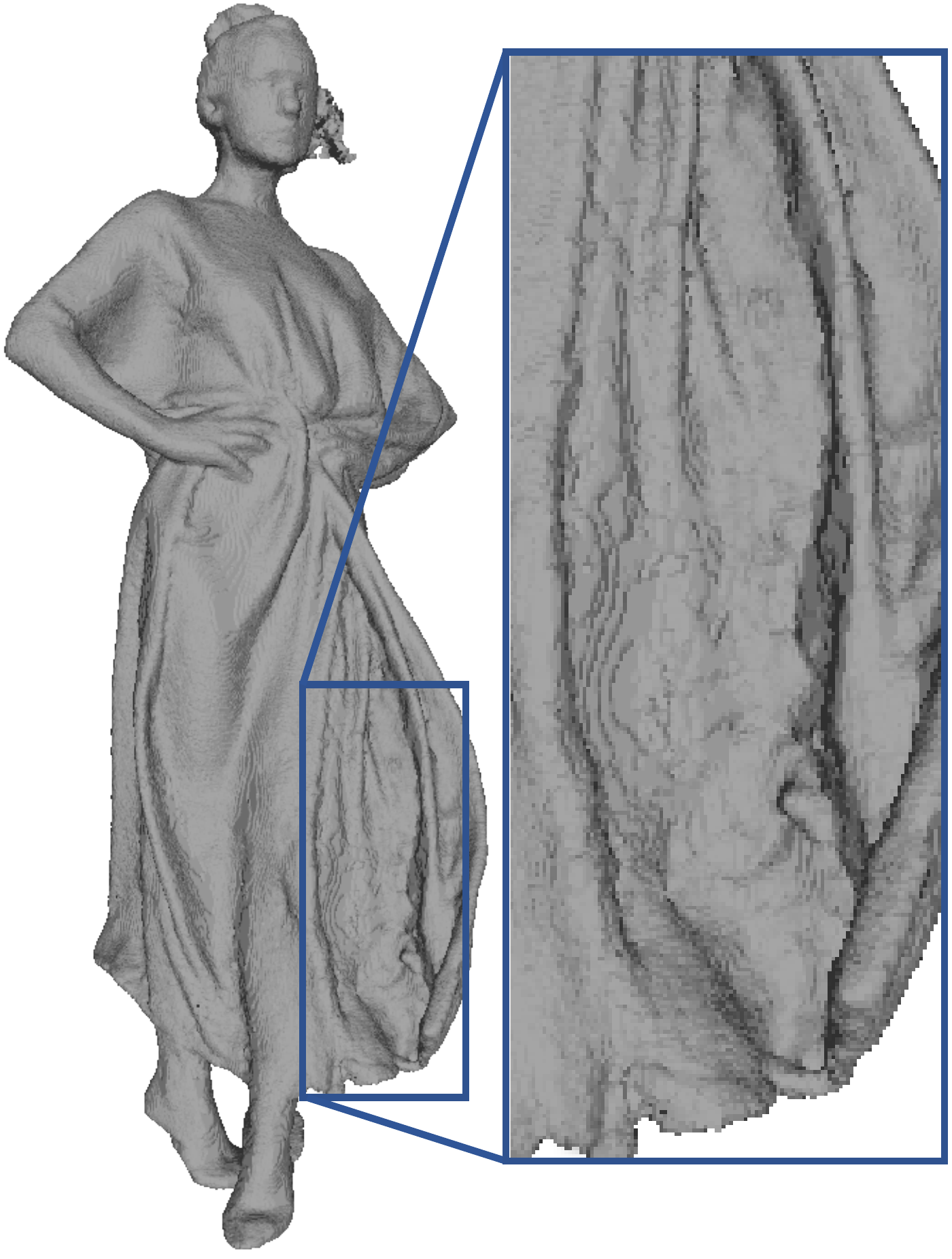} \\
		\vspace{-0.15in}
		\end{minipage}
	}
	\hspace{-0.05in}
	\subfigure[f=8]{
		\begin{minipage}[b]{0.095\textwidth}
			\includegraphics[height=0.09\textheight]{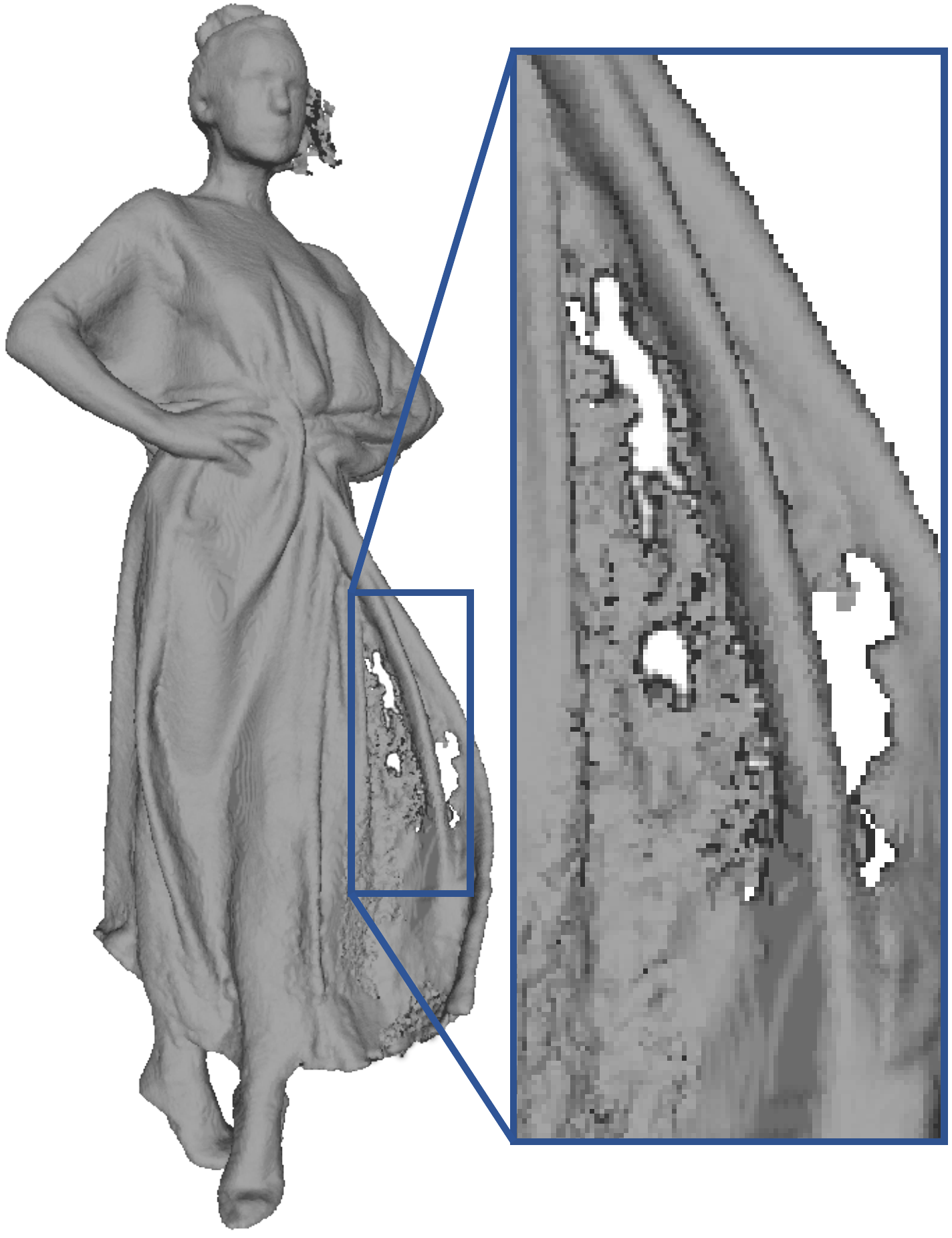} \\
			\includegraphics[height=0.09\textheight]{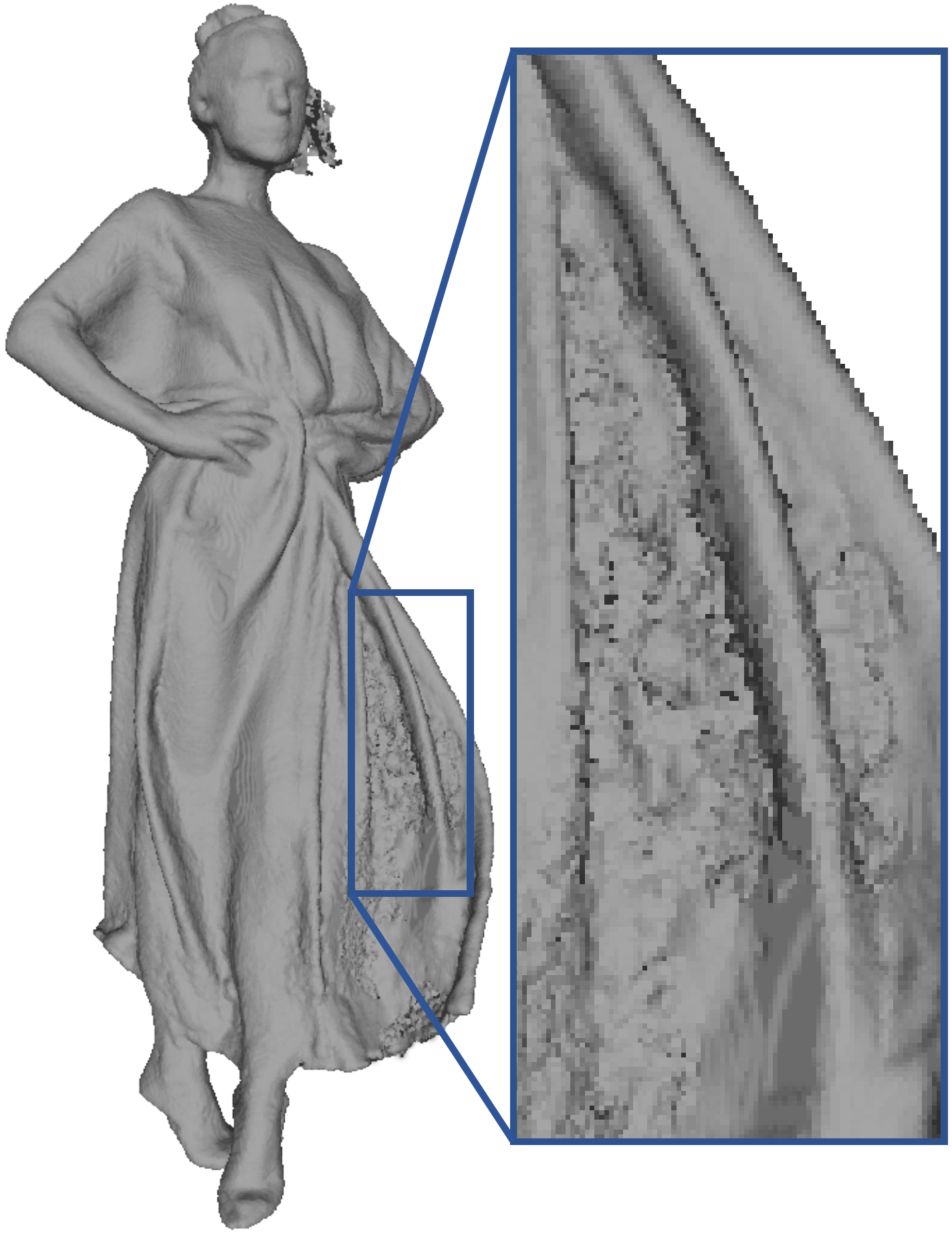} \\
			\includegraphics[height=0.09\textheight]{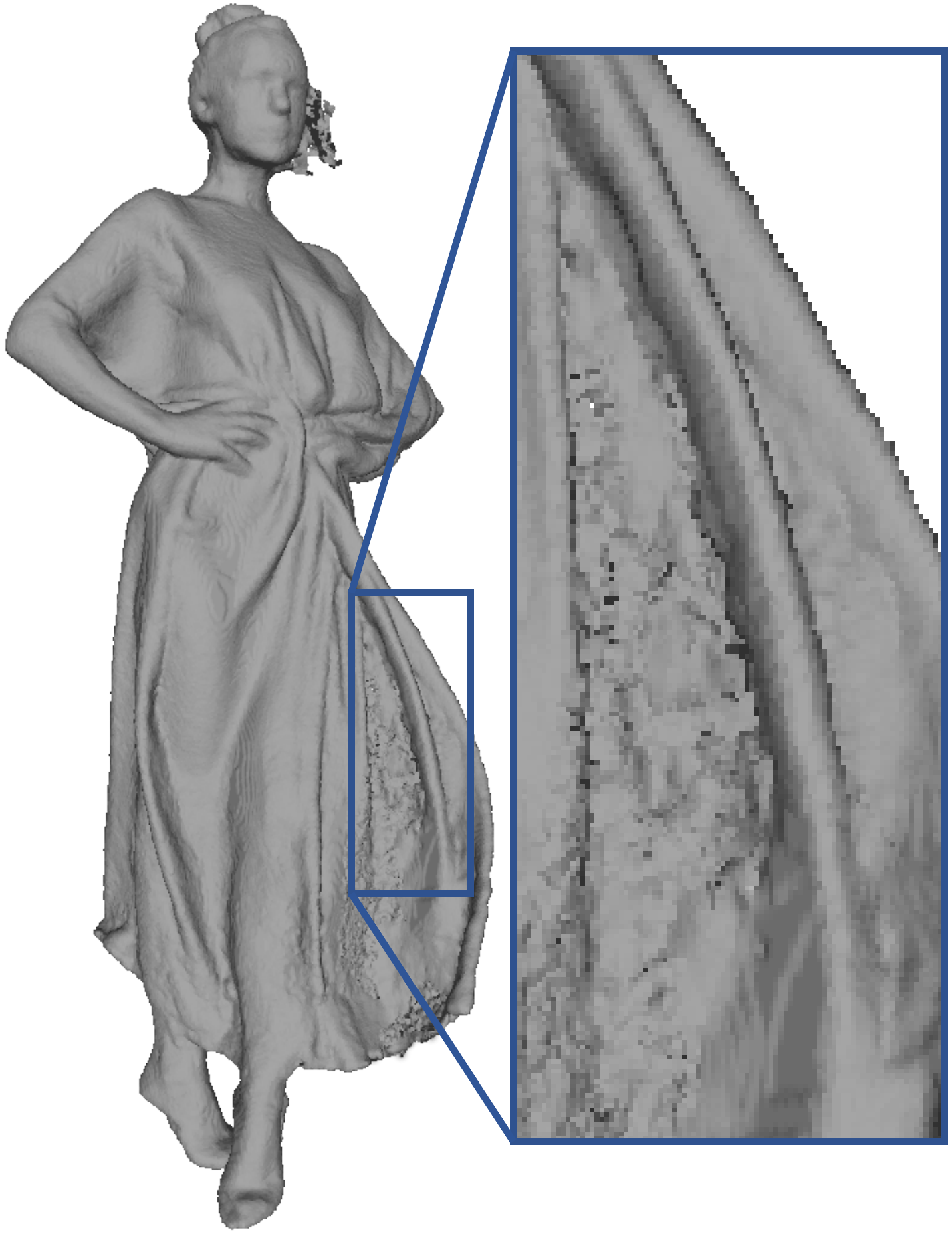} \\
		\vspace{-0.15in}
		\end{minipage}
	}
	\hspace{-0.05in}
	\subfigure[f=9]{
		\begin{minipage}[b]{0.096\textwidth}
			\includegraphics[height=0.09\textheight]{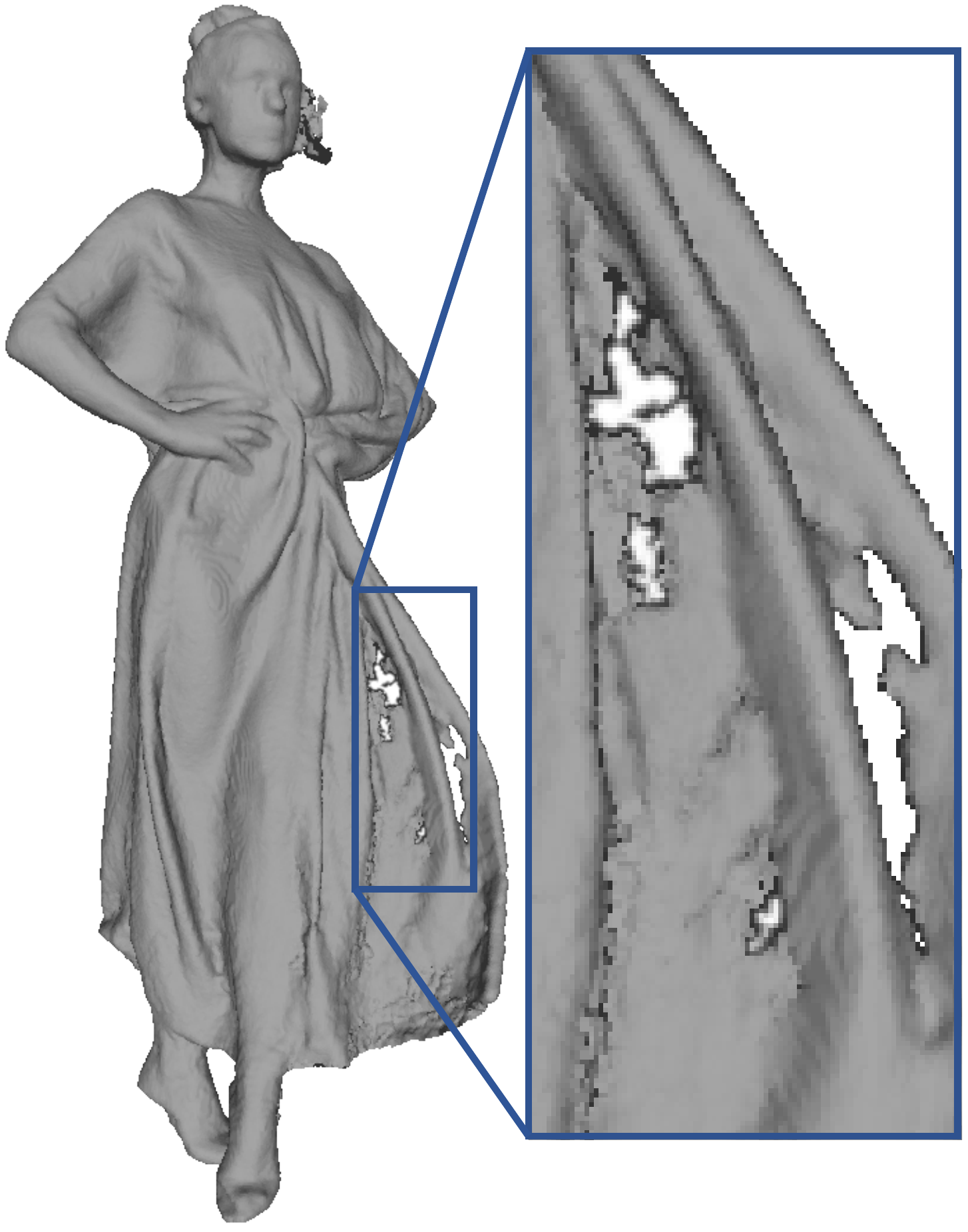} \\
			\includegraphics[height=0.09\textheight]{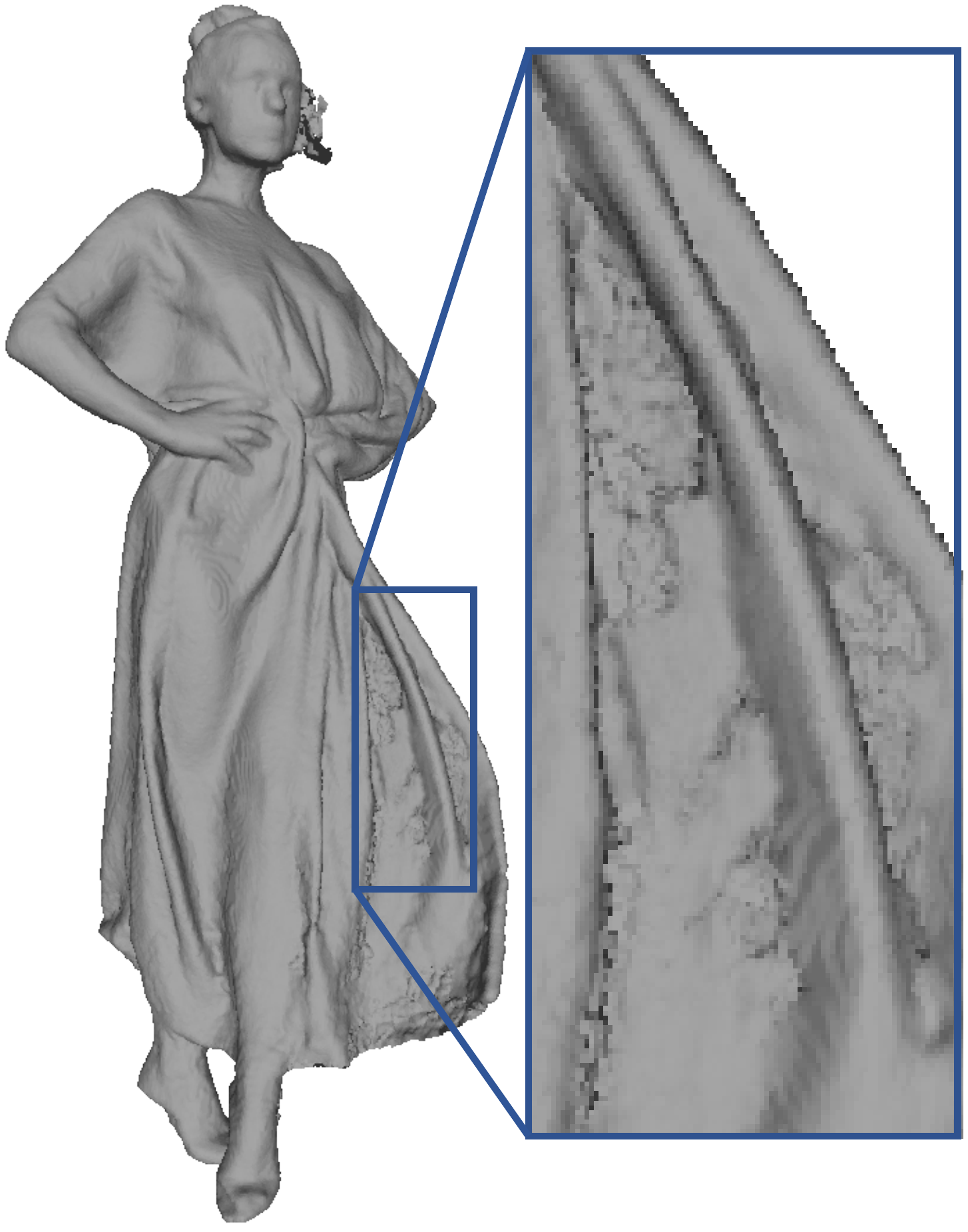} \\
			\includegraphics[height=0.09\textheight]{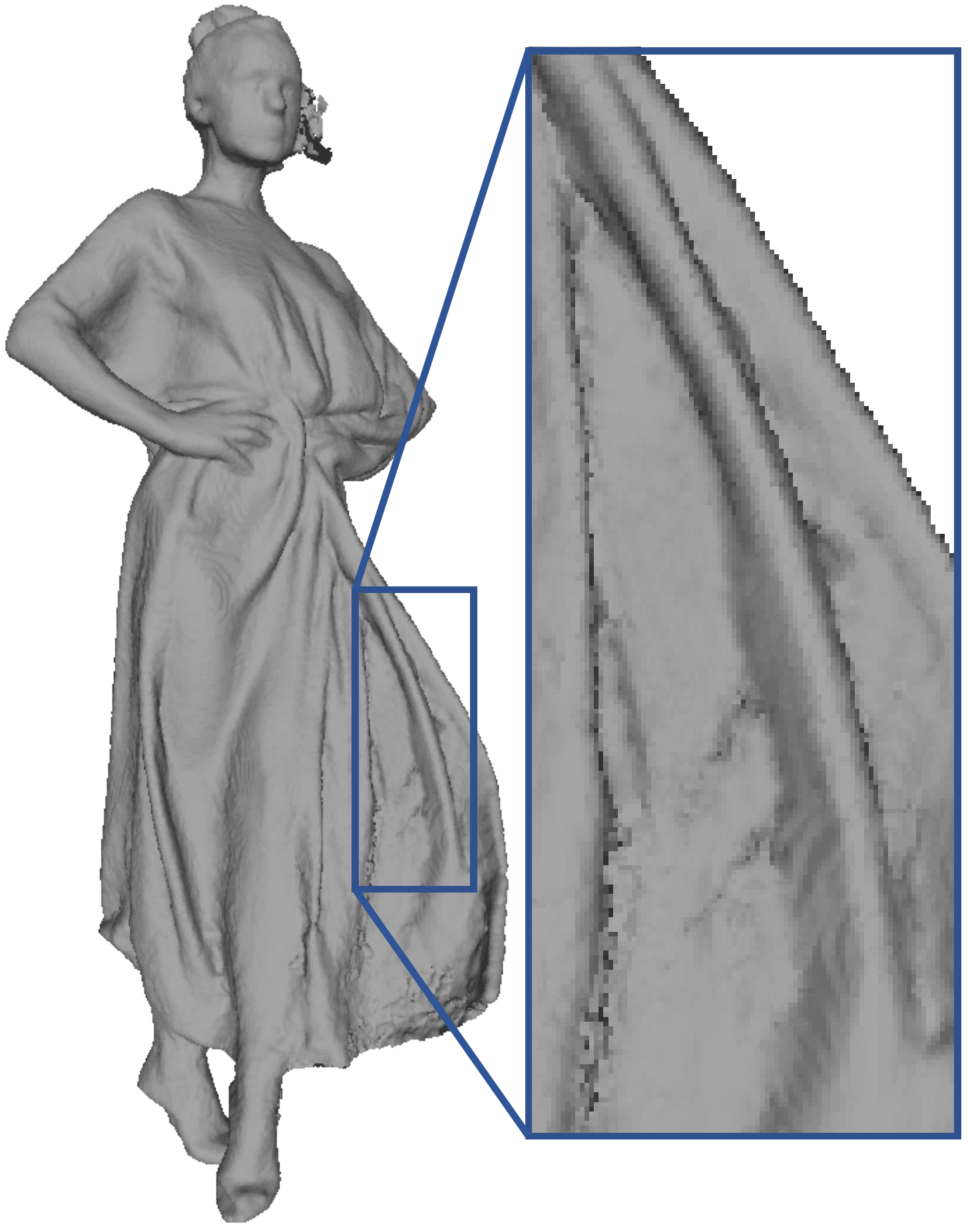} \\
		\vspace{-0.15in}
		\end{minipage}
	}
	\hspace{-0.05in}
	\subfigure[f=10]{
		\begin{minipage}[b]{0.096\textwidth}
			\includegraphics[height=0.09\textheight]{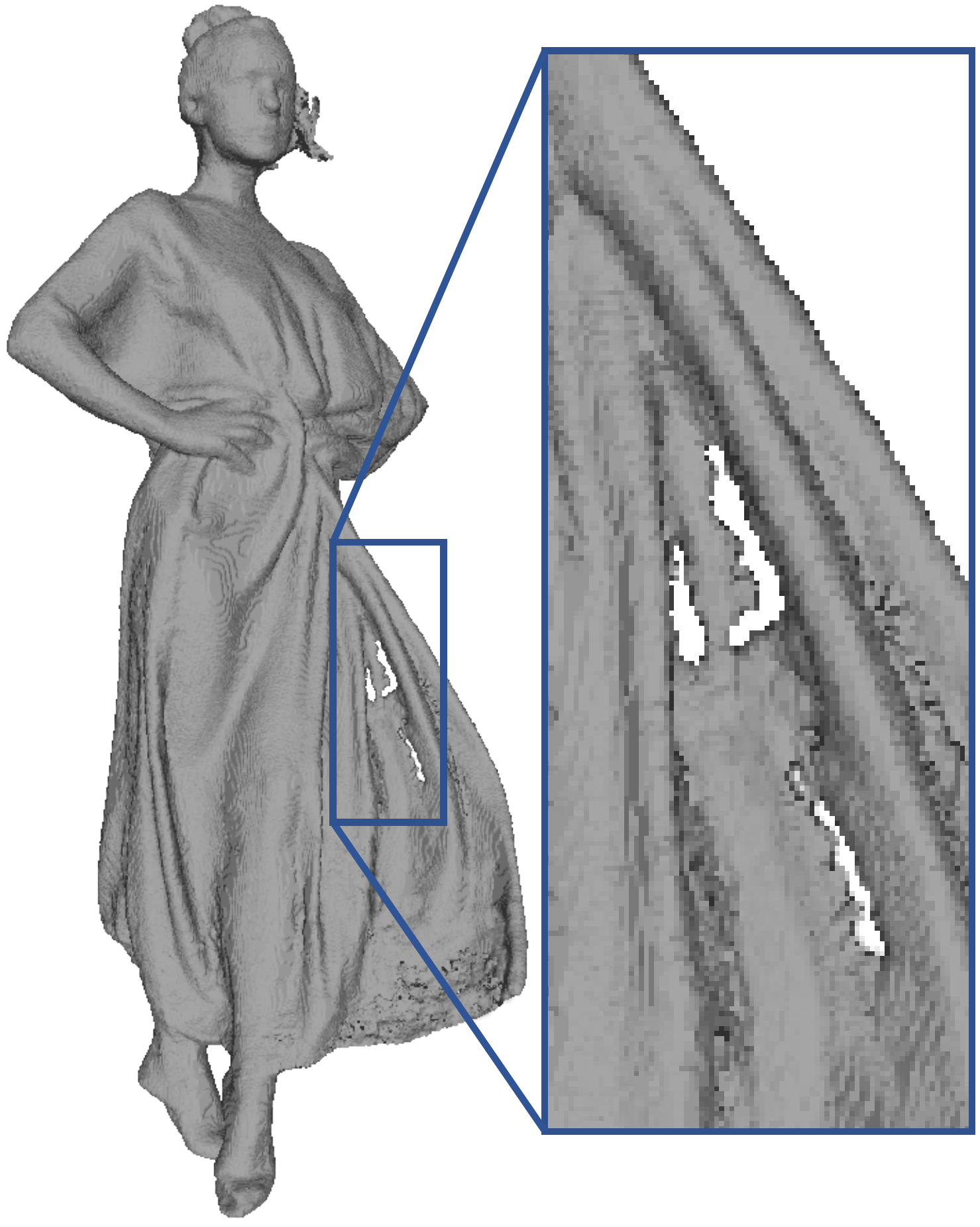} \\
			\includegraphics[height=0.09\textheight]{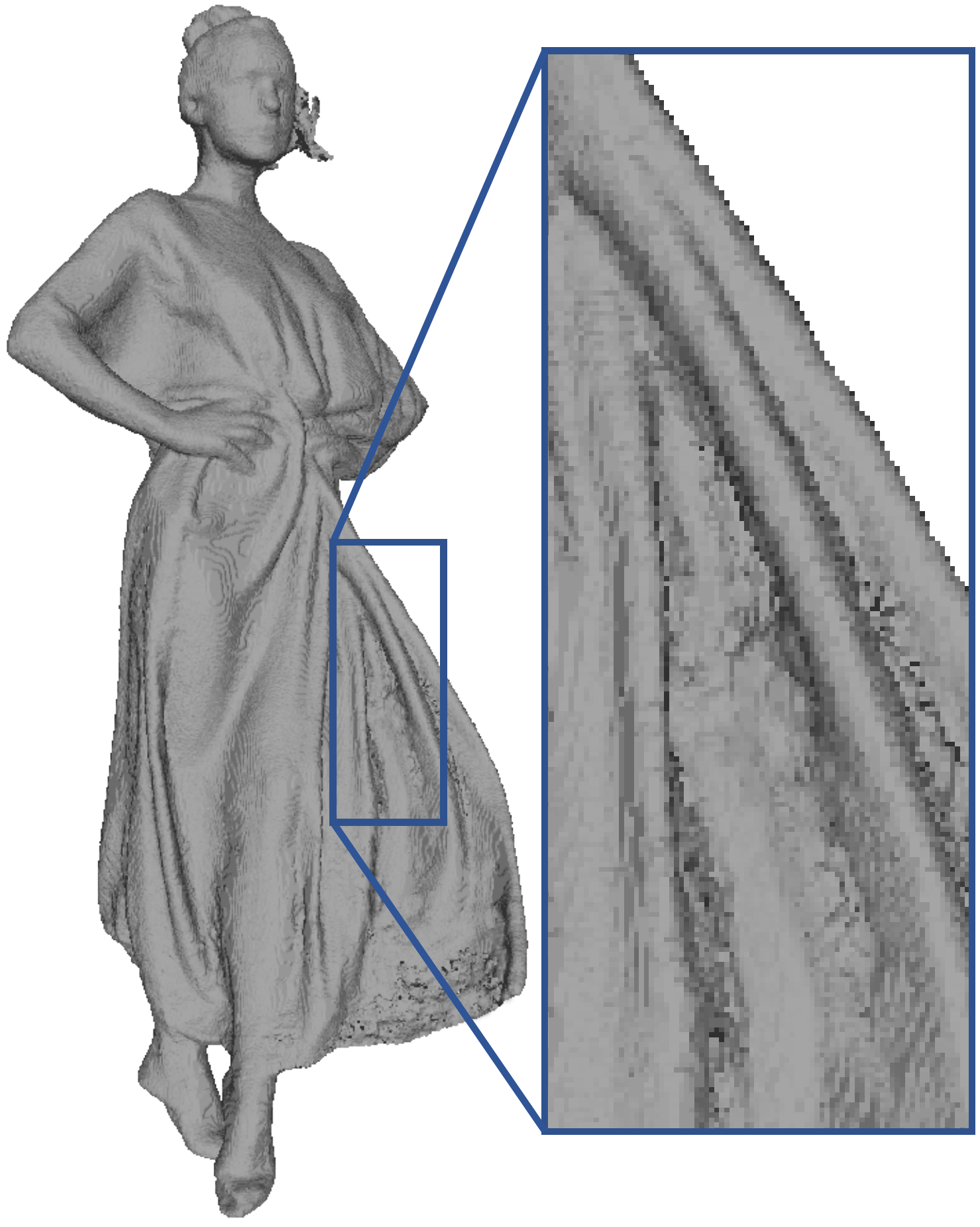} \\
			\includegraphics[height=0.09\textheight]{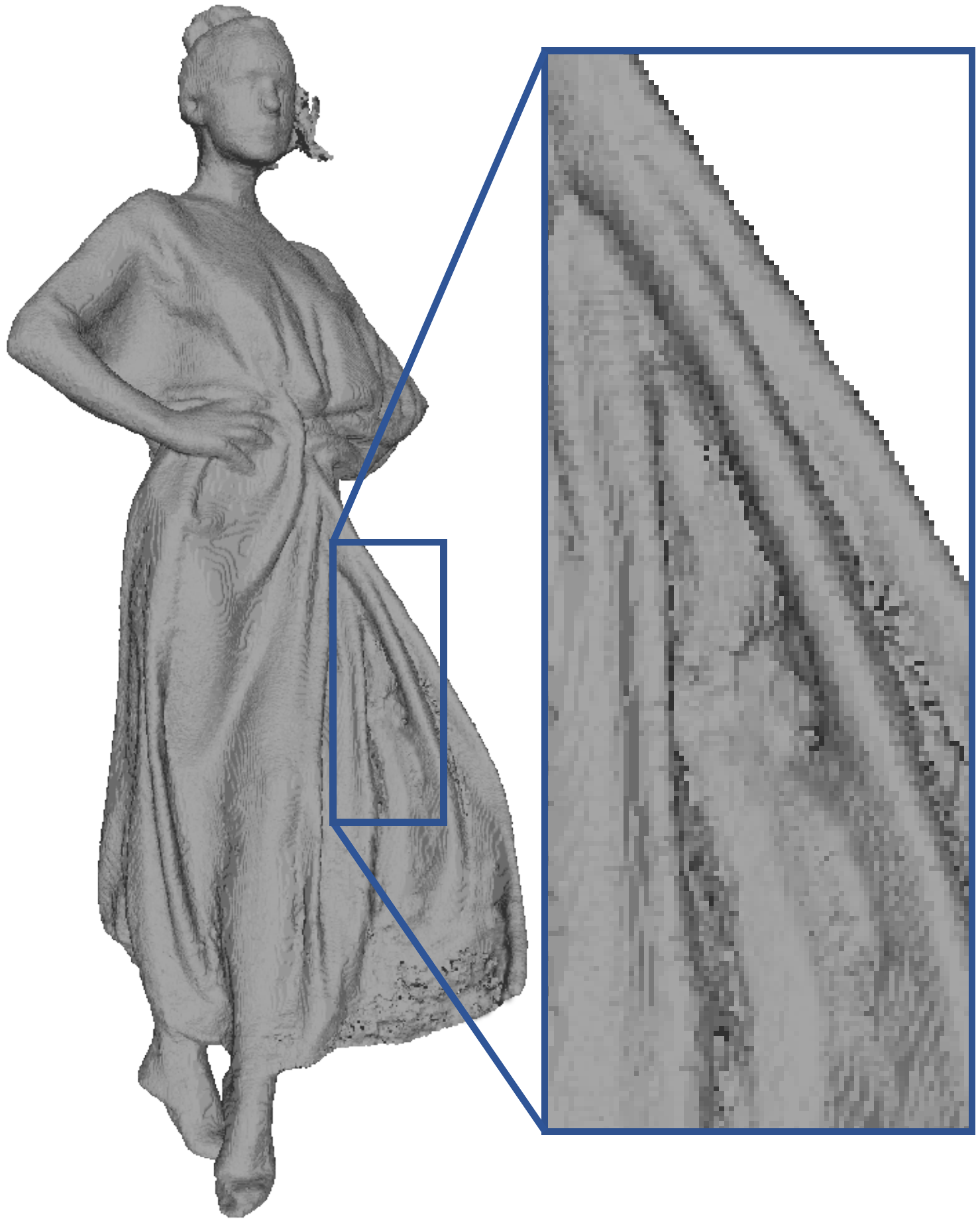} \\
		\vspace{-0.15in}
		\end{minipage}
	}
	\hspace{-0.05in}
	\subfigure[f=11]{
		\begin{minipage}[b]{0.098\textwidth}
			\includegraphics[height=0.09\textheight]{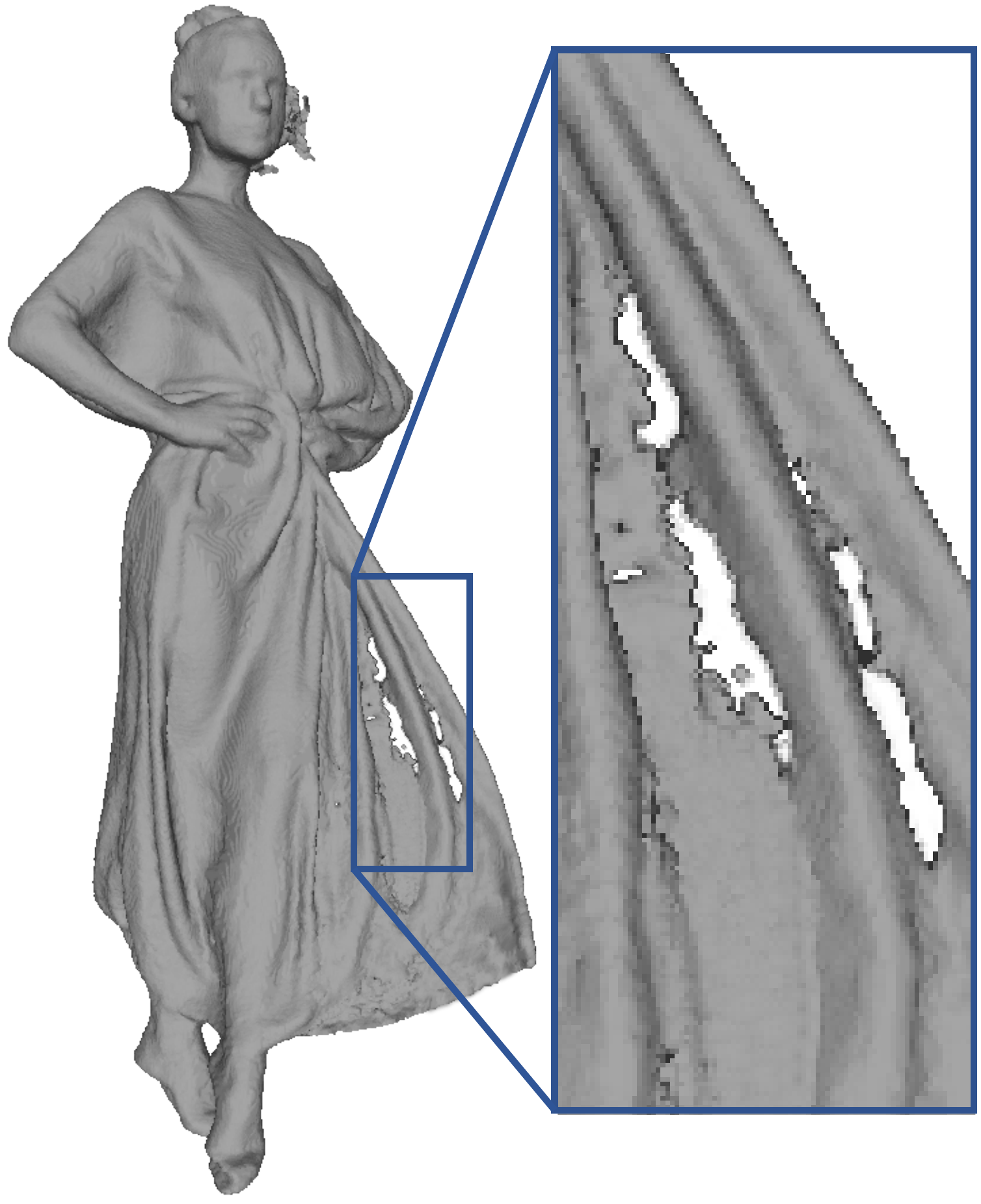} \\
			\includegraphics[height=0.09\textheight]{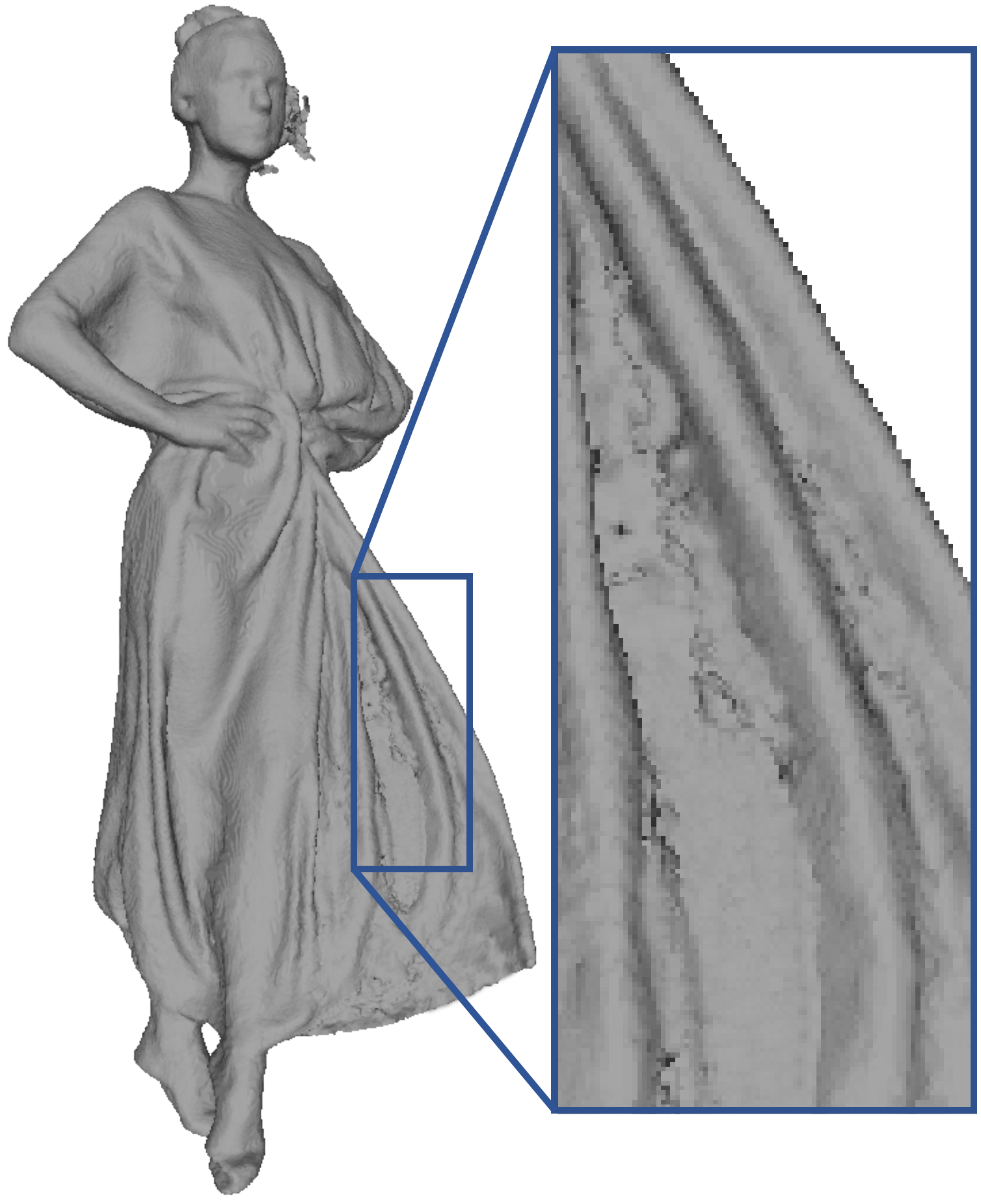} \\
			\includegraphics[height=0.09\textheight]{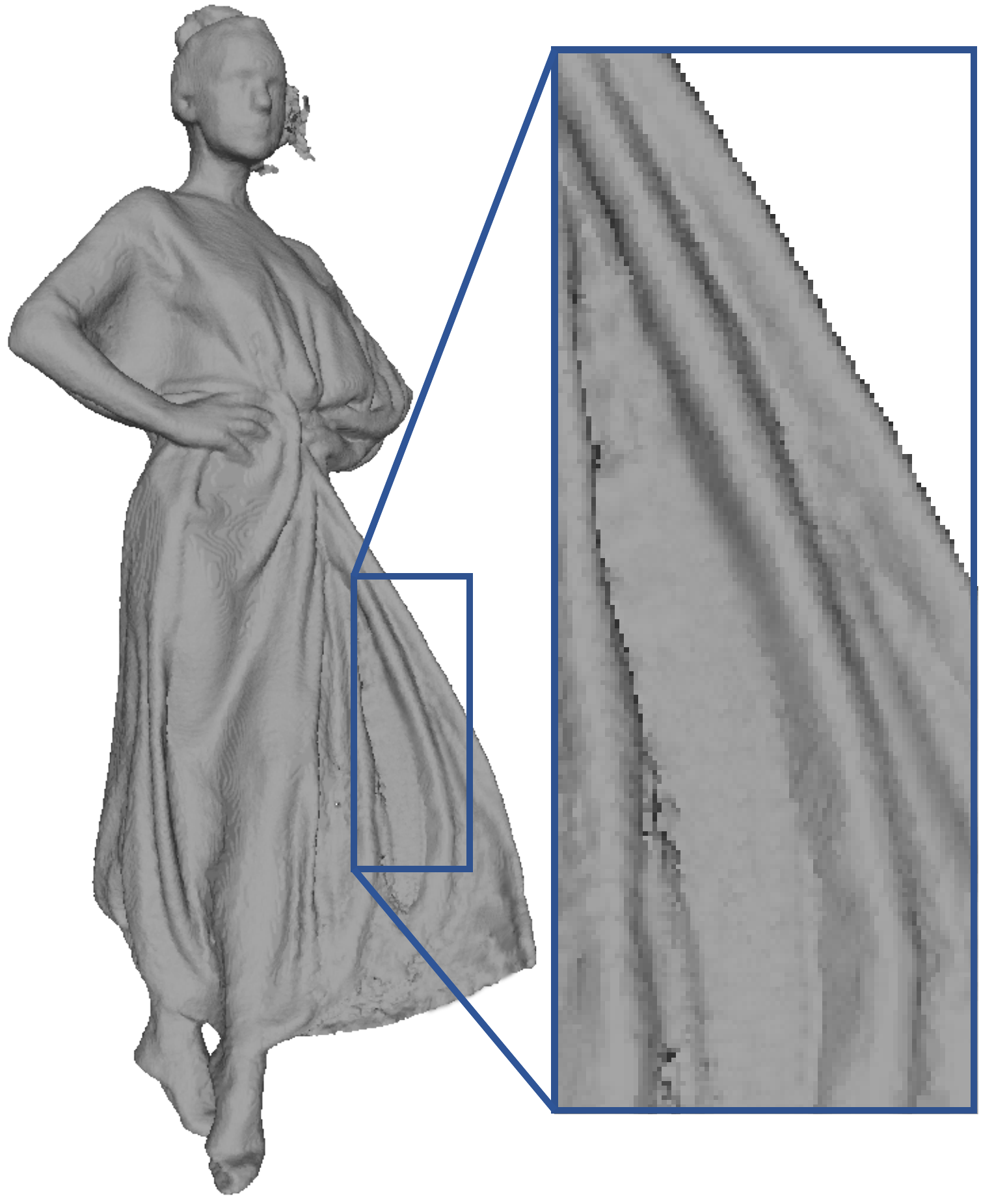} \\
		\vspace{-0.15in}
		\end{minipage}
	}
	\hspace{-0.05in}
	\subfigure[f=12]{
		\begin{minipage}[b]{0.098\textwidth}
			\includegraphics[height=0.09\textheight]{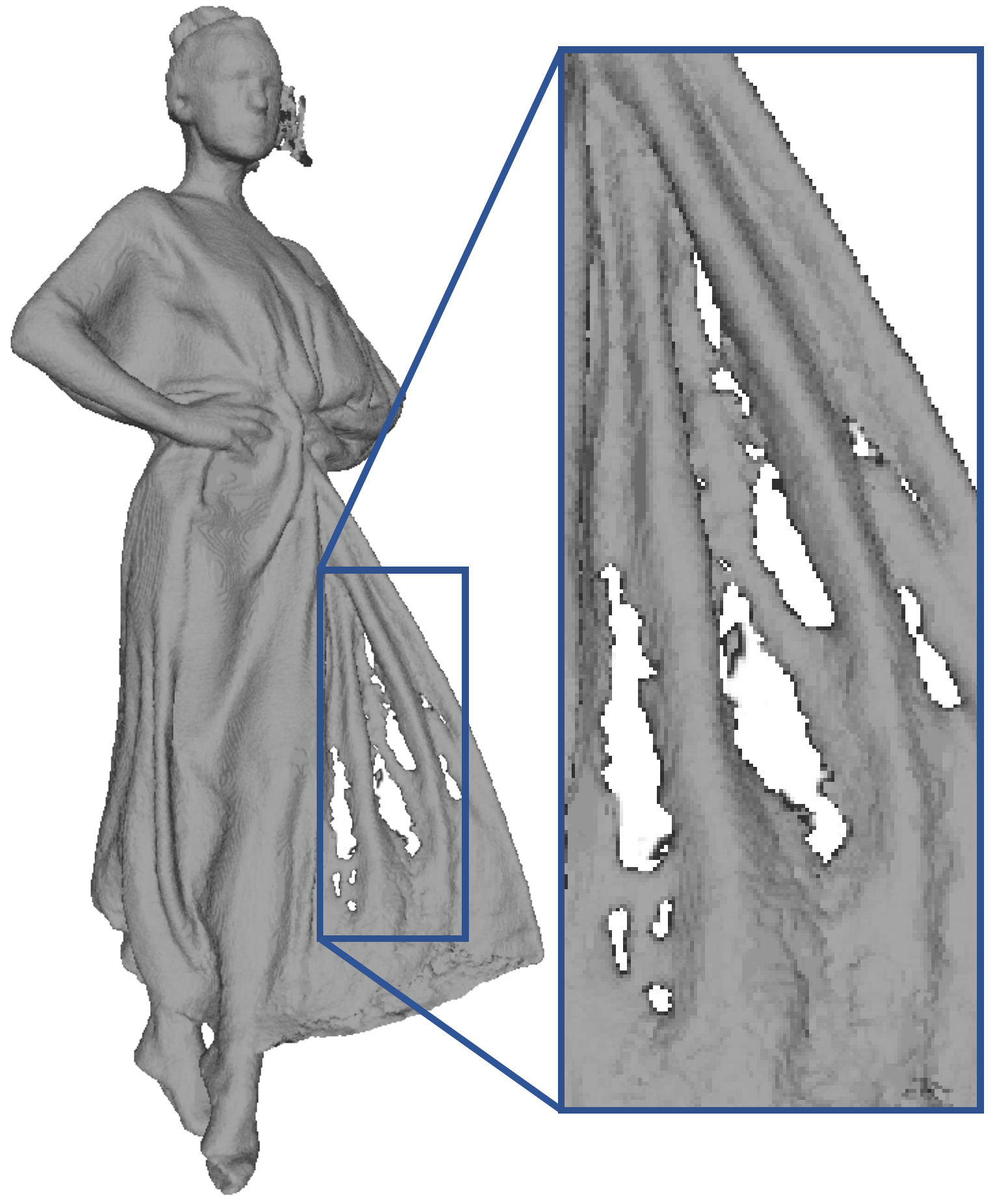} \\
			\includegraphics[height=0.09\textheight]{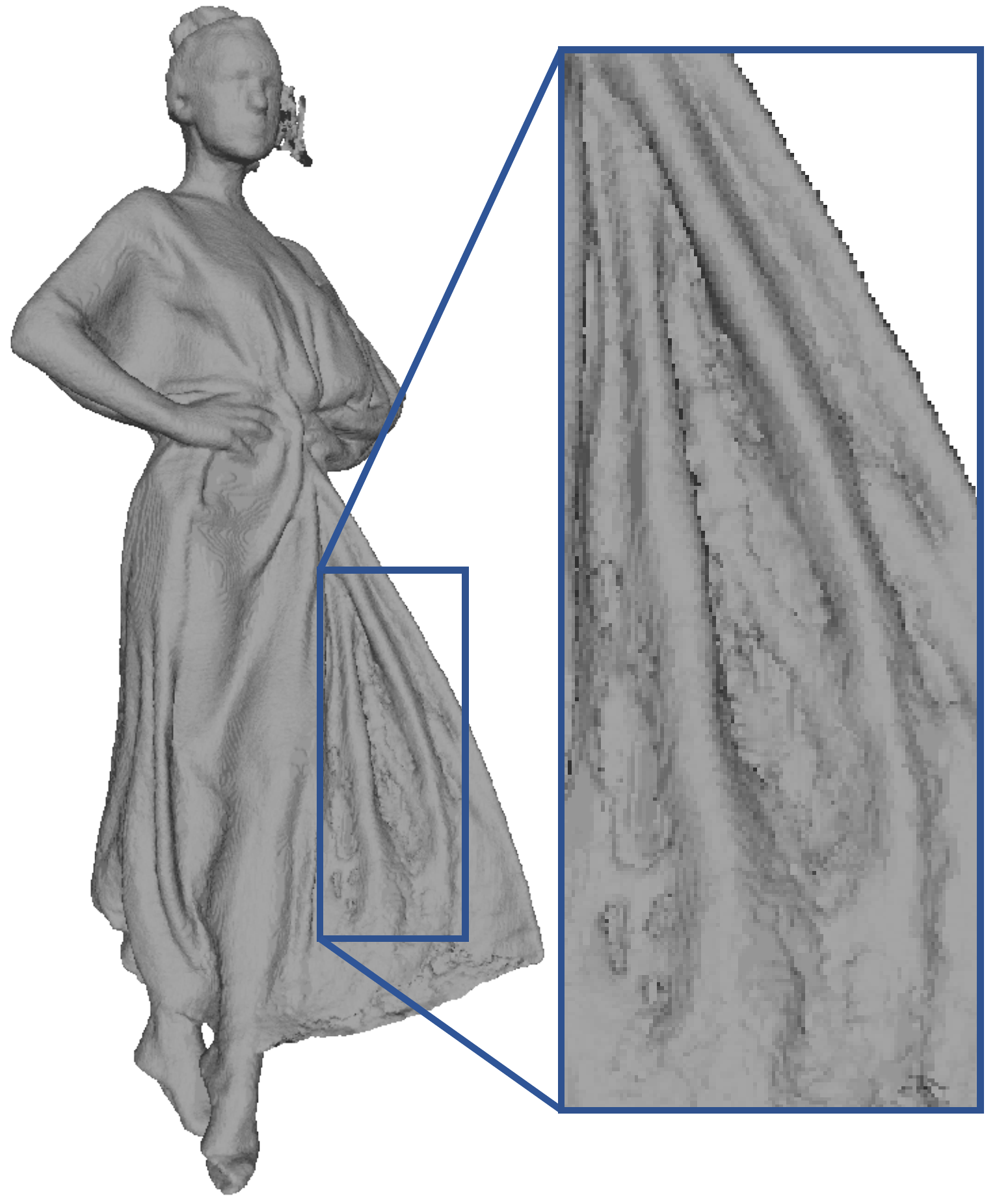} \\
			\includegraphics[height=0.09\textheight]{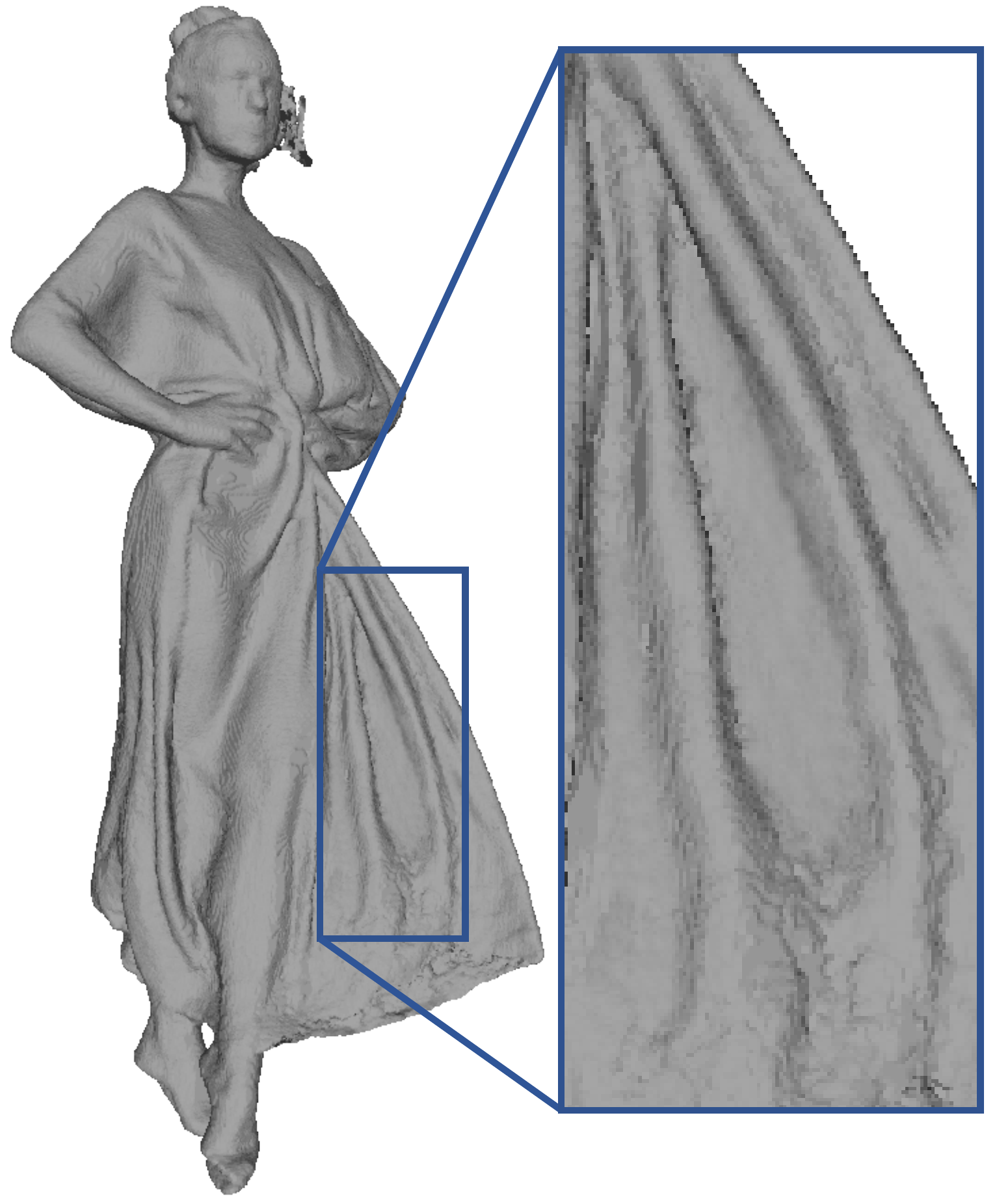} \\
		\vspace{-0.15in}
		\end{minipage}
	}
	\vspace{-0.1in}
	\caption{Several frames of the inpainting results from different methods for \textit{Longdress} with the real holes magnified.}
	\label{fig:result1}
	\vspace{-0.2in}
\end{figure*}

Further, we define a weight matrix $ \mathbf{W}_{f-1} \in M^3\times M^3 $ to encode the temporal connectivities between $ \vec{c}_t $ and $ \hat{\vec{c}}_t^{f-1} $. 
The rows of $\mathbf{W}_{f-1}$ correspond to points in $ \vec{c}_t $ and columns correspond to points in $ \hat{\vec{c}}_t^{f-1} $.   
Specifically, the weight in $ \mathbf{W}_{f-1} $ between nodes $k$ and $l$ is assigned as
\vspace{-0.12in}
\begin{equation}
	w_{k,l}=
    	\begin{cases}
        	~1, & k \sim l\\
        	~0, & \text{otherwise}
    	\end{cases}
	\label{eq:noweight}
\vspace{-0.1in}
\end{equation}
where $k \sim l$ means nodes $k$ and $l$ are temporally corresponding points and thus connected. 
We set weight $1$ to each pair of temporally connected points as the temporal correlation is strong for corresponding points. 

It is the same for the construction of temporal connectivities between $ \vec{c}_t $ and $ \hat{\vec{c}}_t^{f+1} $. Similarly, we define a weight matrix $ \mathbf{W}_{f+1} $ to describe the temporal connections.  

\vspace{-0.12in}
\subsection{Problem Formulation}
\label{subsec:formula}
\vspace{-0.06in}
Finally, we cast dynamic point cloud inpainting as an optimization problem, which is regularized by the graph-signal smoothness prior as mentioned in Section~\ref{sec:graph} and temporal consistency. It is formulated as
\vspace{-0.08in}
\begin{equation}
\begin{split}
	\min_{\vec{c}_r}~~ &
		\norm{\overline{\mathbf{\Omega}}\vec{c}_r-\overline{\mathbf{\Omega}}\vec{c}_t}_2^2
		+\alpha\norm{\mathbf{\Omega}\vec{c}_r-\mathbf{\Omega}\hat{\vec{c}}_s}_2^2
		+\gamma\vec{c}_r^T\mathcal{L}\vec{c}_r+\\
		& \beta\norm{\vec{c}_r-\mathbf{W}_{f-1}\hat{\vec{c}}_t^{f-1}}_2^2
		  +\beta\norm{\vec{c}_r-\mathbf{W}_{f+1}\hat{\vec{c}}_t^{f+1}}_2^2
	\label{eq:formu}
\end{split}
\vspace{-0.13in}
\end{equation}
where $ \vec{c}_r\in \mathbb{R}^{M^3 \times 3} $ is the desired resulting cube. 
$ \mathbf{\Omega} $ is a $ M^3\times M^3 $ diagonal matrix with $ \Omega_{i,i} \in \{0,1\} $, where 0 indicates known points and 1 indicates missing points. Thus $ \mathbf{\Omega}\vec{c}_r $ and $ \mathbf{\Omega}\hat{\vec{c}}_s $ represent the missing region in $ \vec{c}_r $ and $ \hat{\vec{c}}_s $ respectively. $ \overline{\mathbf{\Omega}} $ is complementary to $ \mathbf{\Omega} $, which extracts the known region.
$ \mathcal{L} $ is the Laplacian matrix of the spatial graph constructed over $ \vec{c}_t $ as described in Section~\ref{subsec:constru}. 
$ \alpha $, $ \beta $ and $ \gamma $ are three weighting parameters (we empirically set $ \alpha=1 $, $ \beta=0.5 $ and $ \gamma=0.5 $ in the experiments). %Specifically, $ \alpha $ will affect the likeliness between the inpainted region and the corresponding region in the intra-source cube. $ \beta $ and $ \gamma $ will affect the smoothness along consecutive frames and the smoothness of the inpainted cube, respectively.

The first term in (\ref{eq:formu}) is a data fidelity term, which ensures the desired cube to be close to $ \vec{c}_t $ in the known region. The second term constraints the missing region of $ \vec{c}_r $ to be similar to that of $ \hat{\vec{c}}_s $. The last two terms aim to make the structure of $ \vec{c}_r $ mimic that of $ \hat{\vec{c}}_t^{f-1} $ and $ \hat{\vec{c}}_t^{f+1} $, which enforces the temporal consistency. Further, the third term is the graph-signal smoothness prior, which enforces the internal structure of $ \vec{c}_r $ to be smooth with respect to the constructed spatial graph when merging information from three source cubes.

(\ref{eq:formu}) is a quadratic programming problem. Taking derivative of (\ref{eq:formu}) with respect to $ \vec{c}_r $ and setting the derivative to $0$, we have the closed-form solution:
\vspace{-0.12in}
\begin{equation}
\begin{split}
	\vec{c}_r^{\text{opt}}=
		& (\overline{\mathbf{\Omega}}^2
		  +\alpha\mathbf{\Omega}^2
		  +2\beta\mathbf{I}
		  +\gamma\mathcal{L})^{-1}\\
		& (\overline{\mathbf{\Omega}}^2\vec{c}_t
		  +\alpha\mathbf{\Omega}^2\hat{\vec{c}}_s
		  +\beta\mathbf{W}_{f-1}\hat{\vec{c}}_t^{f-1}
		  +\beta\mathbf{W}_{f+1}\hat{\vec{c}}_t^{f+1}).
	\label{eq:solution}
\end{split}
\end{equation}
\vspace{-0.13in}

(\ref{eq:formu}) is thus solved optimally and efficiently. 
We replace the target cube with the resulting cube in the target frame $ \mathbf{P}_f $, which serves as the output.% After each hole is inpainted, the target frame will be updated, {\it i.e.}, the information of new points will be considered in the inpainting of the subsequent holes and subsequent frames. This provides more opportunities to find more suitable cubes for the subsequent holes and better corresponding information for the subsequent frames, and thus helps improve the performance.

\vspace{-0.1in}
\section{Experiments}
\label{sec:results}
\vspace{-0.12in}
\subsection{Experimental Setup}
\vspace{-0.06in}
We evaluate the proposed method by testing on several 3D dynamic point cloud datasets from MPEG~\cite{MPEG} and JPEG Pleno~\cite{JPEG}, including \textit{Longdress, Loot, Redandblack, Soldier}, and \textit{UlliWegner}. We test on two types of holes: 1) real holes generated during the capturing process, which have no ground truth; 2) synthetic holes on point clouds so as to compare with the ground truth.
%In particular, the number of nearest neighbors $ K $ is considered to be related to $ m $, the number of existing points in the cube. Empirically, $ K =\sqrt{m} $ in our experiments. Besides, the size $ H $ of the searching box in Section~\ref{subsec:inter} is considered to be related to $ M $, the size of the unit cube. Empirically, $ H=2.5M $ in our experiments.

Further, we compare our method with three competing algorithms for {\it static} 3D geometry inpainting, including Meshlab~\cite{Meshlab}, Lozes {\it et al.}~\cite{Lozes15} and Hu {\it et al.}~\cite{Fu18TIP}. We test the static methods by performing them on each frame separately. Besides, as Meshlab is based on meshes, we convert point clouds to meshes via the Meshlab software \cite{Meshlab} prior to testing the method, and then convert the inpainted meshes back to point clouds as the final output.

\vspace{-0.12in}
\subsection{Results on Point Cloud Inpainting}
\vspace{-0.06in}

\textbf{Objective results.} It is nontrivial to measure the geometry difference of point clouds objectively. We apply the geometric distortion metrics in~\cite{Tian17} and~\cite{Dinesh17}, referred to as GPSNR and NSHD, respectively, as the metric for evaluation. The higher GPSNR is and the lower NSHD is, the smaller the difference between two point clouds is.

Table~\ref{tb:gpsnr} and Table~\ref{tb:nshd} show the average objective results of the frames for each sequence with synthetic holes in GPSNR and NSHD, respectively. We see that our scheme outperforms all the competing methods in GPSNR and NSHD significantly. Specifically, in Table~\ref{tb:gpsnr} we achieve 26.80 dB gain in GPSNR on average over Meshlab, 16.18 dB over~\cite{Lozes15} and 5.26 dB over~\cite{Fu18TIP}. In Table~\ref{tb:nshd}, we produce much lower NSHD than the other methods, at least three times lower compared to the next best method \cite{Fu18TIP}.

\begin{figure*}[ht]
	\vspace{-0.2in}
	\centering
	\subfigure[f=2]{
		\begin{minipage}[b]{0.109\textwidth}
			\includegraphics[width=\textwidth]{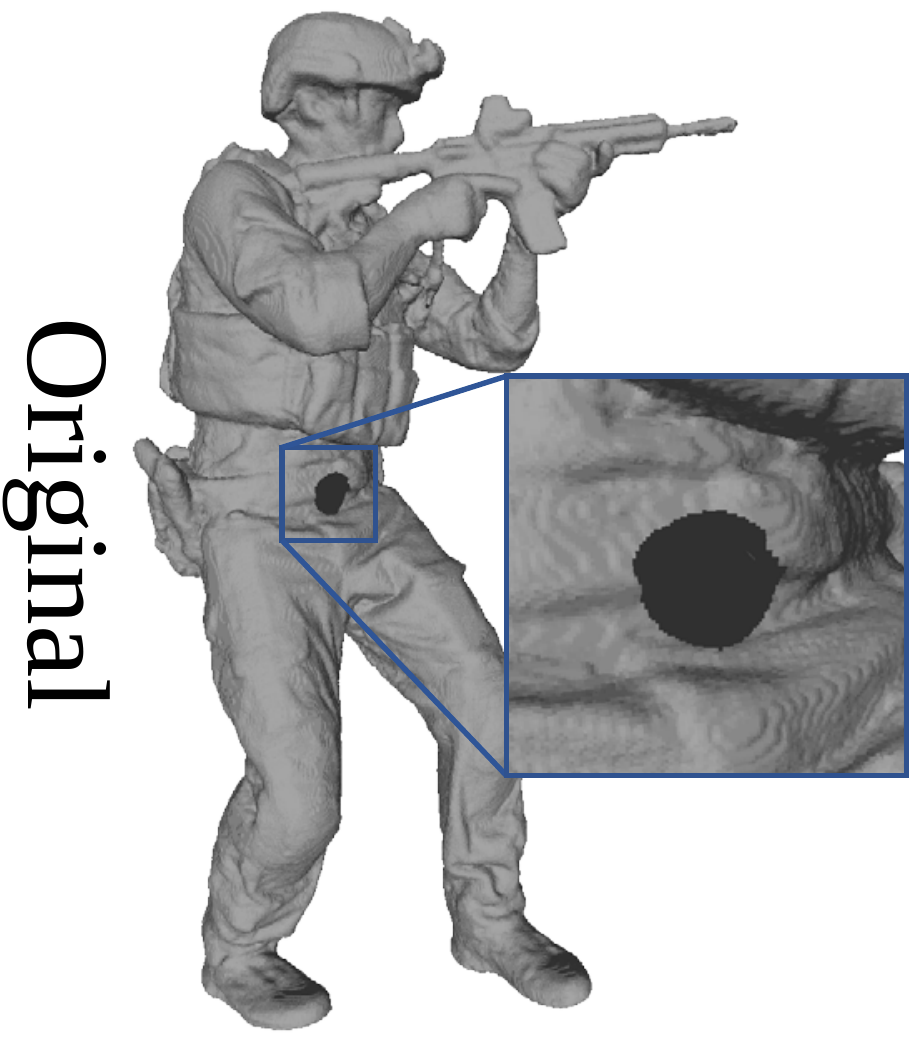} \\
			\includegraphics[width=\textwidth]{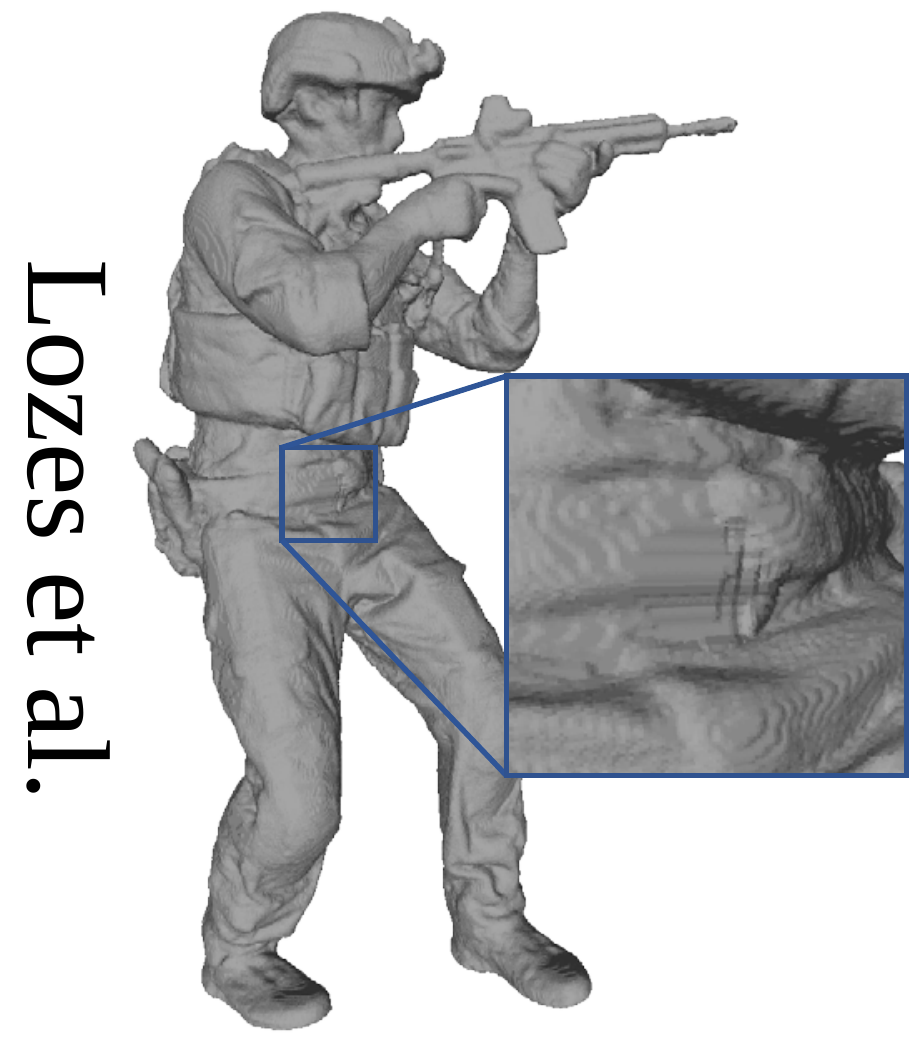} \\
			\includegraphics[width=\textwidth]{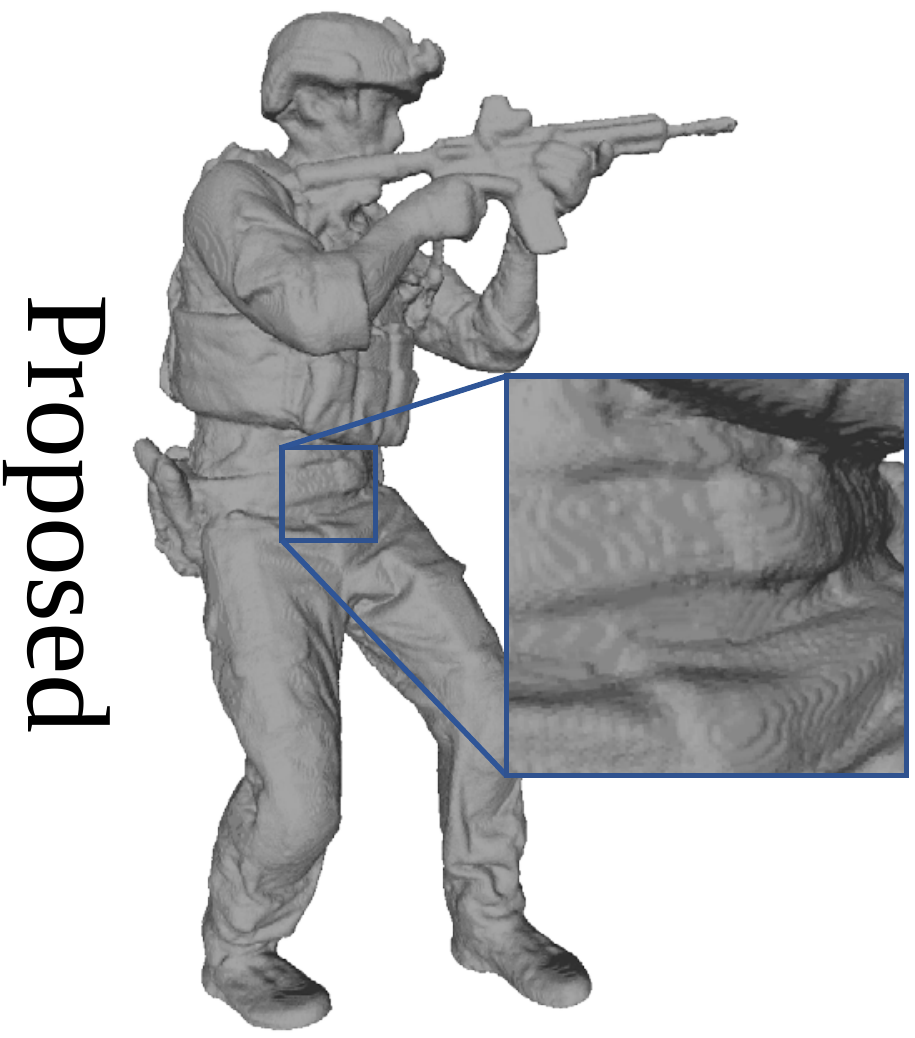} \\
			\includegraphics[width=\textwidth]{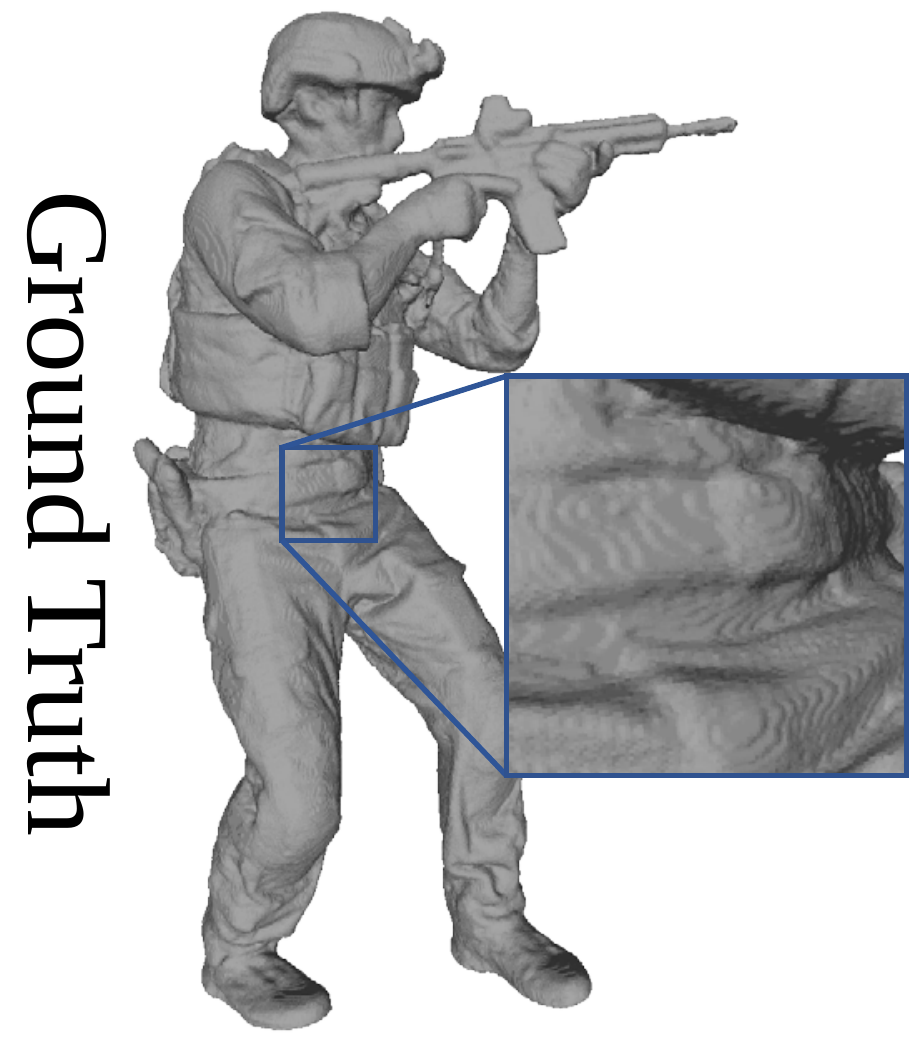} \\
		\vspace{-0.15in}
		\end{minipage}
	}
	%\hspace{-0.1in}
		\subfigure[f=4]{
		\begin{minipage}[b]{0.094\textwidth}
			\includegraphics[width=\textwidth]{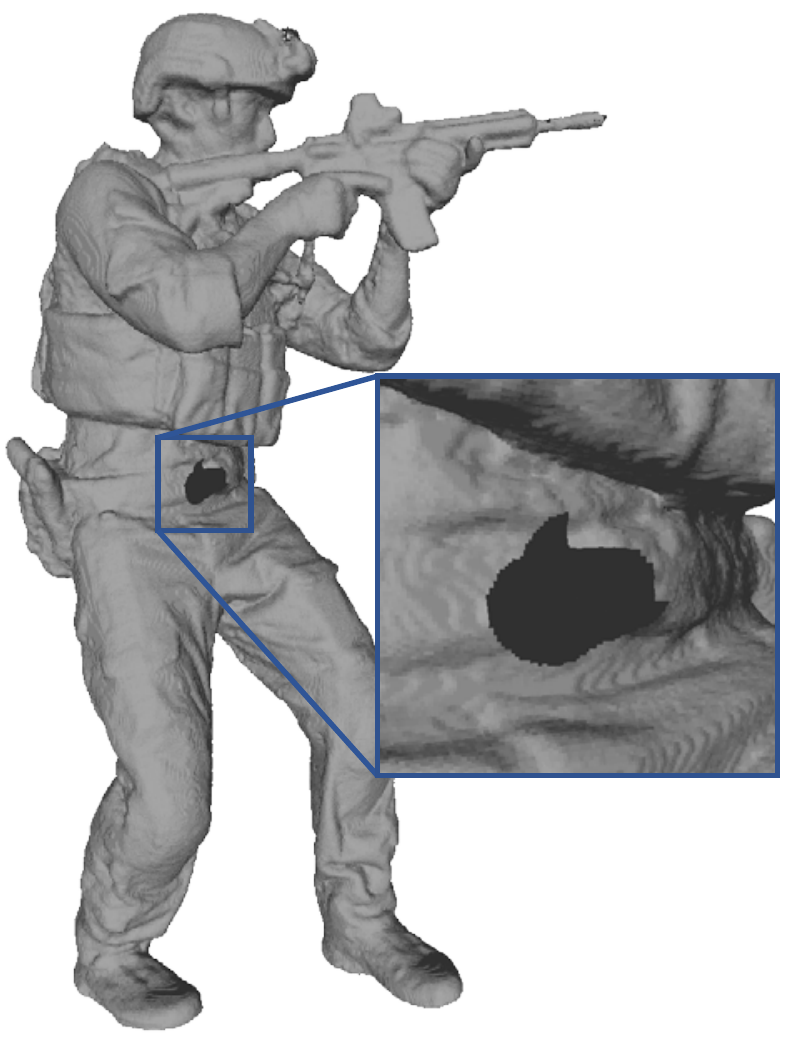} \\
			\includegraphics[width=\textwidth]{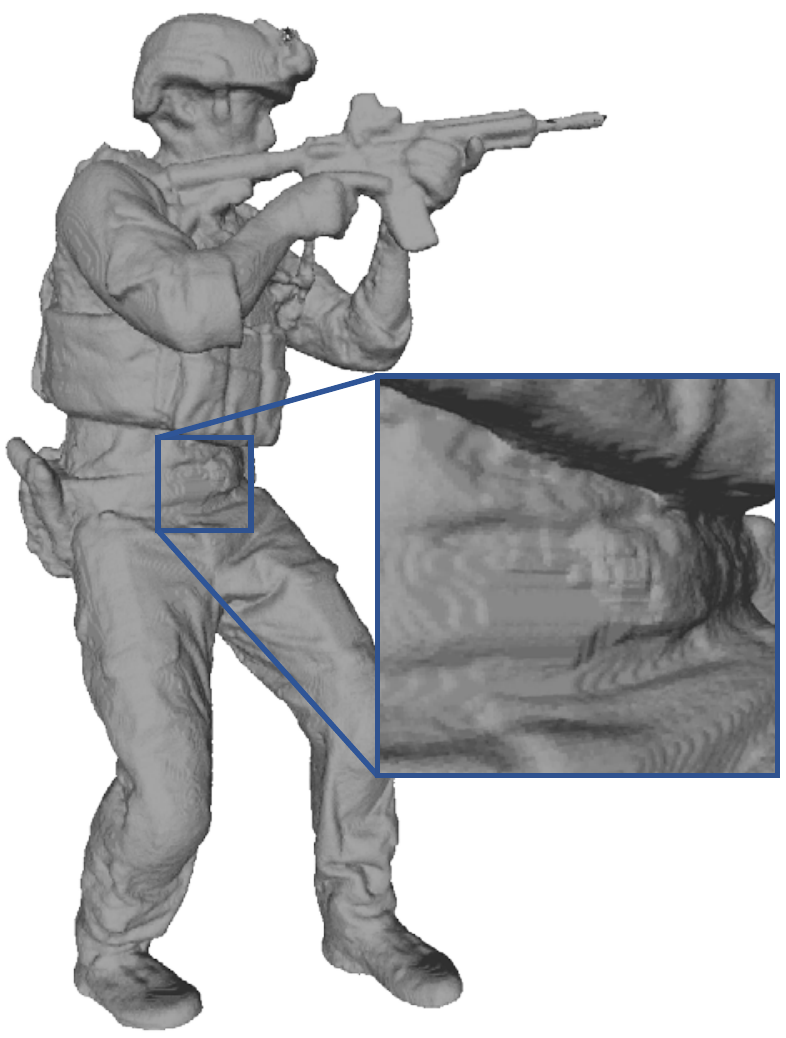} \\
			\includegraphics[width=\textwidth]{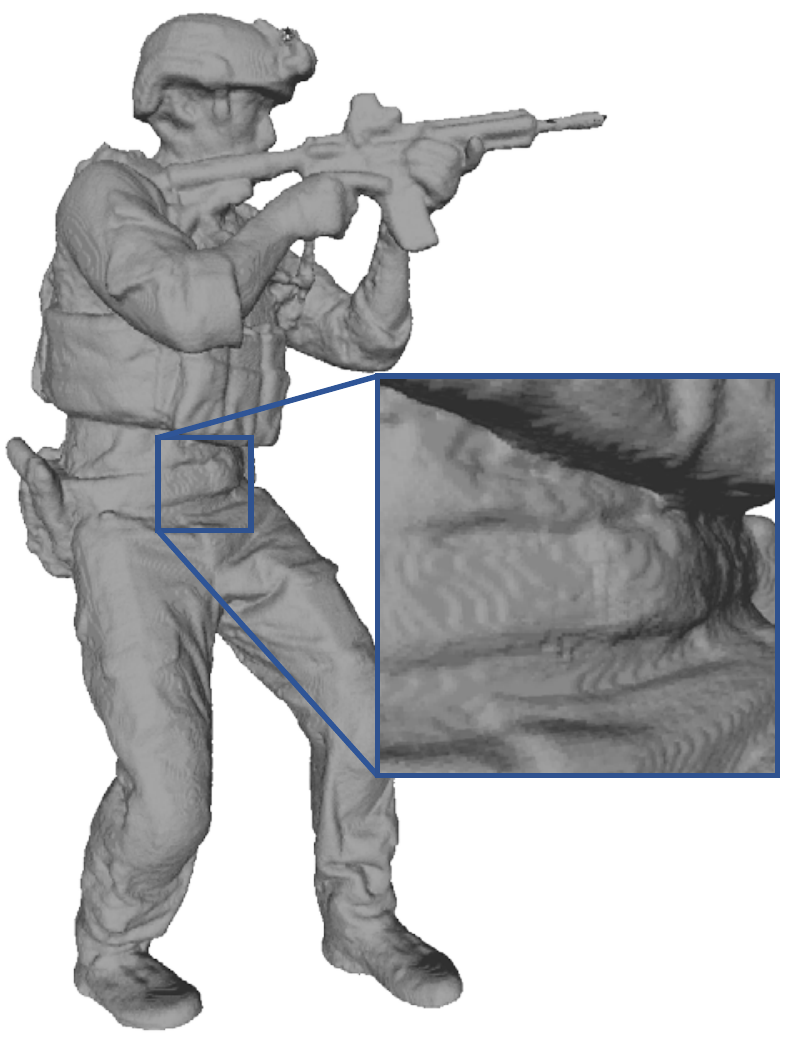} \\
			\includegraphics[width=\textwidth]{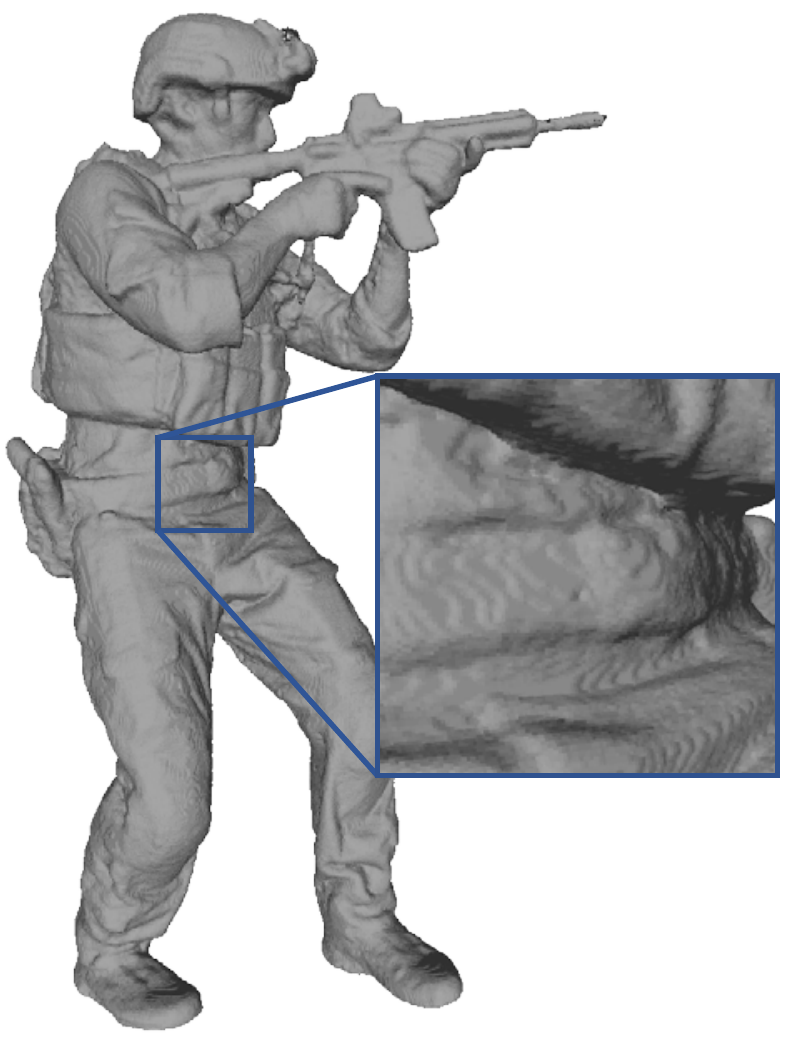} \\
		\vspace{-0.15in}
		\end{minipage}
	}
	%\hspace{-0.1in}
		\subfigure[f=5]{
		\begin{minipage}[b]{0.094\textwidth}
			\includegraphics[width=\textwidth]{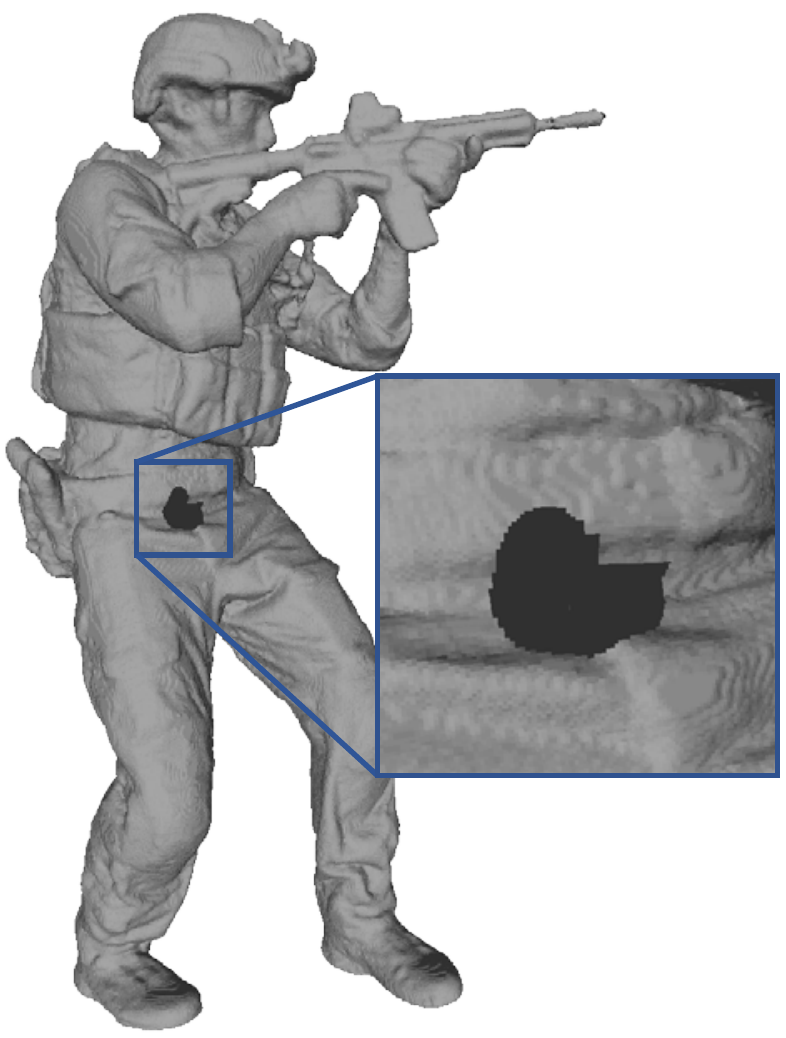} \\
			\includegraphics[width=\textwidth]{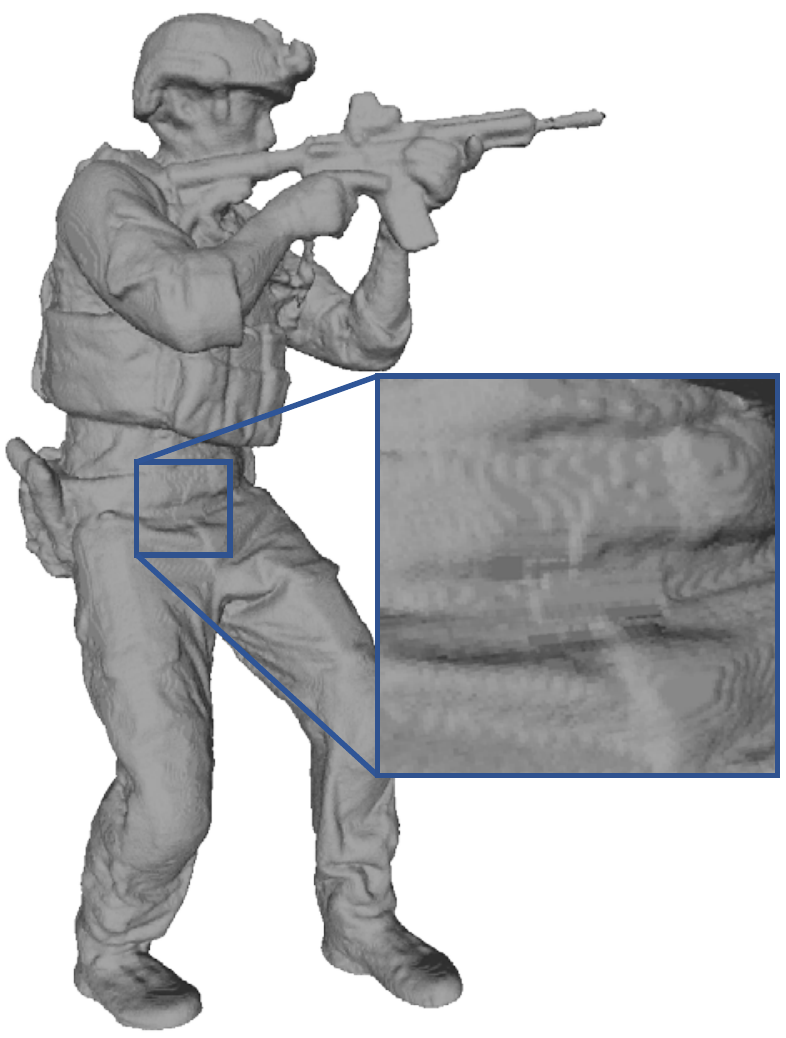} \\
			\includegraphics[width=\textwidth]{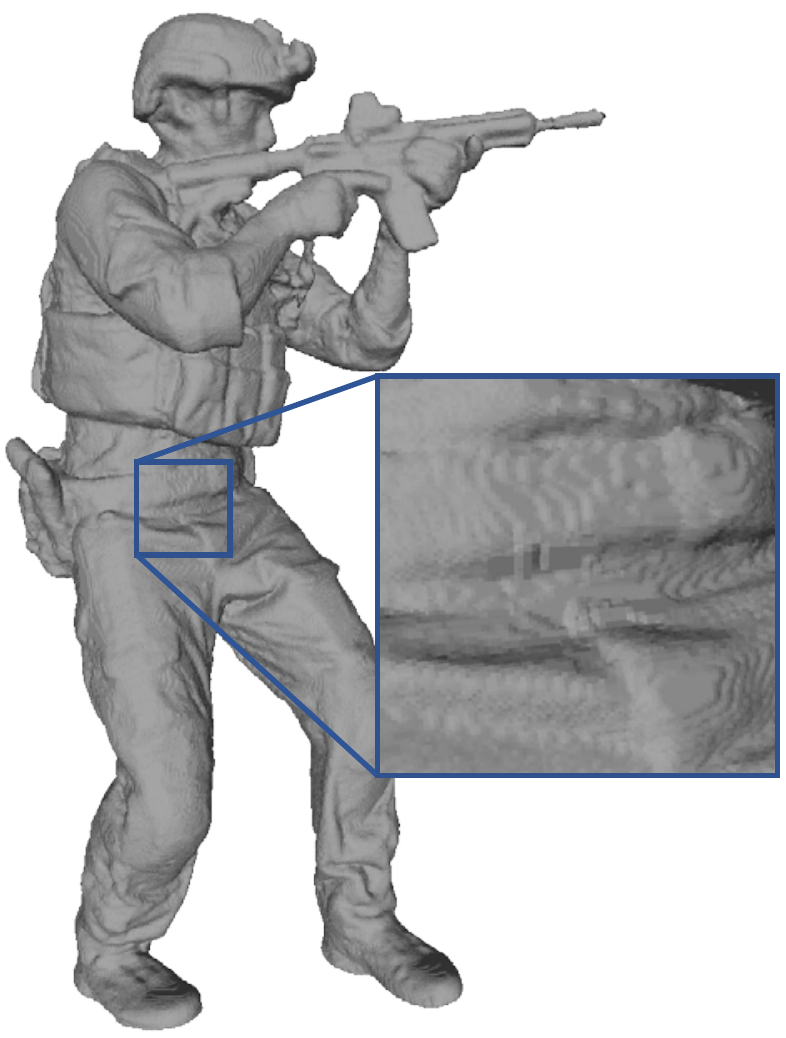} \\
			\includegraphics[width=\textwidth]{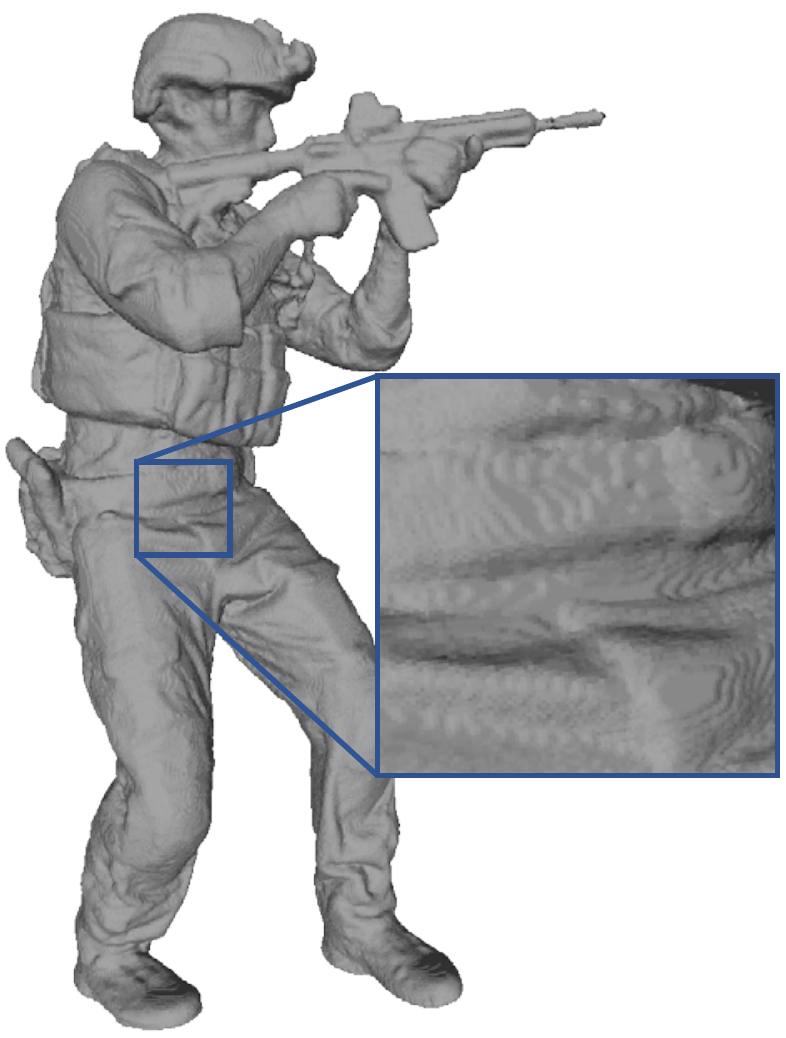} \\
		\vspace{-0.15in}
		\end{minipage}
	}
	%\hspace{-0.1in}
		\subfigure[f=9]{
		\begin{minipage}[b]{0.094\textwidth}
			\includegraphics[width=\textwidth]{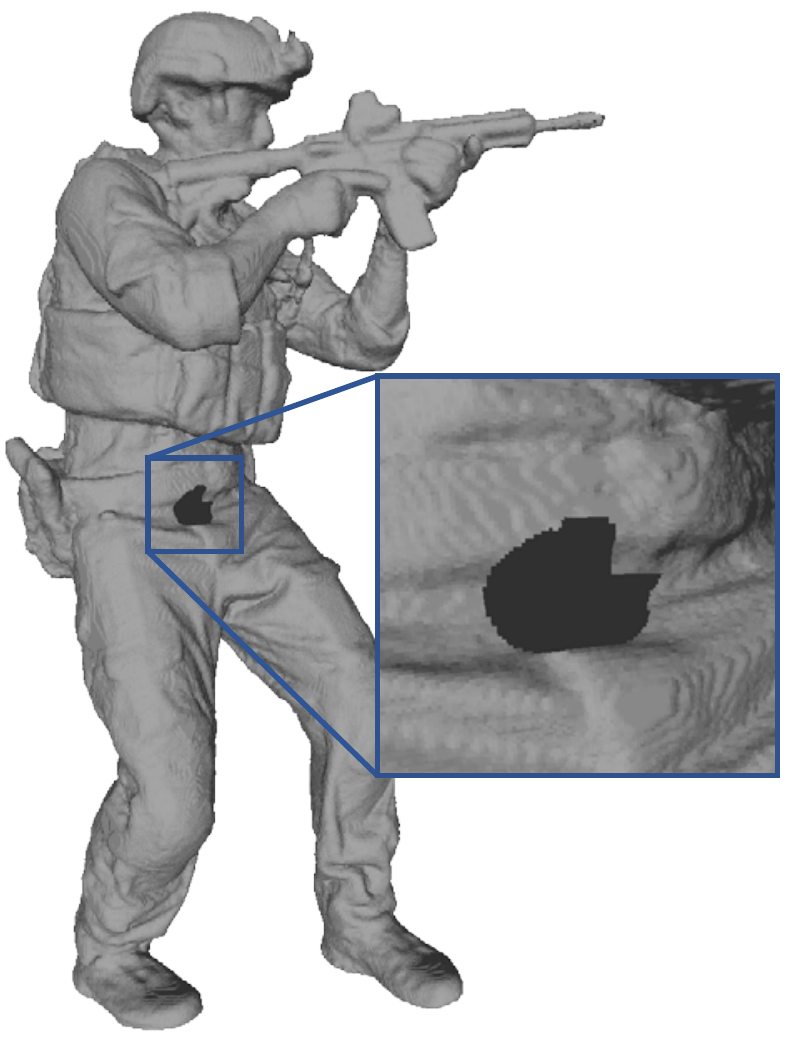} \\
			\includegraphics[width=\textwidth]{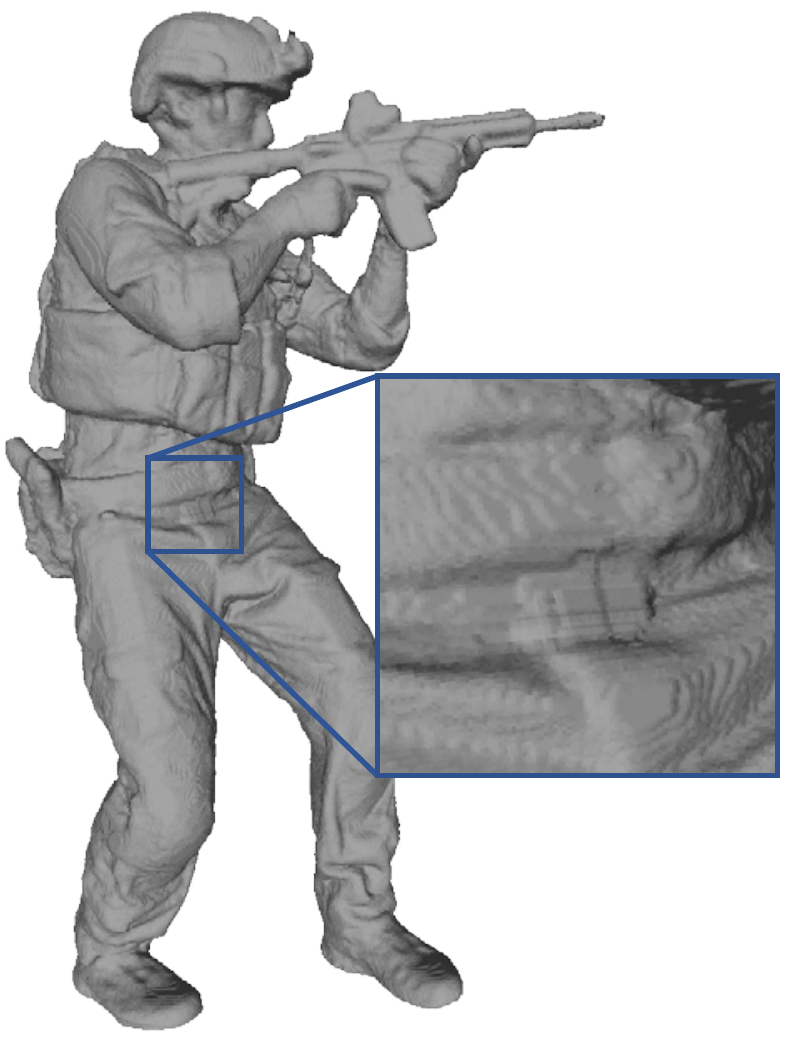} \\
			\includegraphics[width=\textwidth]{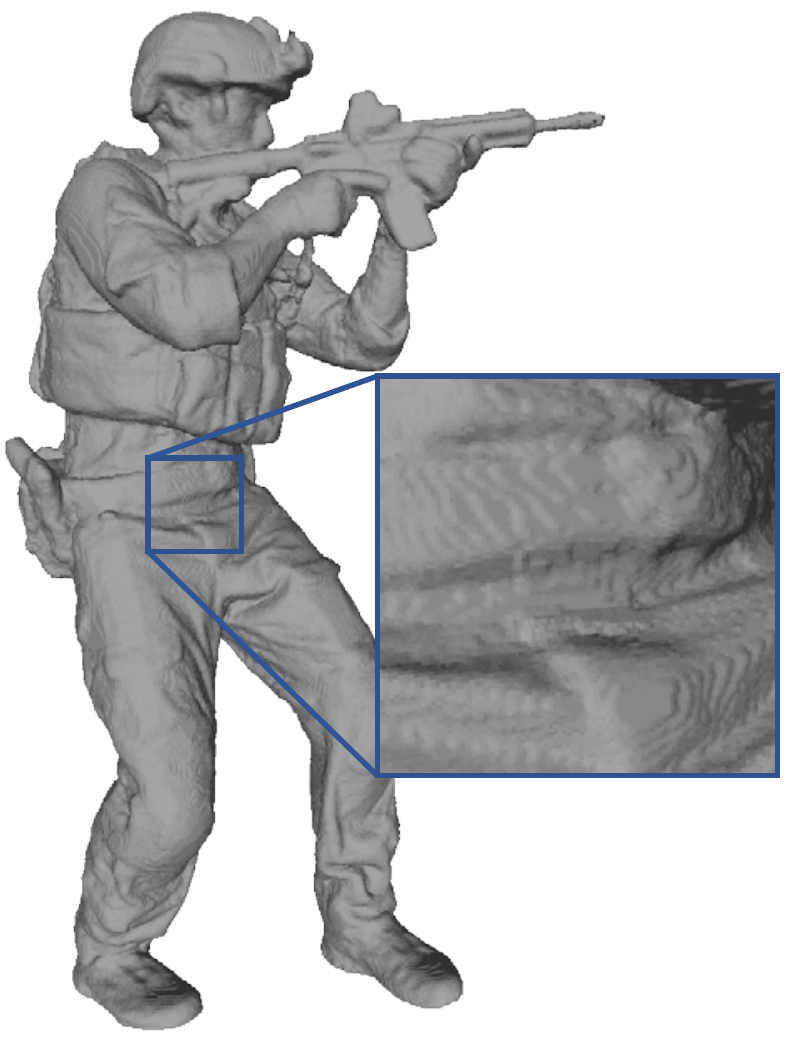} \\
			\includegraphics[width=\textwidth]{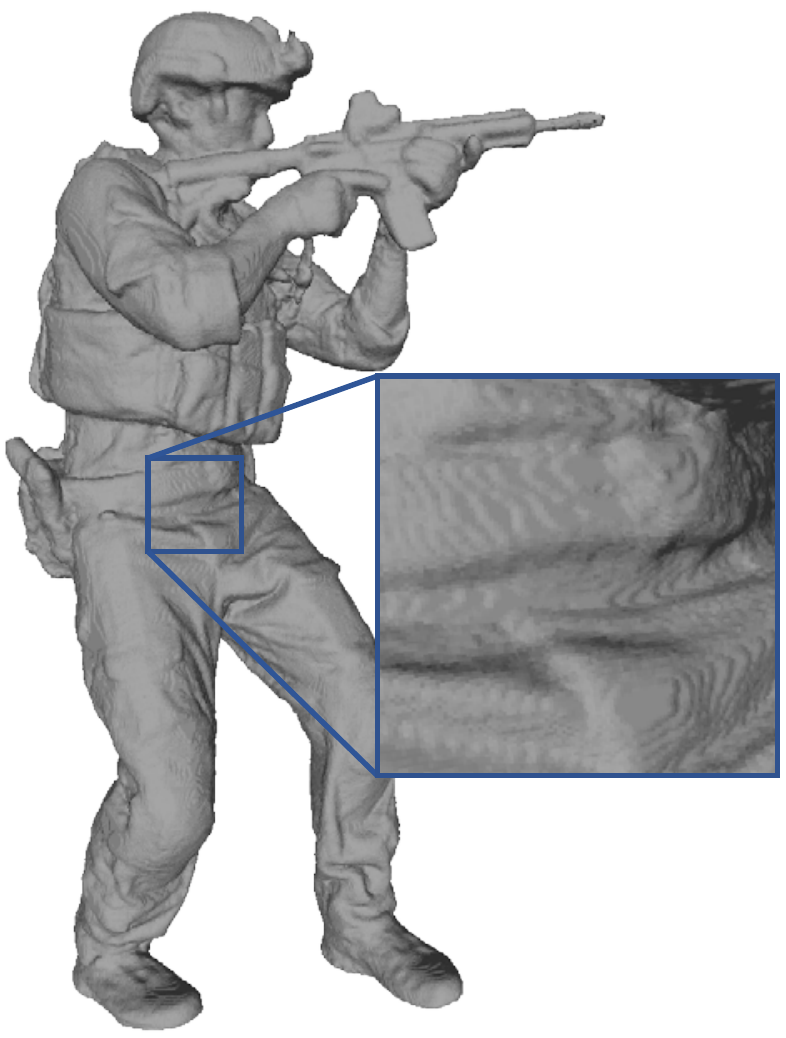} \\
		\vspace{-0.15in}
		\end{minipage}
	}
	%\hspace{-0.1in}
		\subfigure[f=11]{
		\begin{minipage}[b]{0.094\textwidth}
			\includegraphics[width=\textwidth]{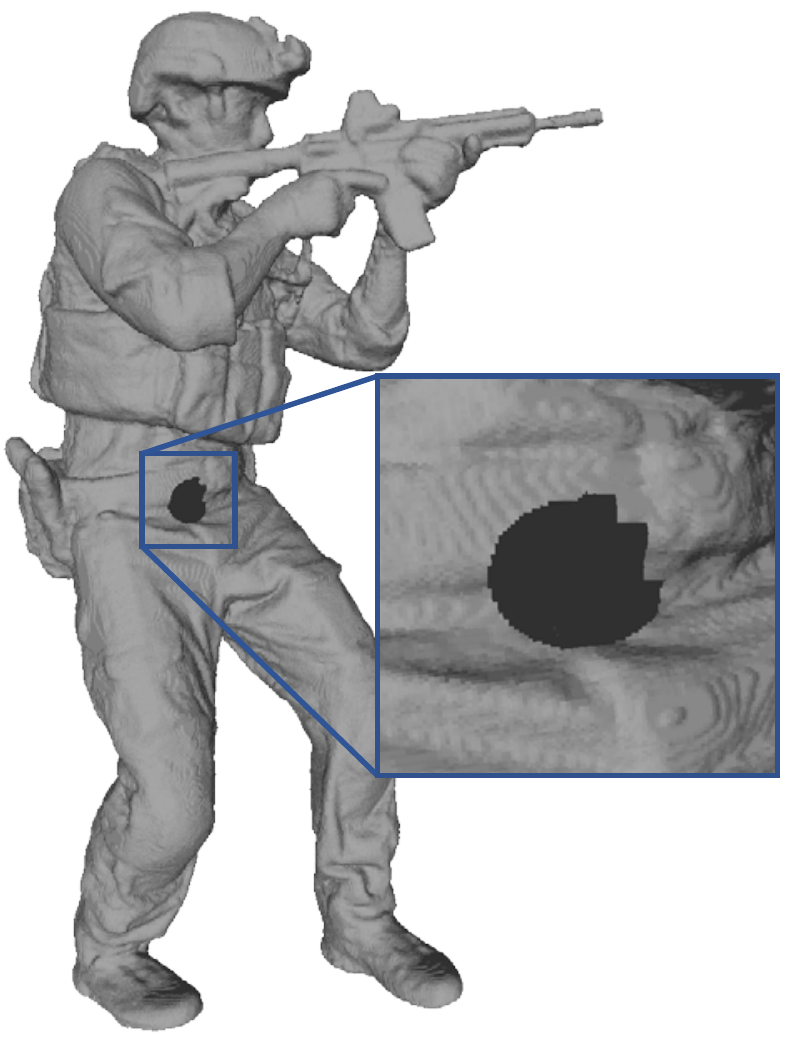} \\
			\includegraphics[width=\textwidth]{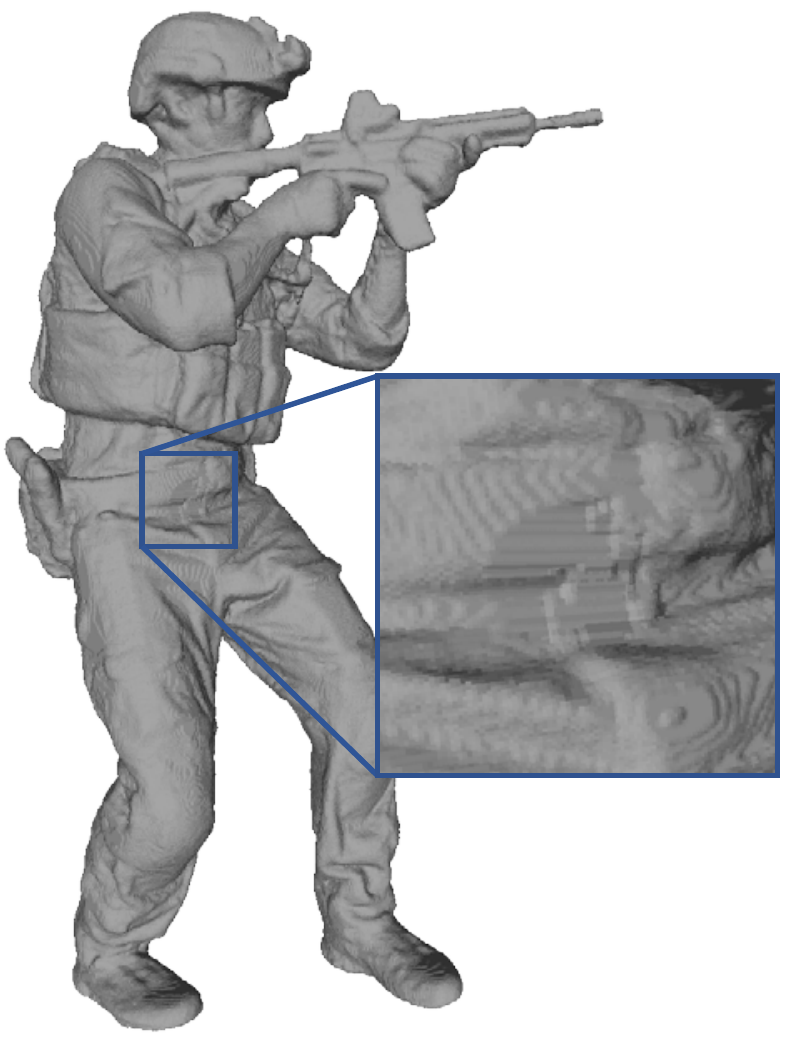} \\
			\includegraphics[width=\textwidth]{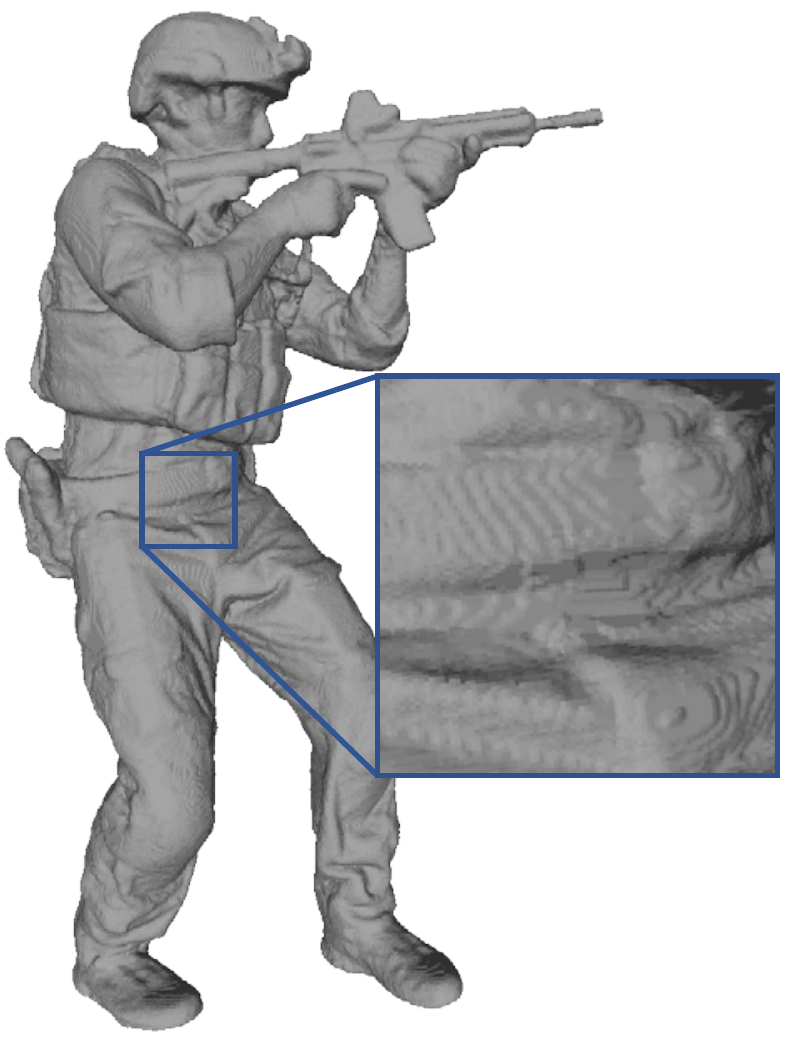} \\
			\includegraphics[width=\textwidth]{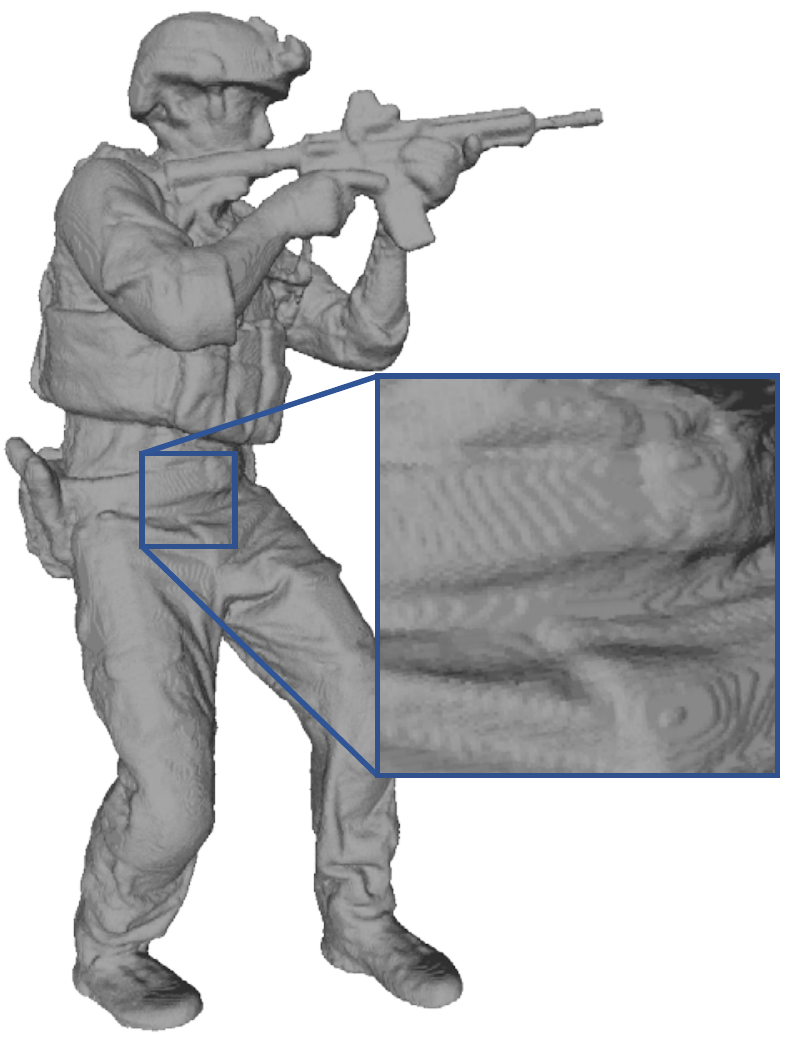} \\
		\vspace{-0.15in}
		\end{minipage}
	}
	%\hspace{-0.1in}
		\subfigure[f=13]{
		\begin{minipage}[b]{0.094\textwidth}
			\includegraphics[width=\textwidth]{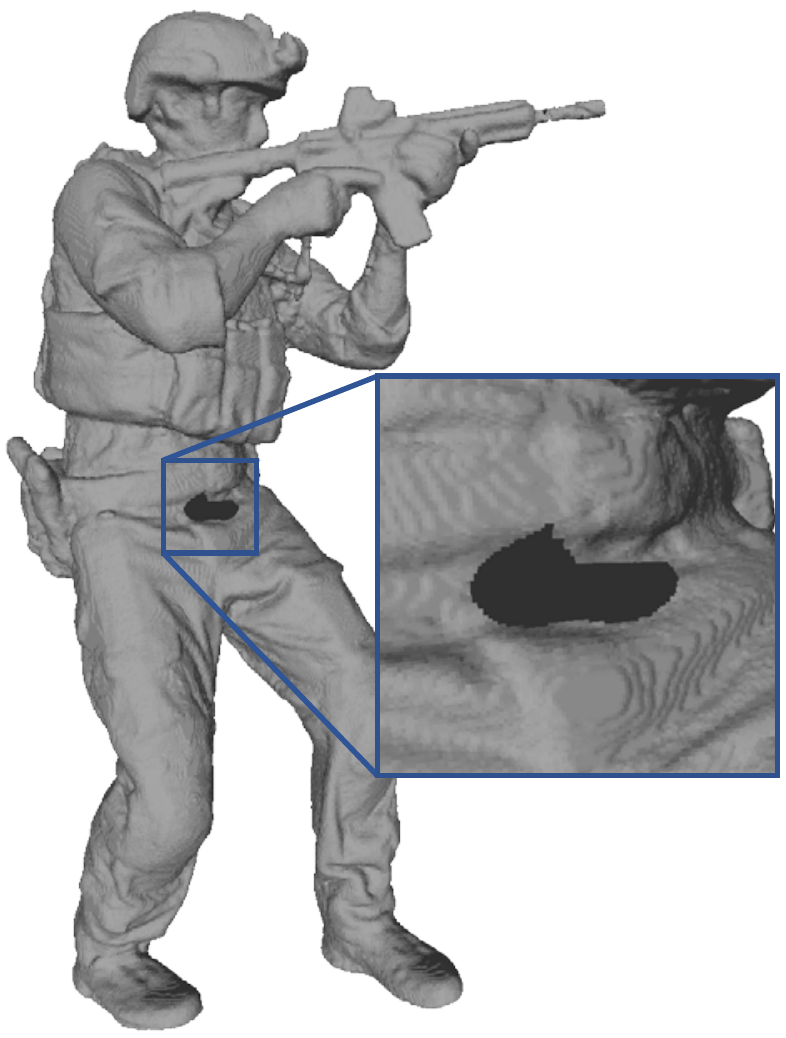} \\
			\includegraphics[width=\textwidth]{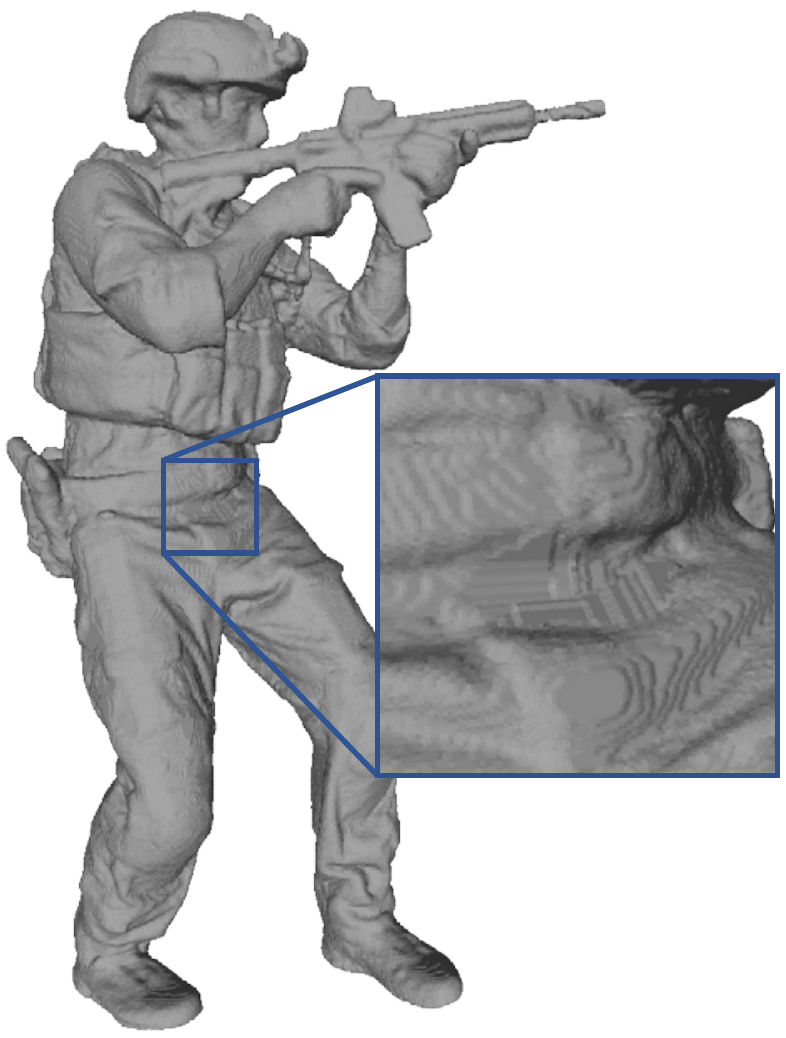} \\
			\includegraphics[width=\textwidth]{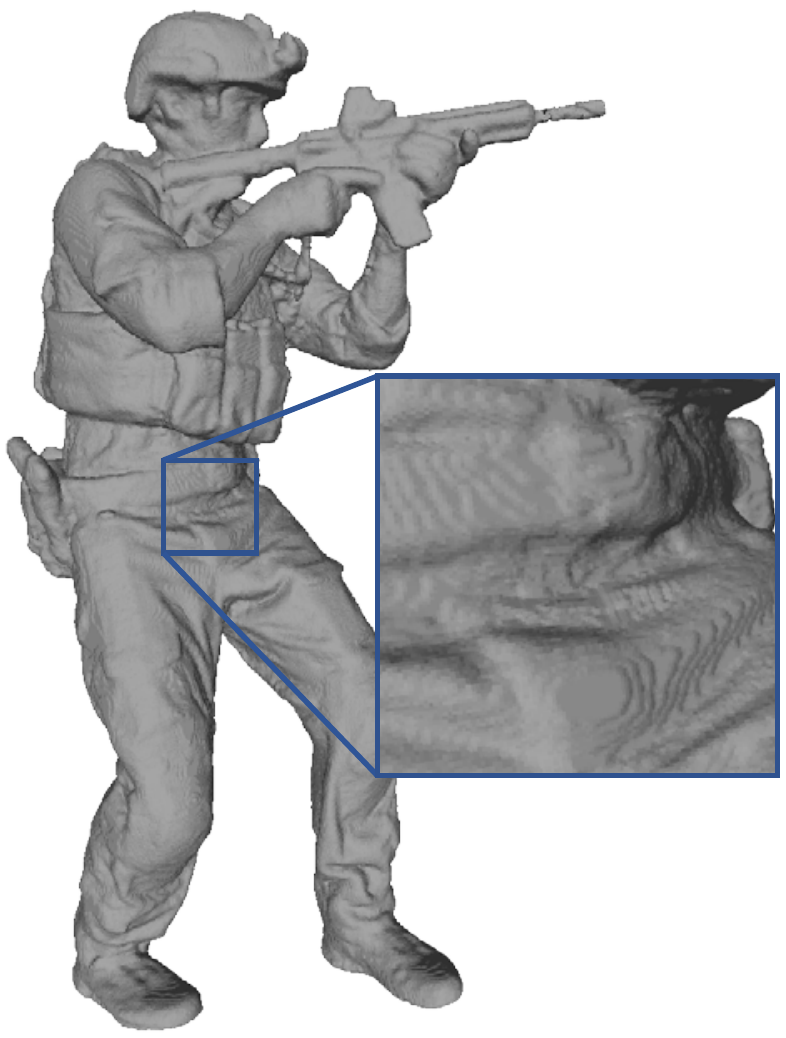} \\
			\includegraphics[width=\textwidth]{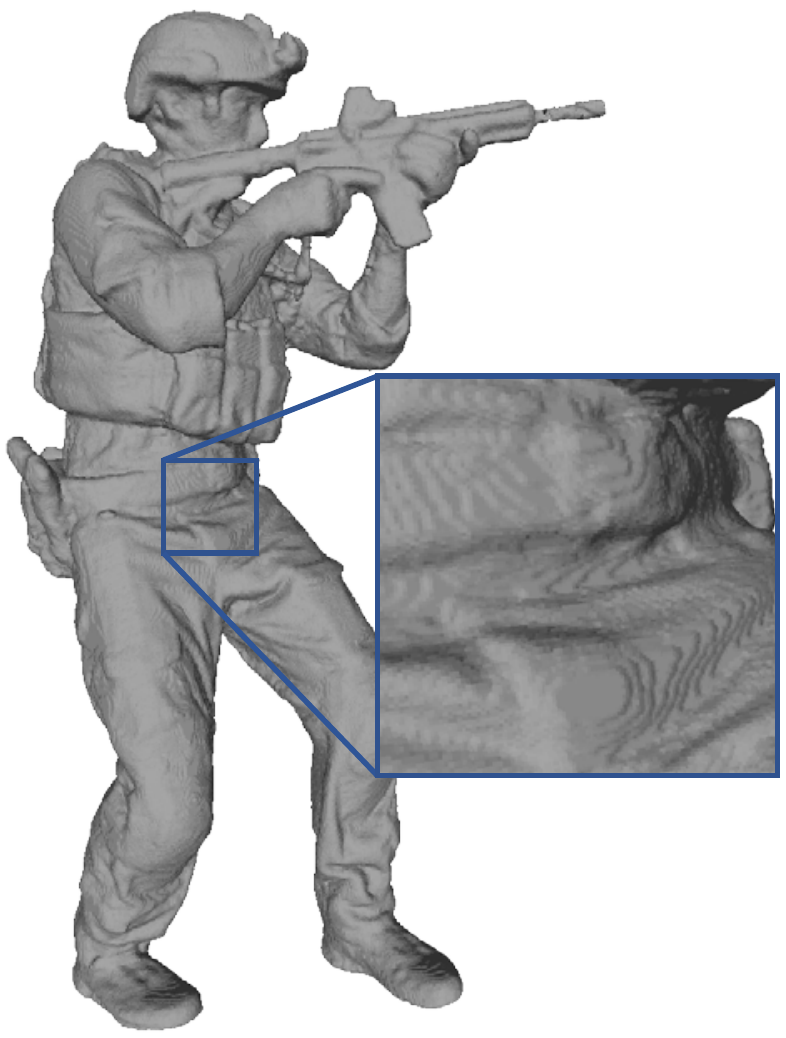} \\
		\vspace{-0.15in}
		\end{minipage}
	}
	%\hspace{-0.1in}
		\subfigure[f=16]{
		\begin{minipage}[b]{0.094\textwidth}
			\includegraphics[width=\textwidth]{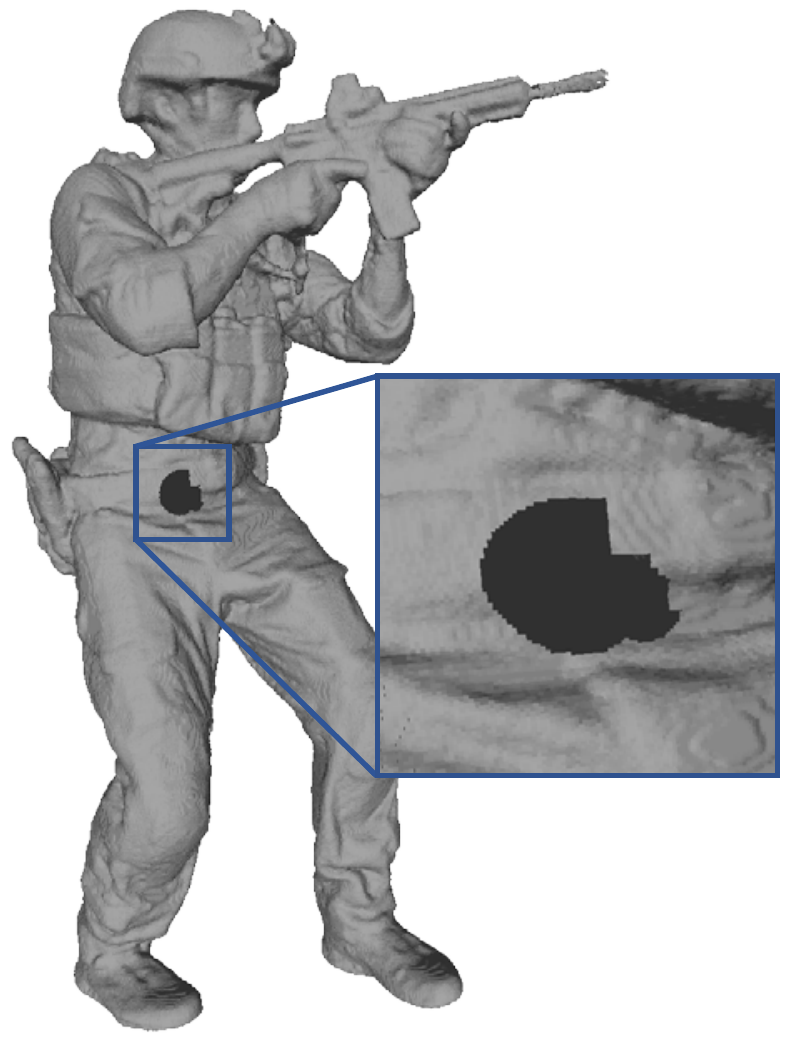} \\
			\includegraphics[width=\textwidth]{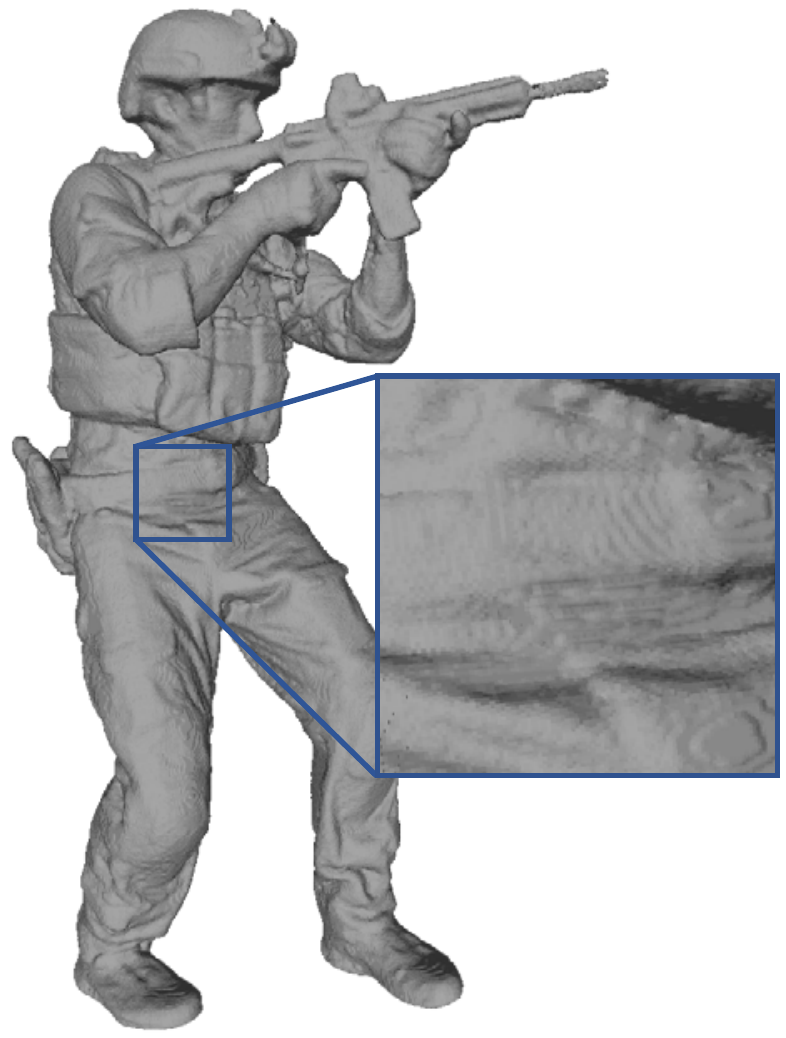} \\
			\includegraphics[width=\textwidth]{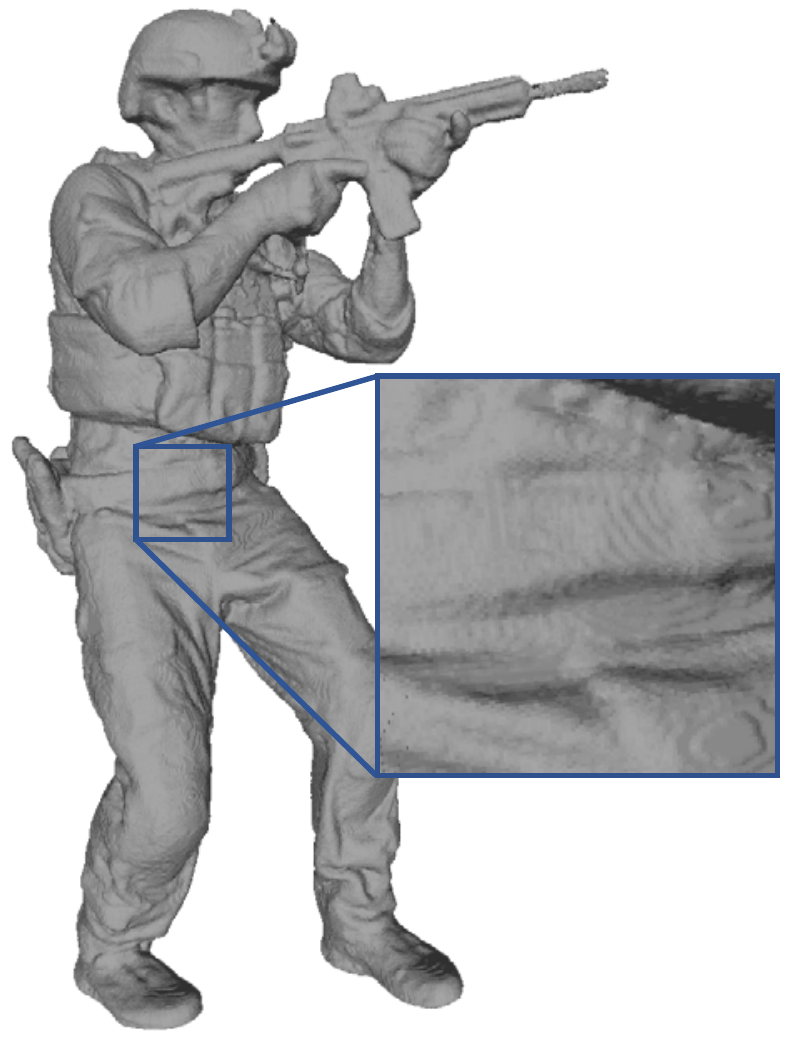} \\
			\includegraphics[width=\textwidth]{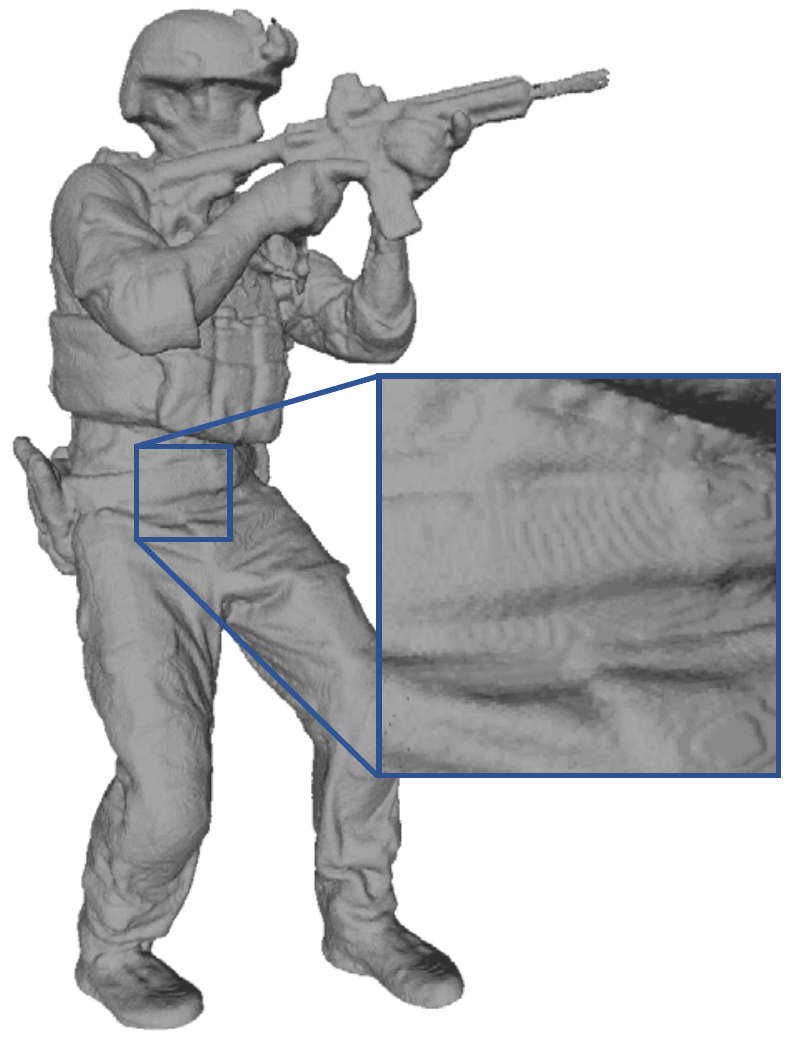} \\
		\vspace{-0.15in}
		\end{minipage}
	}
	%\hspace{-0.1in}
		\subfigure[f=18]{
		\begin{minipage}[b]{0.094\textwidth}
			\includegraphics[width=\textwidth]{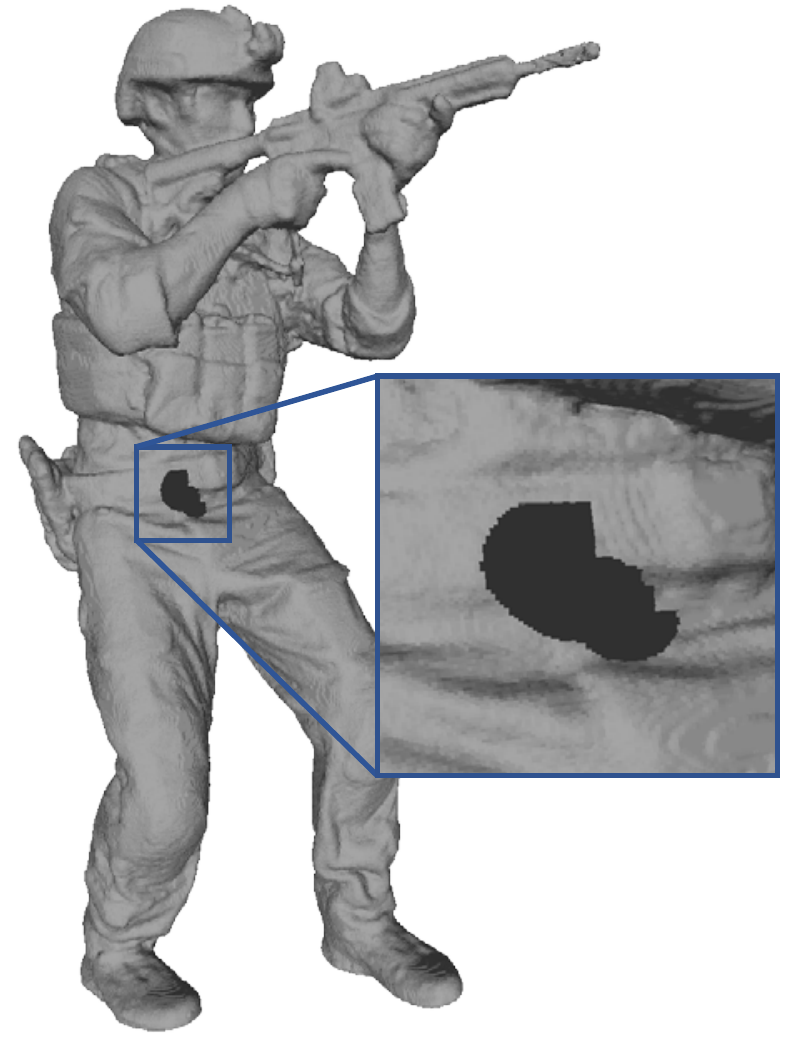} \\
			\includegraphics[width=\textwidth]{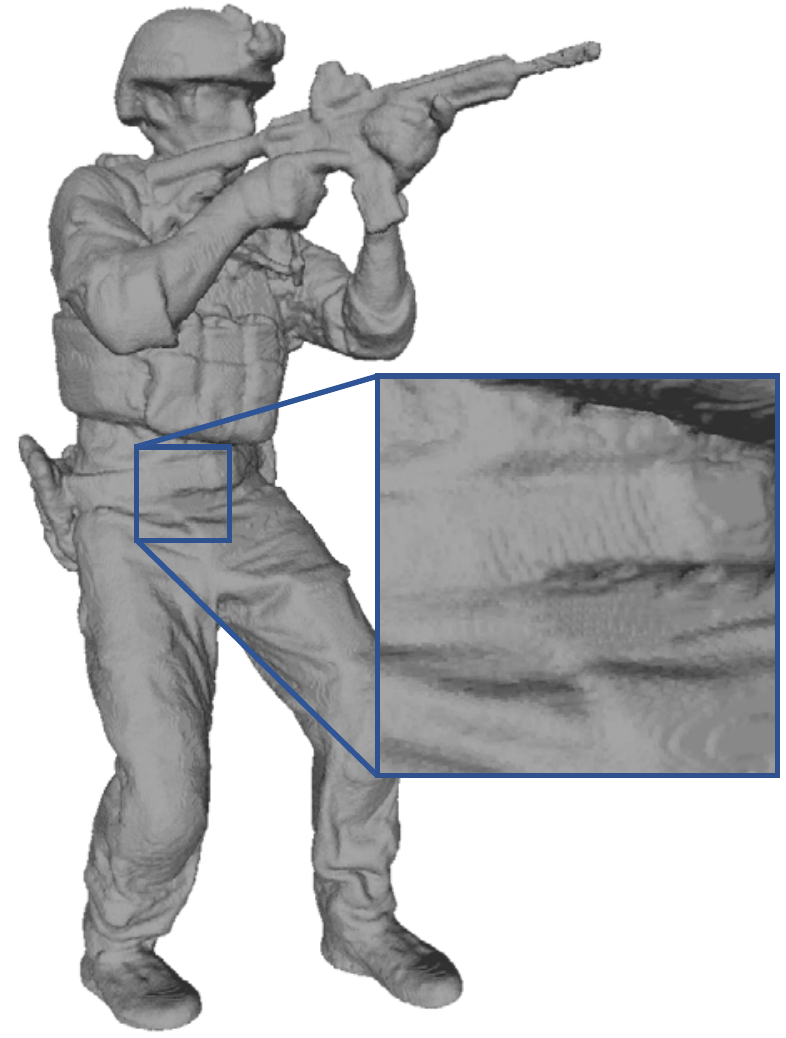} \\
			\includegraphics[width=\textwidth]{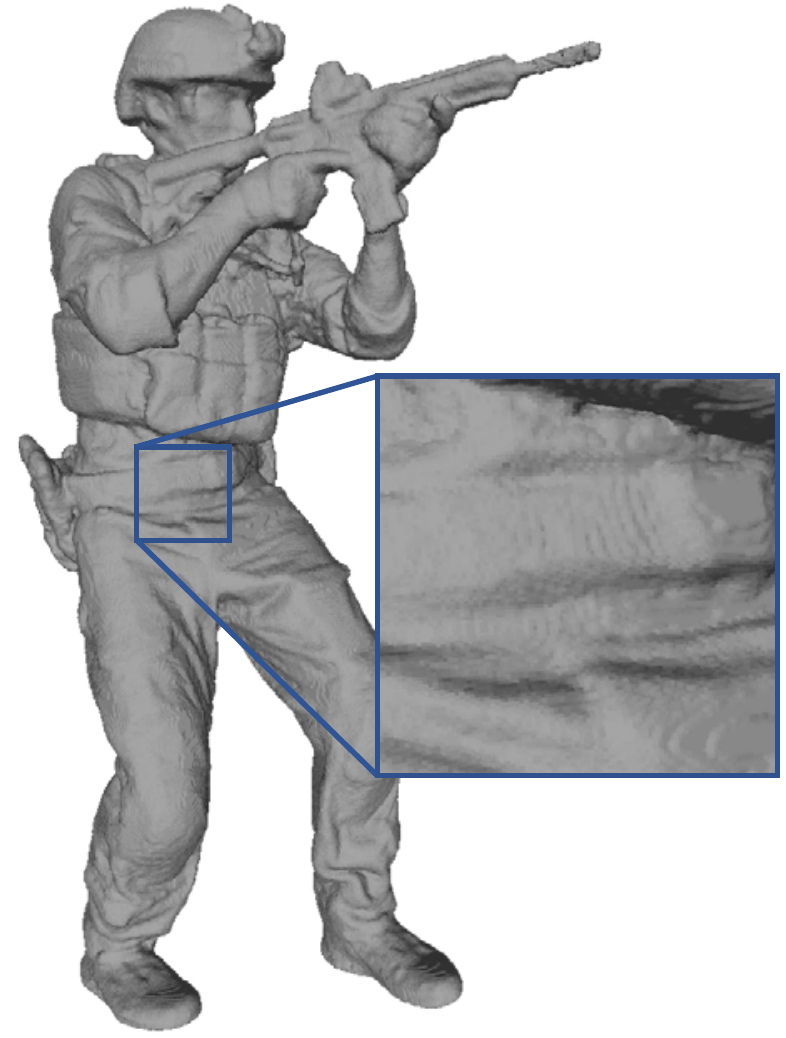} \\
			\includegraphics[width=\textwidth]{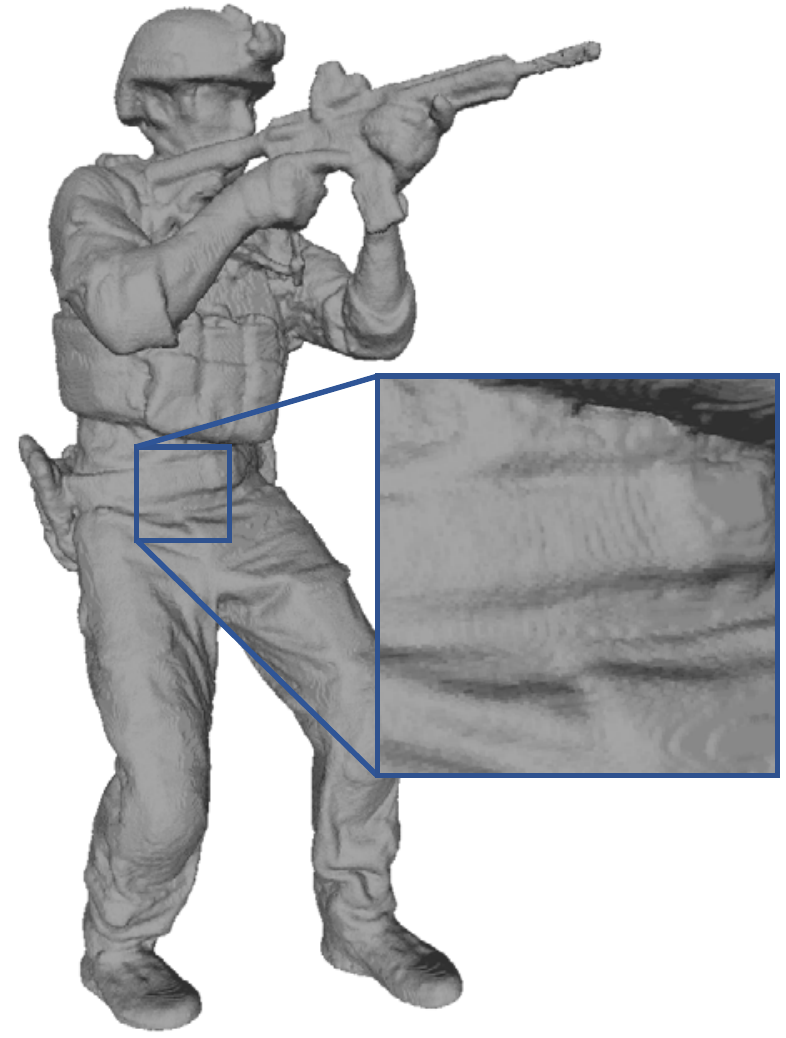} \\
		\vspace{-0.15in}
		\end{minipage}
	}
	\hspace{-0.05in}
		\subfigure[f=21]{
		\begin{minipage}[b]{0.094\textwidth}
			\includegraphics[width=\textwidth]{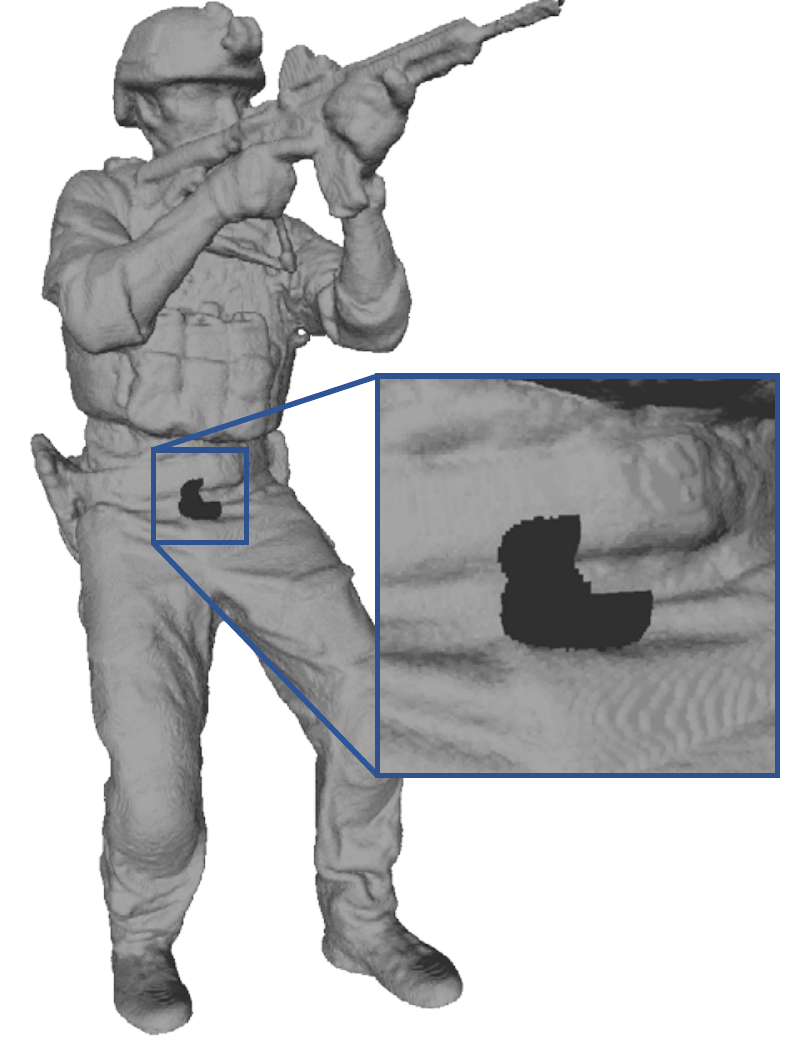} \\
			\includegraphics[width=\textwidth]{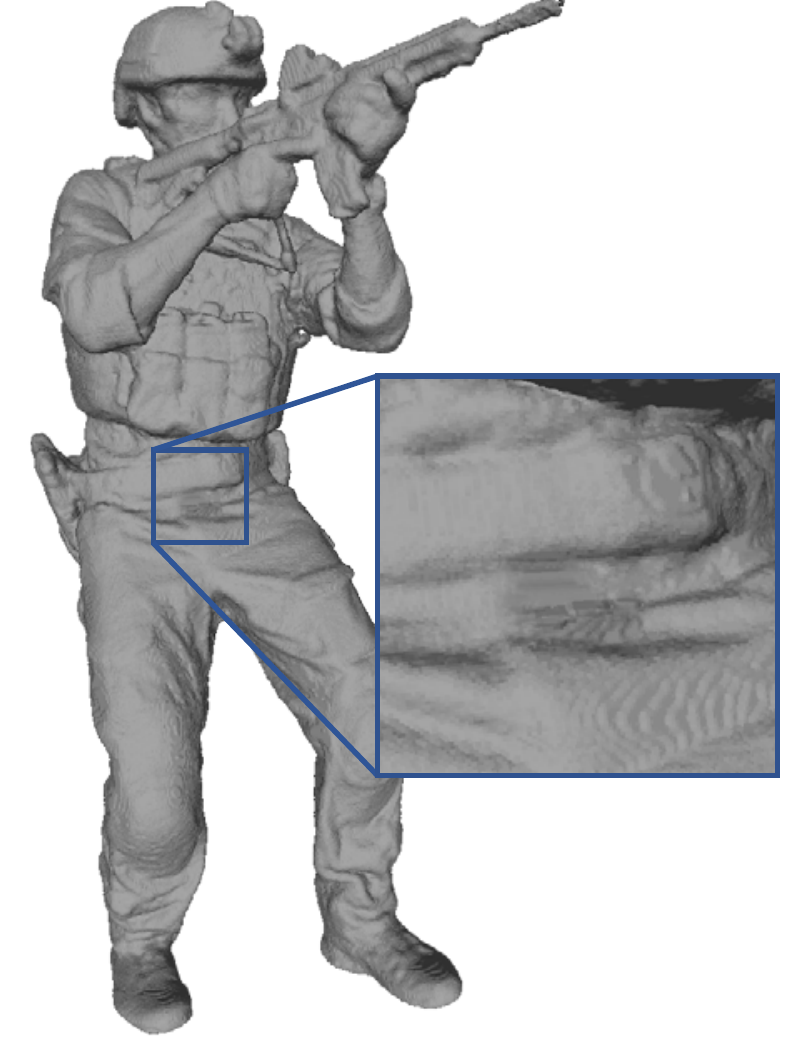} \\
			\includegraphics[width=\textwidth]{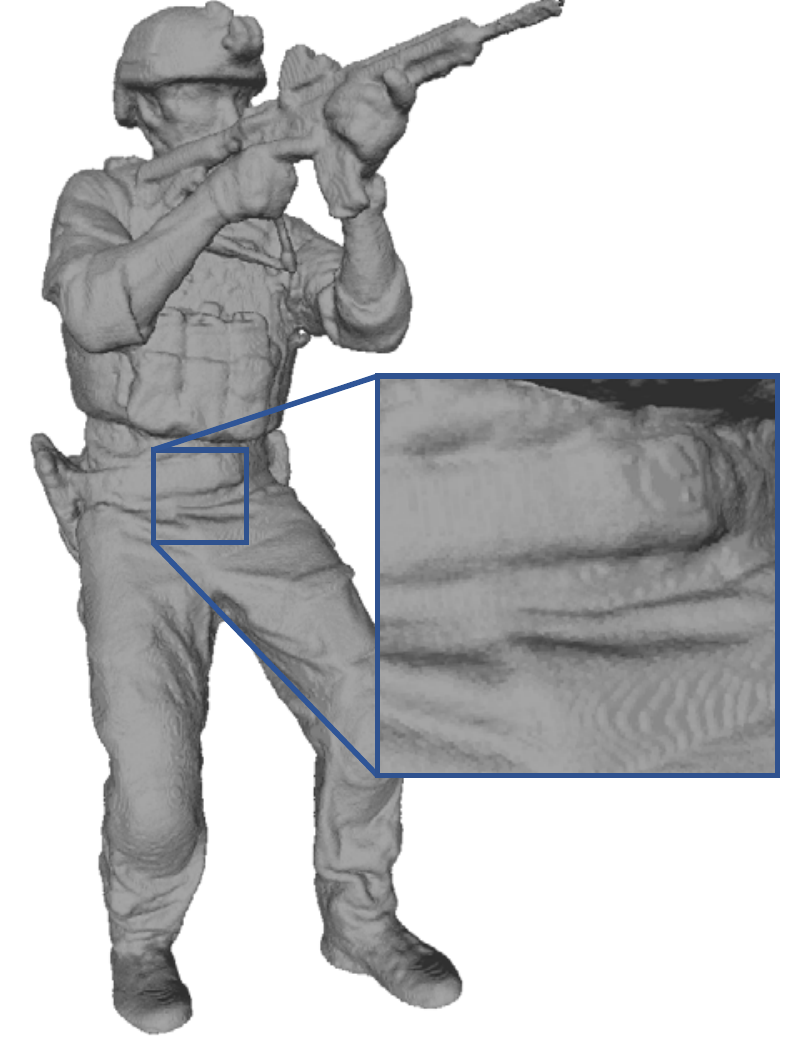} \\
			\includegraphics[width=\textwidth]{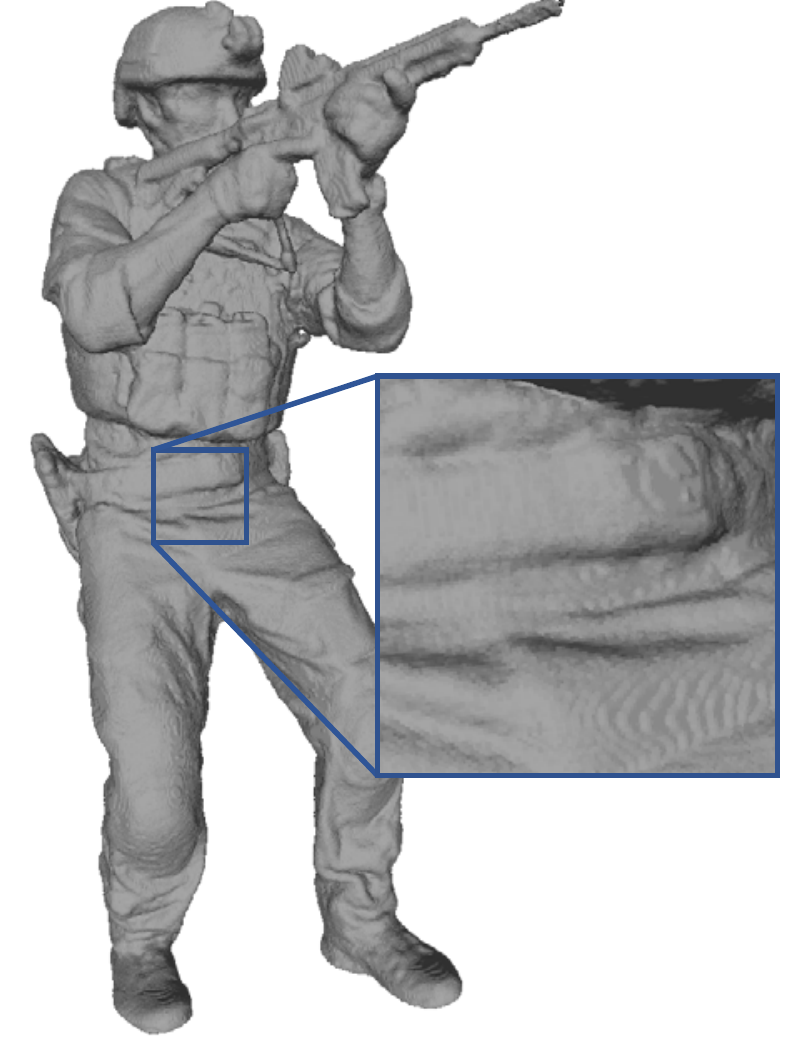} \\
		\vspace{-0.15in}
		\end{minipage}
	}
	\vspace{-0.1in}
	\caption{Several frames of the inpainting results from different methods for \textit{Soldier} with the synthetic holes magnified.}
	\label{fig:result2}
	\vspace{-0.13in}
\end{figure*}

\renewcommand\arraystretch{1.3}
\begin{table}[h]
	\vspace{-0.2in}
	\caption{Performance Comparison in GPSNR (dB)}
	\label{tb:gpsnr}
	\vspace{-0.08in}
	\begin{center}
	\small
		\begin{tabular}{|p{1.53cm}<{\centering}|p{1.68cm}<{\centering}|p{1.35cm}<{\centering}|p{0.98cm}<{\centering}|p{1.11cm}<{\centering}|}
			\hline
			{}				&  Meshlab\cite{Meshlab}	&  Lozes\cite{Lozes15}	&  Hu\cite{Fu18TIP}	&  Proposed\\
			\hline
			Longdress		&  11.7926	&  30.3883			&  41.5686			&  \textbf{43.1301}\\
			\hline
			Loot			&  16.4451	&  27.3715			&  40.1546			&  \textbf{47.5648}\\
			\hline
			Redandblack		&  13.1772	&  24.4810			&  33.9921			&  \textbf{39.0103}\\
			\hline
			Soldier			&  17.4697	&  23.1571			&  34.5062			&  \textbf{42.2980}\\
			\hline
			UlliWegner		&  24.9424	&  31.5455			&  41.3037			&  \textbf{45.8411}\\
			\hline
		\end{tabular}
	\end{center}
	\vspace{-0.2in}
\end{table}

\begin{table}[h]
	\vspace{-0.2in}
	\caption{Performance Comparison in NSHD ($ \times10^{-7} $)}
	\label{tb:nshd}
	\vspace{-0.08in}
	\begin{center}
	\small
		\begin{tabular}{|p{1.53cm}<{\centering}|p{1.68cm}<{\centering}|p{1.35cm}<{\centering}|p{0.98cm}<{\centering}|p{1.11cm}<{\centering}|}
			\hline
			{}				&  Meshlab\cite{Meshlab}	&  Lozes\cite{Lozes15}	&  Hu\cite{Fu18TIP}	&  Proposed\\
			\hline
			Longdress		&  24.8631	&  7.1362			&  2.9174			&  \textbf{0.9131}\\
			\hline
			Loot			&  14.1925	&  8.9410			&  3.1102			&  \textbf{0.3549}\\
			\hline
			Redandblack		&  22.8300	&  9.6233			&  5.9856			&  \textbf{1.9324}\\
			\hline
			Soldier			&  17.1376	&  10.0044			&  5.2145			&  \textbf{1.2057}\\
			\hline
			UlliWegner		&  11.2509	&  6.7370			&  1.9658			&  \textbf{0.6362}\\
			\hline
		\end{tabular}
	\end{center}
	\vspace{-0.1in}
\end{table}

\textbf{Subjective results.} Further, Fig.~\ref{fig:result1} and Fig.~\ref{fig:result2} demonstrate the subjective inpainting results for real and synthetic holes, respectively. Due to the page limit, we show several representative frames compared with one competitive method~\cite{Lozes15}. For the real holes in Fig.~\ref{fig:result1} (a), which are fragmentary, the results of \cite{Lozes15} exhibit artificial contours, since it attempts to connect the boundary of the hole region with planar structures without smoothing. However, the inpainted results are not smooth in the local region, which indicates the temporal inconsistency to some extent. In comparison, our results shown in row 3 of Fig.~\ref{fig:result1} demonstrate that our method is able to inpaint holes with appropriate geometry structure and smoothness over the hole region. Besides, since we leverage the inter-frame correlation, our inpainted regions show good consistency across neighboring frames.

In Fig.~\ref{fig:result2}, we synthesize holes in the point cloud sequence \textit{Soldier}, with more complex and bigger holes than the real holes in Fig.~\ref{fig:result1}. We observe that \cite{Lozes15} covers the missing area with stripy geometry, which introduces wrong geometry around the holes compared to the ground truth. Also, the contents look incoherent among the consecutive frames. In comparison, our results shown in row 3 of Fig.~\ref{fig:result2} are almost the same as the ground truth, and exhibit consistency between neighboring frames. This gives credits to the intra-frame self-similarity, the inter-frame consistency and graph-signal smoothness prior.

\vspace{-0.1in}
\section{Conclusion}
\label{sec:conclude}
\vspace{-0.1in}
We propose a novel 3D dynamic point cloud inpainting method. The key idea is to enforce both intra-frame self-similarity and inter-frame consistency in point cloud sequences. 
Given a target cube with holes inside, we propose to efficiently search for its intra-frame self-similar cube and inter-frame corresponding cubes.  
We then cast dynamic point cloud inpainting as a quadratic programming problem, based on the searched source cubes and regularized by the graph-signal smoothness prior. 
Experimental results show that our algorithm significantly outperforms three competing methods. Future works include the extension to inpainting the attributes of dynamic point clouds.

\begin{footnotesize}
\begin{spacing}{0.4}
\bibliographystyle{IEEEbib}
\bibliography{ref}
\end{spacing}
\end{footnotesize}

\end{document}